\documentclass[aps,pre,twocolumn,10pt]{revtex4-2}

\usepackage{times}
\usepackage{graphicx}
\usepackage{placeins}
\usepackage{setspace}
\usepackage{calc}
\usepackage{calrsfs}
\usepackage{mathtools}
\usepackage{ulem} % only load once
\usepackage{amssymb,amsmath,amsfonts}
\usepackage{graphics,epsfig}
\usepackage{color,bm}
\usepackage{xcolor}
\usepackage{tikz}
\usetikzlibrary{calc}
\usetikzlibrary{arrows.meta,shapes.geometric,arrows,decorations.markings}
\usepackage{yfonts}
\usepackage{comment}

\usepackage[colorlinks=true, linkcolor=blue, citecolor=blue, breaklinks=true]{hyperref}

\DeclareMathAlphabet{\pazocal}{OMS}{zplm}{m}{n}

\definecolor{lime}{HTML}{A6CE39}
\DeclareRobustCommand{\orcidicon}{\hspace{-4pt}
\begin{tikzpicture}
\draw[lime, fill=lime] (0,0)
circle [radius=0.16]
node[white] {\hspace{0.1mm}{\fontfamily{qag}\selectfont \tiny ID}};
\draw[white, fill=white] (-0.07,0.1)
circle [radius=0.01];
\end{tikzpicture}
\hspace{-3.2mm}
}
\foreach \x in {A, ..., Z}{\expandafter\xdef\csname
orcid\x\endcsname{\noexpand\href{https://orcid.org/\csname orcidauthor\x\endcsname}
{\noexpand\orcidicon}}
}
\begin{document}
%================================================================================
%=========================  Author Information ==================================
%================================================================================

\title{Synchronization with Annealed Disorder and Higher-Harmonic Interactions in Arbitrary Dimensions: When Two Dimensions Are Special}

\author{Rupak Majumder\orcidA{}}
\email{rupak.majumder@tifr.res.in}
\affiliation{Department of Theoretical Physics, Tata Institute of Fundamental Research, Homi Bhabha Road, Mumbai 400005, India}

\author{Shamik Gupta\orcidC{}}
\email{shamik.gupta@theory.tifr.res.in}
\affiliation{Department of Theoretical Physics, Tata Institute of Fundamental Research, Homi Bhabha Road, Mumbai 400005, India}

\date{\today}

\begin{abstract}
The impact of disorder on collective phenomena depends crucially on whether it is quenched or annealed. In synchronization problems, quenched disorder in higher-dimensional Kuramoto models is known to produce unconventional dimensional effects, including a striking odd–even dichotomy: synchronization transitions are continuous in even dimensions and discontinuous in odd dimensions. By contrast, the impact of annealed disorder has received comparatively little attention. Here we study a $D$-dimensional Kuramoto model with both fundamental and higher-harmonic interactions under annealed disorder, and develop an arbitrary-dimensional center-manifold framework to analyze the nonlinear dynamics near the onset of collective behavior. We show that annealed disorder fundamentally alters the role of dimensionality. With fundamental coupling alone, it completely removes the odd–even dichotomy, yielding continuous synchronization transitions with universal mean-field scaling in all dimensions. Higher-harmonic interactions preserve this universality while rendering the synchronization transition tunable between continuous and discontinuous. At the same time, they give rise to a novel, correlation-driven transition between a symmetry-protected incoherent phase and a symmetry-broken state lacking global synchronization, which is therefore invisible to the conventional Kuramoto order parameter. This transition is continuous in two dimensions but discontinuous in higher dimensions, revealing an emergent and previously-unrecognized special role of two dimensions.
\end{abstract}

\maketitle

\tableofcontents

\begin{figure*}
    \centering
    \includegraphics[width=\linewidth]{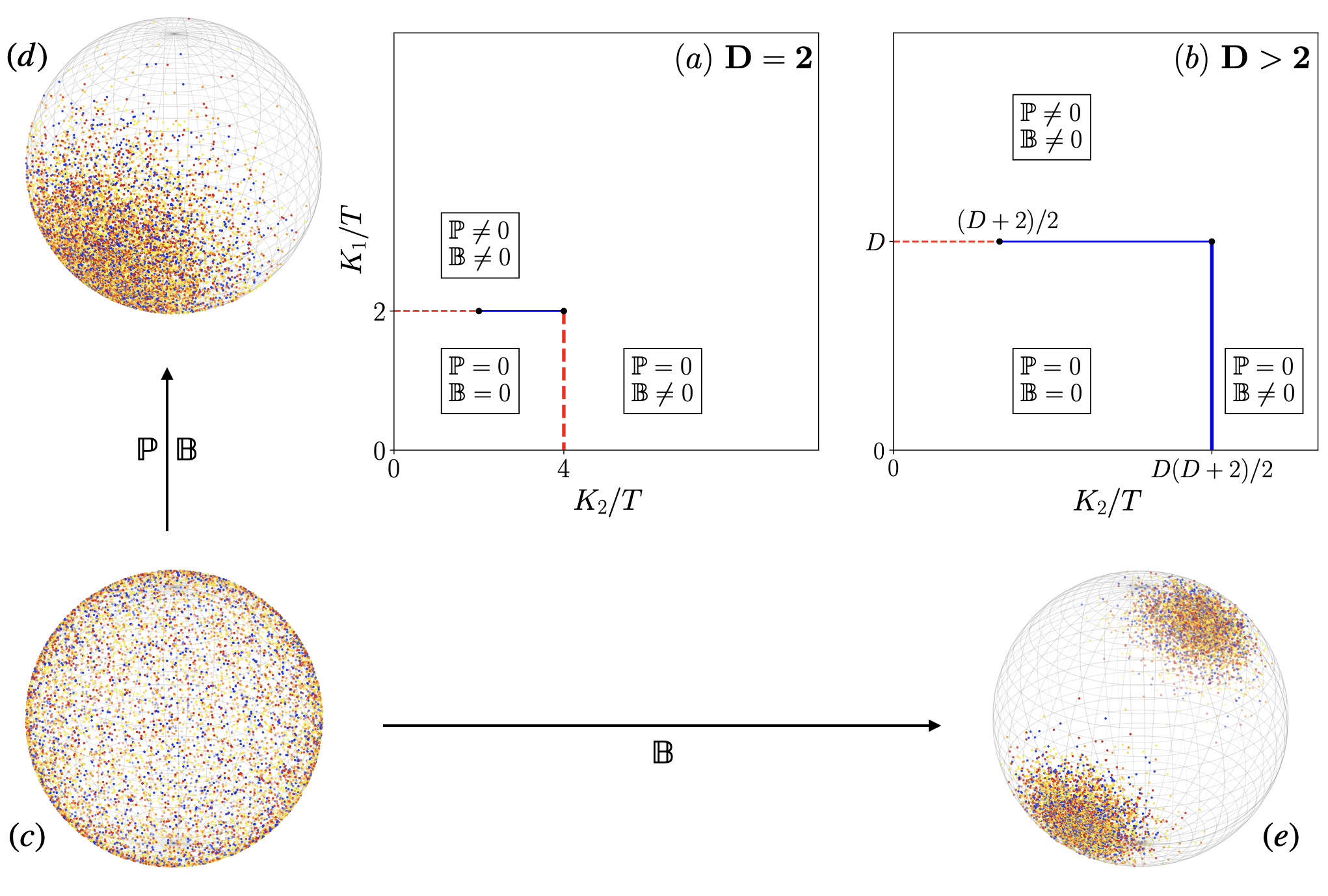}
    \caption{Phase diagram of the model is presented in (a) for $D=2$ and in (b) for $D>2$ in the $K_1$ (strength of the fundamental interaction) -- $K_2$ (strength of the higher harmonic interaction) plane . The red dashed lines denote continuous transition and the blue solid lines denote discontinuous transition. The enclosed region by the blue solid and the red dashed lines and the $K_1$, $K_2$ axes denote the parameter region in which the uniformly incoherent state (c), denoted by $\mathbb{P}=\mathbb{B}=0$, is linearly stable. The horizontal red dashed and the blue solid lines denote the transition points at which the order parameters $\mathbb{P}$ and $\mathbb{M}$ shows a transition to a synchronized phase (d). The verticle red dashed /blue solid line denotes the transition points of the uniformly incoherent state to a symmetry broken state without global order (e) denoted by $\mathbb{P}=0$ and $\mathbb{B}\neq0$. }
    \label{fig:schematic}
\end{figure*}

\section{\label{sec:level1}Introduction}
Disorder plays a central role in shaping collective phenomena, with its consequences depending crucially on whether it is quenched or annealed. In systems with quenched disorder, impurities or random fields are frozen on the timescale of the dynamics, whereas in annealed disorder, the disorder variables fluctuate dynamically alongside the system’s intrinsic degrees of freedom. Quenched disorder is known to profoundly alter thermodynamic, critical, and dynamical behavior, influencing universality classes and even the existence and nature of ordered phases in both classical and quantum systems~\cite{PhysRevLett.35.1399, PhysRevLett.37.1364, PhysRevLett.69.534, PhysRevLett.111.097202}. It can give rise to striking phenomena such as glassy dynamics~\cite{edwards1975theory,parisi1979infinite,bernaschi2024quantum} and non-self-averaging behavior~\cite{aharony1996absence,wiseman1995lack}. By contrast, determining how annealed disorder influences collective phenomena has remained an enduring theoretical challenge~\cite{belitz2000annealed,um2024validity,ferraro2025exact}. The distinction between quenched and annealed disorder has thus emerged as a unifying theme across a wide range of contemporary research areas, including quantum gravity~\cite{baldwin2020quenched}, condensed matter physics~\cite{PhysRevB.90.024202,xu2025quenched}, active matter~\cite{PhysRevLett.129.188004}, biophysics~\cite{liu2005quantification}, and sociophysics~\cite{jkedrzejewski2017person, nowak2021discontinuous}.

Synchronization among interacting nonlinear oscillators provides a paradigmatic and broadly relevant setting in which both quenched and annealed disorder arise naturally~\cite{buck1966biology,Gray1989,neda2000sound}. As a prototypical form of collective order emerging from disorder, synchronization has been observed across an extraordinary range of systems, including biological~\cite{winfree1980geometry,Cobb1995,Rubchinsky2012,Sarfati2020FireflySync,Weber2020YeastSync,10.7554/eLife.78908} and chemical oscillators~\cite{zaikin1970concentration,doi:10.1126/science.1070757,doi:10.1126/science.1166253}, power grids~\cite{PhysRevLett.102.074101,motter2013spontaneous,sajadi2022synchronization}, quantum platforms~\cite{PhysRevLett.111.234101,PhysRevResearch.1.033012,PhysRevLett.125.013601,PhysRevResearch.5.033209}, and even financial systems~\cite{WALTI201196,GUO2021101934,Baltakiene2022}. Over the years, the Kuramoto model has emerged as a minimal and unifying theoretical framework for understanding the onset of synchronization in large populations of interacting non-linear oscillators~\cite{Kuramoto}. The model describes phase-only oscillators with distributed natural frequencies that are globally coupled through the sine of their phase differences. Its analytical tractability has made it a cornerstone for the study of nonequilibrium phase transitions driven by interactions and disorder~\cite{Strogatz1990,st,STROGATZ20001,RevModPhys77137,OA,PhysRevResearch.2.023057,Gupta_2014,pikovsky2015dynamics,gupta2018statistical,PhysRevLett.134.197201,Laing2009}. A central result is the emergence of a phase transition between incoherent and synchronized states at a finite coupling strength, whose nature depends sensitively on the form of disorder.

Recent work has shown, however, that even minimal generalizations of the Kuramoto model can give rise to qualitatively new phenomena. In particular, higher-dimensional extensions have been introduced in which each oscillator is represented by a $D$-dimensional unit vector rather than a single phase, with the conventional Kuramoto model corresponding to $D=2$. As argued in Ref.~\cite{Sarthak2019}, these models naturally arise in the context of flocking birds,  fish schools, flying drones, opinion dynamics as well as in a surprising context of neutrino oscillations~\cite{PhysRevD.58.073002,PhysRevD.74.105010, PhysRevD.65.053011}. Analyses of these models in the presence of quenched disorder have revealed a striking odd–even dimensional dichotomy: synchronization transitions are continuous in even dimensions, while in odd dimensions, they become discontinuous and occur at vanishing coupling strength~\cite{Sarthak2019,ZouWei2025,FARIELLO2024114431,chandra2019observing}. This unexpected result highlights the profound role of dimensionality in shaping collective dynamics~\cite{chandra2019complexity,BARIONI2021111090,barioni2021ott,lipton2021kuramoto}. Dimensional effects have since been explored in numerous extensions, including models with three-body interactions~\cite{huh2024critical,cz3j-lwcg,PhysRevLett.127.258301,doi:10.1137/20M1369002,Lohe_2022}, nonuniform~\cite{PhysRevE.109.034215,Lee_2024,WEI2023105554,markdahl2021almost}, adaptive~\cite{PhysRevLett.125.194101,PENG2025134448,PhysRevLett.130.107202}, and matrix-valued couplings~\cite{buzanello2022matrix,Wang_2024,PhysRevE.107.044205}, among many others~\cite{zou2025synchronization,PhysRevResearch.7.023103,ling2022effects,DAI2021110888,MARKDAHL2020108736,JACIMOVIC2025134953}. Notably, the overwhelming majority of these studies focus on quenched disorder, leaving open the fundamental question of whether dimensional dichotomies persist when disorder becomes annealed. In a related class of swarming models, annealed disorder has recently been studied both in the presence of quenched disorder in $D=3$~\cite{PhysRevE.104.014216} and in its absence in arbitrary dimensions~\cite{PhysRevE.106.064601}.

A second, equally fundamental gap concerns the role of interaction structure. In two-dimensional synchronization models, higher-harmonic interactions are known to qualitatively modify phase transitions, alter critical behavior, and induce multistability~\cite{PhysRevLett.77.1406,DAIDO199624,KOMAROV201418,rupakfinite2025}. Whether such effects persist, compete with, or fundamentally reshape dimensional phenomena in higher-dimensional systems remains unknown, particularly when higher harmonics and annealed disorder act simultaneously.

From a theoretical standpoint, a key challenge is development of analytical tools that go beyond linear stability analysis and are capable of addressing nonlinear dynamics in higher-dimensional synchronization models. In two dimensions, the Kuramoto model admits a powerful center-manifold framework~\cite{Crawford1994,CrawfordDavies1999,Chiba2015,BarreMetivier2016,ChibaMedvedevMizuhara2018,MetivierGupta2019,MetivierWetzelGupta2020,rupakfinite2025}, which enables a systematic characterization of phase transitions beyond linear stability analysis. In higher dimensions and in the absence of annealed disorder, substantial progress has been achieved through generalized Ott–Antonsen-type ansatz~\cite{chandra2019complexity,barioni2021ott}, allowing for exact low-dimensional reductions. However, such approaches do not extend naturally to systems with annealed disorder. Developing an analytical framework capable of addressing this gap is essential for understanding the nonlinear structure of the dynamics, determining the nature of phase transitions, and characterizing behavior near criticality.

In this work, we address these challenges by studying a $D$-dimensional Kuramoto model with both fundamental and higher-harmonic couplings in the presence of annealed disorder. The paper is organized as follows. In Sec.~\ref{sec:level1b}, we present a concise overview of our principal findings, highlighting the key physical insights and their implications for synchronization in the presence of annealed disorder and higher-harmonic interactions. In Sec. ~\ref{sec: model 1 formulation}, we introduce the model in detail, starting from its formulation and proceeding to define the relevant dynamical observables that characterize macroscopic order. We also discuss the underlying symmetries of the system in Sec.~\ref{sec: model 1 formulation}B and elucidate their important consequences for the structure of the dynamics and the form of the effective description.

 In Sec.~\ref{sec: FP model 1}, we reformulate the microscopic dynamics in terms of a Fokker–Planck equation governing the evolution of the probability density in the thermodynamic limit, thereby providing a closed macroscopic description of the system. Building on this framework, in Sec.~\ref{sec: CM} we develop a nontrivial generalization of the center-manifold reduction to arbitrary spatial dimension $D$, enabling a systematic analysis of the nonlinear dynamics near the onset of collective behavior.

The resulting reduced evolution equations for the two order parameters, capturing, respectively, global synchronization and correlation-driven ordering, are derived and analyzed in Secs.~\ref{sec: CM}D and~\ref{sec: CM}E. These sections further establish the nature of the associated phase transitions, including their dependence on dimensionality and interaction structure. We conclude in Sec.~\ref{sec: conclusion}. Technical details are provided in Appendices~\ref{app: -1}–\ref{app: 21}.

\section{\label{sec:level1b}Main Results}

Motivated by the need to systematically characterize phase transitions in higher-dimensional Kuramoto models with both fundamental and higher-harmonic interactions in the presence of annealed disorder, we develop a nontrivial generalization of the two-dimensional Kuramoto center-manifold analysis to arbitrary dimension. This approach goes beyond linear stability theory and provides direct access to the nonlinear growth of instabilities near criticality, allowing for a rigorous determination of the nature of the associated phase transitions.

Our first main result concerns the fate of the odd–even dimensional dichotomy previously reported for higher-dimensional Kuramoto models with quenched disorder. We show that, in the presence of annealed disorder and fundamental coupling alone, this dichotomy disappears entirely. The transition from incoherence to synchronization is strictly continuous in all dimensions, and the synchronization order parameter exhibits universal mean-field scaling with critical exponent $\beta=1/2$, independent of dimensionality. Annealed disorder thus restores universality across dimensions, consistent with recent findings in related swarming models~\cite{PhysRevE.104.014216,PhysRevE.106.064601}. We further find that on introducing higher-harmonic interactions, the transition between incoherent and synchronized phases can be tuned from continuous to discontinuous in any dimension by varying the strength of such interactions up to a dimension-dependent critical value.

Our second main result is that higher-harmonic interactions give rise to a transition from a symmetry-protected incoherent phase to a symmetry-broken state that does not exhibit global synchronization. This transition appears to be a novel feature of the $D$-dimensional Kuramoto model, not previously observed in higher dimensions. Across this transition, the conventional synchronization order parameter remains identically zero and therefore fails to detect the ordering. Instead, the transition is characterized by a correlation matrix encoding ensemble-averaged correlations between different components of the oscillator vectors. This correlation-driven transition is continuous in two dimensions but becomes discontinuous in all higher dimensions, leading to a rich phase diagram summarized in Fig.~\ref{fig:schematic}. 

Taken together, these results show that within the framework of $D$-dimensional Kuramoto models, annealed disorder gives rise to a significantly richer phase-transition landscape than might be expected. In its presence, and with only fundamental interaction, the standard synchronization transition remains continuous across all dimensions. However, the inclusion of higher-harmonic interactions fundamentally alters this picture: the transition may become either continuous or discontinuous, depending on both the oscillator dimensionality and the strength of the higher harmonics. In this sense, while annealed disorder suppresses discontinuous transitions in the case of purely fundamental coupling, it is no longer sufficient to prevent them once higher-harmonic interactions are introduced. Another remark is that as we consider only annealed disorder, frustration effects associated with quenched randomness are absent.

This dimensional dependence of the correlation-driven phase transition is the third main result of our work. Thus, in presence of higher-harmonic interactions, dimensional effects reemerge in a qualitatively new manner, making $D=2$ uniquely distinguished from higher dimensions. At the nonlinear level, the special role of dimensionality is further reflected in the scaling of the leading nonlinear terms near the correlation-driven transition. In two dimensions, the instability is governed by a quadratic nonlinearity, whereas in higher dimensions, the leading contribution scales with a non-integer exponent equal to $3/2$. Taken together, these results identify $D=2$ as uniquely special in the presence of higher-harmonic interactions, in sharp contrast to the fully universal behavior observed when such interactions are absent.

 We now comment on potential applications of our results. One physical way to motivate inclusion of annealed disorder in the dynamics is to consider the Kuramoto-like models to be in interaction with the external environment modeled as a heat bath in equilibrium at a given temperature. Indeed, no real system evolves in isolation and is inevitably under non-negligible interaction with its
environment, making an otherwise deterministic dynamics stochastic. On the theoretical front, a major challenge is to carefully analyse the influence of the environment and to devise schemes to
minimize such an influence if it proves to be detrimental. Study of such issues lies at the heart of statistical mechanics~\cite{Huang}, and has been pursued extensively since early days. One simplifying assumption regarding the evolution of the environment variables is that of time scale separation, whereby such variables are assumed to evolve on a much faster time scale with respect to the system, and hence, the environment may be considered to be in equilibrium at all times. The noise or the disorder introduced into the dynamics of the system by the environment qualifies as annealed disorder. Our work demonstrates that environmental effects, manifested as annealed disorder, play a decisive role in governing both the static and dynamic behavior of Kuramoto-like models, with particular emphasis on how the dimensionality of the oscillators fundamentally shapes these dynamics. It is interesting to note that noisy models such as ours, albeit for the case $D=2$, have been invoked recently in modeling  experimental data of human neuronal
cultures~\cite{lucente2025novelkuramotomodelinhibition}.

From a theoretical perspective, our generalized center-manifold reduction yields closed evolution equations for the macroscopic order parameters. More broadly, our work establishes a unified and versatile dynamical framework for analyzing collective phenomena in higher-dimensional Kuramoto-type models, opening the door to systematic studies of disorder, finite-size effects, and further extensions in higher-dimensional Kuramoto-like models.

 In summary, our results establish three key conclusions: (i) annealed disorder removes the odd–even dichotomy and restores universal mean-field behavior; (ii) higher-harmonic interactions reintroduce tunability between continuous and discontinuous transitions; and (iii) a novel correlation-driven phase transition emerges, revealing a uniquely special role of two dimensions.

\section{Formulation of the model \label{sec: model 1 formulation}}

We first recall the set-up of the usual two-dimensional Kuramoto model. The model in presence of annealed disorder and higher-harmonic interactions involves a system of $N$ phase-only oscillators. The phase of the $j$-th oscillator at time $t$ is denoted by the angle variable $\theta_j(t) \in [0,2\pi)$, with $j=1,2,\ldots,N$. Considering the interaction between the oscillators to be all-to-all, the time evolution is given by~\cite{sakaguchi1988cooperative}:
\begin{eqnarray}
    \frac{d \theta_j}{dt} &=& \omega_j + \frac{K_1}{N}\sum_{k = 1}^N \sin(\theta_k-\theta_j) + \frac{K_2}{N}\sum_{k = 1}^N \sin2(\theta_k-\theta_j)\nonumber \\
    &&+ \sqrt{2T}~\eta_j(t),\label{eq: 2D Kuramoto 2}
\end{eqnarray}
where the annealed disorder $\eta_j(t)$ is a Gaussian, white noise acting upon the oscillator $j$, with the properties $\langle \eta_j(t)\rangle = 0$~and~$\langle \eta_i(t)\eta_j(t')\rangle = \delta_{ij}\delta(t-t')~\forall~i,j$. The notation $\langle \cdot \rangle$ denotes averaging over noise realizations. The coefficient $K_1$ is the strength of the fundamental coupling, i.e., the strength of the first-harmonic interaction, while $K_2$ is the strength of the second-harmonic interaction, which is a representative higher-harmonic interaction. In Eq.~\eqref{eq: 2D Kuramoto 2}, the natural frequencies $\omega_j \in (-\infty,\infty)$ of the oscillators are quenched-disordered random variables distributed according to a given probability distribution $g(\omega)$, which has a finite mean $\omega_0 \geq 0$ and a finite width $\sigma \geq 0$. Examples of such distributions are uniform, Lorentzian, Gaussian, etc. The above model may also be interpreted as a model of continuous classical XY spins with mean-field interactions, driven out-of-equilibrium, and with $T$ playing the role of the temperature (in suitable units) of a heat bath in contact~\cite{GuptaCampaRuffo2018}.

Putting $T = 0$ and $K_2=0$ in Eq.~\eqref{eq: 2D Kuramoto 2} reduces the system to the original Kuramoto model~\cite{Kuramoto} with only first-harmonic interaction. Reference~\cite{Sarthak2019} has recently generalized this limit of the model to general $D$ dimensions, where $D=2$ corresponds to the Kuramoto model. In line with this work, in this section, we will generalize the noisy Kuramoto model with first-harmonic and second-harmonic interaction, as described by Eq.~\eqref{eq: 2D Kuramoto 2}, to a $D$-dimensional noisy Kuramoto model. Before considering the general $D$-dimensional case, let us first write Eq.~\eqref{eq: 2D Kuramoto 2} in a convenient vector form in two dimensions. In the context of Eq.~\eqref{eq: 2D Kuramoto 2}, we may consider each oscillator as a point particle moving on a circle with unit radius. Let the position vector of the $j$-th oscillator from the centre of the  circle be denoted by $\vec{\boldsymbol{\sigma}}_j$. The phase $\theta_j$ thus becomes the angle that the vector $\vec{\boldsymbol{\sigma}}_j$ forms with a given axis, say, the $x$-axis. The vector $\vec{\boldsymbol{\sigma}}_j$ may then be written as $\vec{\boldsymbol{\sigma}}_j = \cos {\theta_j }~\hat{\textbf{x}}+\sin {\theta_j }~\hat{\textbf{y}}$, where $\hat{\mathrm{\textbf{x}}}$ and $\hat{\mathrm{\textbf{y}}}$ are the unit vector along the $x$- and the $y$-axis, respectively. Clearly, the magnitude of the vector is $|\vec{\boldsymbol{\sigma}}_j| = 1$. Taking derivative with respect to time, we obtain
\begin{equation}
    \frac{d \vec{\boldsymbol{\sigma}}_j}{dt} = \left(-\sin {\theta_j }~\hat{\textbf{x}}+\cos {\theta_j }~\hat{\textbf{y}}\right)\frac{d \theta_j}{dt}. \label{eq: 2D vectorisation 1}
\end{equation}

From the transformation relation between the Cartesian $(x,y)$ and the polar $(r,\theta)$ coordinates, we may express the unit vector along the radial direction as  $\hat{\textbf{r}}=\cos {\theta }~\hat{\textbf{x}}+\sin {\theta}~\hat{\textbf{y}}$ and the unit vector along the azimuthal direction as $\hat{\boldsymbol{\theta}}=-\sin {\theta }~\hat{\mathrm{\textbf{x}}}+\cos {\theta}~\hat{\mathrm{\textbf{y}}}$. On using Eq.~\eqref{eq: 2D Kuramoto 2}, Eq.~\eqref{eq: 2D vectorisation 1} may then be written as
\begin{align}
     & \frac{d \vec{\boldsymbol{\sigma}}_j}{dt} =  \mathbf{W}_j\vec{\boldsymbol{\sigma}}_j
     +\frac{K_1}{N}\sum_{k = 1}^N \sin(\theta_k-\theta_j) ~\hat{\boldsymbol{\theta}}_j \nonumber \\
     &+\frac{2K_2}{N}\sum_{k = 1}^N \cos(\theta_k-\theta_j) \sin(\theta_k-\theta_j) ~\hat{\boldsymbol{\theta}}_j+ \sqrt{2T}\mathbf{N}_j(t) \vec{\boldsymbol{\sigma}}_j,\label{eq: 2D vectorisation model1 1}
\end{align}
where $\textbf{W}_j$ is the antisymmetric frequency matrix and $\mathbf{N}_j(t)$ is the antisymmetric noise matrix, given respectively by
\begin{equation}
    \mathbf{W}_j \equiv 
    \begin{bmatrix}
        0 & -\omega_j \\
        \omega_j & 0
    \end{bmatrix},~~
    \mathbf{N}_j(t) \equiv 
    \begin{bmatrix}
        0 & -\eta_j(t) \\
        \eta_j(t) & 0
    \end{bmatrix}. \label{eq: freq noise matrix 2d}
\end{equation}
Using the properties of $\eta_j(t)$, we obtain
\begin{equation}
    \langle\mathbf{N}_j(t)\rangle = \mathbb{O}_2,~~\langle\mathbf{N}_i(t)\mathbf{N}_j(t')\rangle = -\delta_{ij}\delta(t-t')\mathbf{I}_2~\forall~i,j,
\end{equation}
where $\mathbb{O}_2$ is a $2\times2$ matrix with all elements equal to zero, while $\mathbf{I}_2$ is a $2\times2$ identity matrix. Using basic trigonometry, one can show that~\cite{Sarthak2019} $\cos(\theta_k-\theta_j) = \vec{\boldsymbol{\sigma}}_k \cdot \vec{\boldsymbol{\sigma}}_j$ and $\sin(\theta_k-\theta_j)\,\hat{\boldsymbol{\theta}}_j = \vec{\boldsymbol{\sigma}}_k-\left(\vec{\boldsymbol{\sigma}}_j \cdot \vec{\boldsymbol{\sigma}}_k\right)\vec{\boldsymbol{\sigma}}_j$. Equation~\eqref{eq: 2D vectorisation model1 1} then yields
\begin{align}
     \frac{d \vec{\boldsymbol{\sigma}}_j}{dt} &=  \mathbf{W}_j\vec{\boldsymbol{\sigma}}_j+\frac{K_1}{N}\sum_{k = 1}^N  \vec{\boldsymbol{\sigma}}_k-\frac{K_1}{N}\sum_{k = 1}^N \left(\vec{\boldsymbol{\sigma}}_j \cdot \vec{\boldsymbol{\sigma}}_k\right)\vec{\boldsymbol{\sigma}}_j \nonumber\\
     & + \frac{2 K_2}{N}\sum_{k = 1}^N \left(\vec{\boldsymbol{\sigma}}_j \cdot \vec{\boldsymbol{\sigma}}_k\right)\vec{\boldsymbol{\sigma}}_k-\frac{2 K_2}{N}\sum_{k = 1}^N \left(\vec{\boldsymbol{\sigma}}_j \cdot \vec{\boldsymbol{\sigma}}_k\right)^2\vec{\boldsymbol{\sigma}}_k\nonumber \\
     &+\sqrt{2T} ~\mathbf{N}_j(t) \vec{\boldsymbol{\sigma}}_j.\label{eq: 2D vectorisation model1 2}
    \end{align}

We may now define two order parameters, characterizing macroscopic order in the system, in the following manner:
\begin{align}
    \vec{\mathbb{P}} &\equiv  \frac{1}{N}\sum_{k = 1}^N  \vec{\boldsymbol{\sigma}}_k, \label{eq: order parameter 1}\\
    \mathbb{M} &\equiv \frac{1}{N}\sum_{k = 1}^N  \vec{\boldsymbol{\sigma}}_k \otimes \vec{\boldsymbol{\sigma}}_k \label{eq: order parameter 2}.
\end{align}
Using the above definition of the order parameters, we may rewrite Eq.~\eqref{eq: 2D vectorisation model1 2} as
\begin{align}
     \frac{d \vec{\boldsymbol{\sigma}}_j}{dt} &=  \mathbf{W}_j\vec{\boldsymbol{\sigma}}_j \nonumber +K_1  \left[\vec{\mathbb{P}}-\left(\vec{\boldsymbol{\sigma}}_j \cdot \vec{\mathbb{P}}\right)\vec{\boldsymbol{\sigma}}_j \right]\nonumber\\
     &+2 K_2\Big[\mathbb{M}-\left(\vec{\boldsymbol{\sigma}}_j \cdot \mathbb{M}\vec{\boldsymbol{\sigma}}_j\right)\mathbf{I}_2\Big]\vec{\boldsymbol{\sigma}}_j+\sqrt{2T} ~\mathbf{N}_j(t) \vec{\boldsymbol{\sigma}}_j.\label{eq: 2D mean field}
\end{align}
Note that Eq.~\eqref{eq: 2D mean field} has the structure of a single dynamical variable $\vec{\boldsymbol{\sigma}}_j$ evolving in presence of a mean ``field" $\vec{\mathbb{P}}$ and a mean ``rotation" $\mathbb{M}$ produced by all the dynamical variables. 

Before proceeding, let us list down some relevant properties of the order parameters. Clearly, the norm of the vector $\vec{\mathbb{P}}$, given by $\mathbb{P} \equiv |\vec{\mathbb{P}}| = \sqrt{\vec{\mathbb{P}} \cdot \vec{\mathbb{P}}}$, satisfies $0\leq \mathbb{P}\leq 1$, as follows from the fact that $|\vec{\boldsymbol{\sigma}}_k| = 1~\forall~k$. In the fully-synchronized phase, when all the $\vec{\boldsymbol{\sigma}}_k$'s point along the same direction, we have $\mathbb{P} = 1$. In the incoherent/unsynchronized phase, when all the $\vec{\boldsymbol{\sigma}}_k$'s are spread out over the entire circle, we have $\mathbb{P} = 0$. Thus, the quantity $\mathbb{P}$ characterizes qualitatively-different macroscopic behavior and is a measure of global macroscopic order in the system. For the two-dimensional case that is being discussed here, the quantity $\mathbb{P}$ is exactly the Kuramoto order parameter $r_1$~\cite{Kuramoto}, i.e.,
\begin{eqnarray}
    \mathbb{P} = r_1 \equiv \left|\frac{1}{N}\sum_{k=1}^Ne^{i\theta_k}\right|.
\end{eqnarray}
Note that the evolution equation~\eqref{eq: 2D mean field} preserves the norm of each $\vec{\boldsymbol{\sigma}}_j$, as may be seen in the following way:
\begin{eqnarray}
    \frac{d \left(\vec{\boldsymbol{\sigma}}_j\cdot\vec{\boldsymbol{\sigma}}_j\right)}{dt} &=& 2~\vec{\boldsymbol{\sigma}}_j\cdot\frac{d\vec{\boldsymbol{\sigma}}_j}{dt}\nonumber\\
    &=& 2~\vec{\boldsymbol{\sigma}}_j\cdot \mathbf{W}_j\vec{\boldsymbol{\sigma}}_j+2\sqrt{2T}~\vec{\boldsymbol{\sigma}}_j\cdot \mathbf{N}_j(t)\vec{\boldsymbol{\sigma}}_j.
\end{eqnarray}
The anti-symmetric structure of $\mathbf{W}_j$ and $\mathbf{N}_j(t)$ ensures that each term on the right hand side is zero, and hence, $d\left(\vec{\boldsymbol{\sigma}}_j\cdot \vec{\boldsymbol{\sigma}}_j\right)/dt = 0$, implying $|\vec{\boldsymbol{\sigma}}_j|=1~\forall~j$ at all times.

Let us turn to the order parameter $\mathbb{M}$. The unit norm of each $\vec{\boldsymbol{\sigma}}_k$ immediately gives $\mathrm{Tr}\left[\mathbb{M}\right] 
 =1$. The definition~\eqref{eq: order parameter 2} further implies that $\mathbb{M}$ is a symmetric matrix, i.e., $\mathbb{M}^\top = \mathbb{M}$, where $\top$ denotes transpose operation. Defining the second Kuramoto-Daido~\cite{Kuramoto} order parameter $z_2 \equiv r_2 e^{i2\psi_2} \equiv N^{-1} \sum_{k=1}^N e^{i 2\theta_k}$, we may express $\mathbb{M}$ for the two-dimensional case as
 \begin{equation}
    \mathbb{M} = 
    \frac{1}{2}\begin{bmatrix}
       1 + \Re\left(z_2\right) & \Im\left(z_2\right) \\
        \Im\left(z_2\right) & 1 - \Re\left(z_2\right)
    \end{bmatrix},
\end{equation}
where $\Re\left(\cdot\right)$ and $\Im\left(\cdot\right)$ denote the real and the imaginary part of the quantity inside the brackets, respectively. In Section~\ref{sec: dis on M}, we discuss the physical property that is captured by the order parameter $\mathbb{M}$.

In the above backdrop, it is now easy to generalize Eqs.~\eqref{eq: 2D vectorisation model1 2}~and~\eqref{eq: 2D mean field} to the $D$-dimensional case, when each of the $\vec{\boldsymbol{\sigma}}_k$'s may be considered a $D$-dimensional vector, with time evolution given by Eq.~\eqref{eq: 2D vectorisation model1 2}~or~\eqref{eq: 2D mean field} with $\mathbf{I}_2$ replaced by $\mathbf{I}_D$, where $\mathbf{I}_D$ denotes $D\times D$ identity matrix. The vectors $\vec{\boldsymbol{\sigma}}_k$ now have their ends on the surface of a hyper-sphere of unit radius that is embedded in a $D$-dimensional space. The frequency matrix $\mathbf{W}_j$ is now a $D\times D$ antisymmetric matrix, known as the natural rotation~\cite{Sarthak2019}, and given by
\begin{eqnarray}
    \mathbf{W}_j \equiv 
    \begin{bmatrix}
        0 & \omega^{(j)}_{12} & \omega^{(j)}_{13} &\cdots &\omega^{(j)}_{1D}\\
        -\omega^{(j)}_{12} & 0 & \omega^{(j)}_{23} &\cdots &\omega^{(j)}_{2D}\\
      -  \omega^{(j)}_{13} & -\omega^{(j)}_{23} & 0 & \cdots & \omega^{(j)}_{3D}\\
      \vdots & \vdots & &\ddots&\vdots\\
      -\omega^{(j)}_{1D} & -\omega^{(j)}_{2D} & -\omega^{(j)}_{3D} & \cdots & 0
    \end{bmatrix}
\end{eqnarray}
in the Cartesian coordinate system, where each of these $\omega^{(j)}_{lm}$ is sampled independently from a distribution $g(\omega)$. Clearly, there are $D(D-1)/2$ independent elements in each natural rotation matrix $\mathbf{W}_j$. Alternatively, we may consider each matrix $\mathbf{W}_j$ to be sampled from the ensemble of random antisymmetric matrices following the distribution $G(\mathbf{W})$. For the $D$-dimensional case, the noise matrix becomes
\begin{eqnarray}
    \mathbf{N}_j(t) \equiv 
    \begin{bmatrix}
        0 & \eta^{(j)}_{12}(t) & \eta^{(j)}_{13}(t) &\cdots &\eta^{(j)}_{1D}(t)\\
        -\eta^{(j)}_{12}(t) & 0 & \eta^{(j)}_{23}(t) &\cdots &\eta^{(j)}_{2D}(t)\\
      -  \eta^{(j)}_{13}(t) & -\eta^{(j)}_{23}(t) & 0 & \cdots & \eta^{(j)}_{3D}(t)\\
      \vdots & \vdots & &\ddots&\vdots\\
      -\eta^{(j)}_{1D} (t)& -\eta^{(j)}_{2D}(t) & -\eta^{(j)}_{3D} (t)& \cdots & 0
    \end{bmatrix}~~~~~\label{eq: noise matrix}
\end{eqnarray}
in the Cartesian coordinate system, with the properties $\langle \eta^{(j)}_{lm}(t)\rangle = 0$ and $\langle \eta^{(i)}_{lm}(t)\eta^{(j)}_{l'm'}(t')\rangle = \delta_{ij}\delta_{ll'}\delta_{mm'}\delta(t-t')$, with $m>l$ and $m'>l'$. Using these properties, we further get
\begin{equation}
    \langle\mathbf{N}_j(t)\rangle = \mathbb{O}_D,~~\langle\mathbf{N}_i(t)\mathbf{N}_j(t')\rangle = -(D-1)\delta_{ij}\delta(t-t')\mathbf{I}_D, \label{eq: noise relation}
\end{equation}
for all $i,j$, with $\mathbb{O}_D$ being a $D \times D$ matrix with all entries equal to zero. Lastly, the order parameter $\vec{\mathbb{P}}$ is a $D\times 1$ column vector, while the order parameter $\mathbb{M}$ is a $D\times D$ symmetric matrix. Hence, the total number of independent quantities in the two order parameters is $D(D+3)/2-1$.

In this paper, we are going to consider the case without any quenched disorder, i.e., $\mathbf{W}_j = \mathbf{W}~\forall~j$, since our main objective to investigate the interplay of annealed disorder and higher-harmonic interactions, as outlined in the Introduction. The ensuing calculations even in this identical natural-rotation case are quite involved, warranting a separate study of the case with quenched disorder~\cite{note1}. For the case without any quenched disorder, using a trasformation $\vec{\boldsymbol{\sigma}}_j\to e^{-\mathbf{W}t}\vec{\boldsymbol{\sigma}}_j~\forall~j$, Eq.~\eqref{eq: 2D vectorisation model1 2} can be reduced to
\begin{align}
     \frac{d \vec{\boldsymbol{\sigma}}_j}{dt} &=  \frac{K_1}{N}\sum_{k = 1}^N  \vec{\boldsymbol{\sigma}}_k-\frac{K_1}{N}\sum_{k = 1}^N \left(\vec{\boldsymbol{\sigma}}_j \cdot \vec{\boldsymbol{\sigma}}_k\right)\vec{\boldsymbol{\sigma}}_j \nonumber\\
     & + \frac{2 K_2}{N}\sum_{k = 1}^N \left(\vec{\boldsymbol{\sigma}}_j \cdot \vec{\boldsymbol{\sigma}}_k\right)\vec{\boldsymbol{\sigma}}_k-\frac{2 K_2}{N}\sum_{k = 1}^N \left(\vec{\boldsymbol{\sigma}}_j \cdot \vec{\boldsymbol{\sigma}}_k\right)^2\vec{\boldsymbol{\sigma}}_k\nonumber \\
     &+\sqrt{2T} ~\mathbf{N}_j(t) \vec{\boldsymbol{\sigma}}_j.\label{eq: 2D vectorisation model1 2 final}
\end{align}
From now on, we will consider that such a transformation has been implemented, and hence, will study the dynamics~\eqref{eq: 2D vectorisation model1 2 final}.

 In the next section, we discuss in detail the properties of the order parameter $\mathbb{M}$, which will have important bearing on our analysis in the later part of the paper.

\subsection{\texorpdfstring{Discussion on $\mathbb{M}$}{Discussion on M}}
\label{sec: dis on M}

Here we discuss the properties of the order parameter $\mathbb{M}$. For simplicity, let us first focus on three dimensions. If we denote the position vector of the $j$-th oscillator in the Cartesian coordinate as $\vec{\boldsymbol{\sigma}}_j = (\sigma^\mathrm{x}_j,\sigma^\mathrm{y}_j,\sigma^\mathrm{z}_j)$,  then from the definition~\eqref{eq: order parameter 2}, we obtain
\begin{equation}
    \mathbb{M} = 
    \begin{bmatrix}
        \overline{(\sigma^\mathrm{x})^2} & \overline{\sigma^\mathrm{x}\sigma^\mathrm{y}}& \overline{\sigma^\mathrm{x}\sigma^\mathrm{z}}\\
         \overline{\sigma^\mathrm{y}\sigma^\mathrm{x}} & \overline{(\sigma^\mathrm{y})^2} &\overline{\sigma^\mathrm{y}\sigma^\mathrm{z}}\\
         \overline{\sigma^\mathrm{z}\sigma^\mathrm{x}} & \overline{\sigma^\mathrm{z}\sigma^\mathrm{y}} &\overline{(\sigma^\mathrm{z})^2}
    \end{bmatrix},
\end{equation}
where we have defined
\begin{equation}
    \overline{\sigma^\mathrm{a}\sigma^\mathrm{b}} \equiv \frac{1}{N}\sum_{j=1}^N \sigma^\mathrm{a}_j\sigma^\mathrm{b}_j,
\end{equation}
with $\mathrm{a,b}=\mathrm{x,y,z}$. Hence, the quantity $\overline{\sigma^\mathrm{a}\sigma^\mathrm{b}}$ denotes the correlation between the $\mathrm{a}$-th and the $\mathrm{b}$-th Cartesian component of the position vector of the oscillators. Clearly, this is valid in any general dimension $D$, where the matrix element $\mathbb{M}_\mathrm{ab}$ denotes the correlation between the $\mathrm{a}$-th  and the $\mathrm{b}$-th Cartesian component of the position vector of the oscillators, and $\mathbb{M}$ denotes a correlation matrix.

In the particular incoherent state where all the oscillators are uniformly spread out on the surface of the $D$-dimensional hyper-sphere (the uniformly incoherent state), one has $\mathbb{P}=0$. Moreover, there is no correlation between the different components of the position vector. Hence, the off-diagonal terms of the matrix $\mathbb{M}$ will all be zero. Further, since in this state, there is no preferred coordinate axis, the diagonal terms of the matrix $\mathbb{M}$, which denote self-correlation of different Cartesian components, should be all equal. The sum of these diagonal elements gives the trace of the matrix $\mathbb{M}$, which we have already seen to be equal to unity. Hence, all the diagonal terms of the matrix $\mathbb{M}$ have the value $1/D$ in the uniformly incoherent state, yielding $\mathbb{M} = D^{-1}\mathbf{I}_D$. In a synchronized state with $\mathbb{P}\neq0$, when the oscillators form one macroscopic cluster, there will be correlations between the different components of the position vector, making the off-diagonal terms of the matrix $\mathbb{M}$ nonzero. Hence, in this state, $\mathbb{M}$ deviates from $D^{-1}\mathbf{I}_D$.

Surprisingly, there can be states in which the oscillators form multiple clusters without generating any global order, the simplest scenario being two clusters of equal size at diametrically opposite position on the surface of the $D$-dimensional hyper sphere, as shown in Figure~\ref{fig:schematic}(e). In these states, clearly, we have $\mathbb{P}=0$. Hence, such states cannot be distinguished from the incoherent state by the usual synchronization order parameter $\vec{\mathbb{P}}$. Consequently, any phase transition between such states and the incoherent state will be invisible to $\vec{\mathbb{P}}$. However, formation of clusters gives rise to correlations between the different components of the position vector, making $\mathbb{M}$ deviate from $D^{-1}\mathbf{I}_D$. To identify these states with multiple clusters without global order as well as to probe the associated transition, we need the correlation-based order parameter $\mathbb{M}$.
Thus, while 
$\vec{\mathbb{P}}$ captures global synchronization, the matrix $\mathbb{M}$ encodes correlations that remain invisible to conventional order parameters. This distinction becomes crucial in identifying phases with broken symmetry but no global order.

Before moving ahead, let us first understand the symmetry of the model given in Eq.~\eqref{eq: 2D vectorisation model1 2 final}, a task we take up in the following section.

\subsection{Symmetry of the model \label{app: sym}}
Since rotation matrices $\mathbf{R}$ are orthogonal transformations satisfying $\mathbf{R}\mathbf{R}^\top = \mathbf{R}^\top\mathbf{R}=\mathbf{I}_D$, their application on any vector preserves its Euclidean norm. Consequently, acting with any rotation matrix $\mathbf{R}$ on any unit vector $\vec{\boldsymbol{\sigma}}_j$  does not change its length, and the rotated vector $\mathbf{R}\vec{\boldsymbol{\sigma}}_j$ remains on the surface of the $D$-dimensional unit sphere.

We now focus on obtaining the evolution equation of $\mathbf{R}\vec{\boldsymbol{\sigma}}_j$. Let all vectors $\vec{\boldsymbol{\sigma}}_j$ be transformed by a fixed rotation matrix $\mathbf{R}$. Applying $\mathbf{R}$ on both sides of Eq.~\eqref{eq: 2D vectorisation model1 2 final} from the left along with the properties of the rotation matrices, we obtain
\begin{eqnarray}
    &&\frac{d\big( \mathbf{R}\vec{\boldsymbol{\sigma}}_j\big)}{dt} =  \frac{K_1}{N}\sum_{k = 1}^N  \mathbf{R}\vec{\boldsymbol{\sigma}}_k-\frac{K_1}{N}\sum_{k = 1}^N \left(\mathbf{R}\vec{\boldsymbol{\sigma}}_j \cdot \mathbf{R}\vec{\boldsymbol{\sigma}}_k\right) \mathbf{R}\vec{\boldsymbol{\sigma}}_j \nonumber\\
     && + \frac{2 K_2}{N}\sum_{k = 1}^N \left(\mathbf{R}\vec{\boldsymbol{\sigma}}_j \cdot \mathbf{R}\vec{\boldsymbol{\sigma}}_k\right)\mathbf{R}\vec{\boldsymbol{\sigma}}_k-\frac{2 K_2}{N}\sum_{k = 1}^N \left(\mathbf{R}\vec{\boldsymbol{\sigma}}_j \cdot \mathbf{R}\vec{\boldsymbol{\sigma}}_k\right)^2\mathbf{R}\vec{\boldsymbol{\sigma}}_k\nonumber \\
    &&+\sqrt{2T} ~\Big[\mathbf{R}\mathbf{N}_j(t)\mathbf{R}^\top\Big]\mathbf{R} \vec{\boldsymbol{\sigma}}_j.\label{eq: 2D vectorisation model1 2 rotated}
\end{eqnarray}
We may define $\widetilde{\mathbf{N}}_j(t) \equiv \mathbf{R}\mathbf{N}_j(t)\mathbf{R}^\top$ as the new noise matrix acting on the $j$-th oscillator. In Appendix~\ref{app: -1}, we show that the properties of the rotated noise matrix $\widetilde{\mathbf{N}}_j(t)$ are identical to those of $\mathbf{N}_j(t)$, which reads as
\begin{equation}
    \langle\widetilde{\mathbf{N}}_j(t)\rangle = \mathbb{O}_D,~~\langle\widetilde{\mathbf{N}}_i(t)\widetilde{\mathbf{N}}_j(t')\rangle = -(D-1)\delta_{ij}\delta(t-t')\mathbf{I}_D. \label{eq: noise relation rot}
\end{equation}
Thus, we may replace the noise terms $\widetilde{\mathbf{N}}_j(t) \equiv \mathbf{R}\mathbf{N}_j(t)\mathbf{R}^\top$ in Eq.~\eqref{eq: 2D vectorisation model1 2 rotated} with $\mathbf{N}_j(t)$ itself. Further redefining the rotated vectors as $\vec{\widetilde{\boldsymbol{\sigma}}}_j \equiv \mathbf{R}\vec{\boldsymbol{\sigma}}_j~\forall~j$, we may rewrite Eq.~\eqref{eq: 2D vectorisation model1 2 rotated} as
\begin{align}
    \frac{d \vec{\widetilde{\boldsymbol{\sigma}}}_j}{dt} &=  \frac{K_1}{N}\sum_{k = 1}^N  \vec{\widetilde{\boldsymbol{\sigma}}}_k-\frac{K_1}{N}\sum_{k = 1}^N \left(\vec{\widetilde{\boldsymbol{\sigma}}}_j \cdot \vec{\widetilde{\boldsymbol{\sigma}}}_k\right)\vec{\widetilde{\boldsymbol{\sigma}}}_j\nonumber\\
    &+ \frac{2 K_2}{N}\sum_{k = 1}^N \left(\vec{\widetilde{\boldsymbol{\sigma}}}_j \cdot \vec{\widetilde{\boldsymbol{\sigma}}}_k\right)\vec{\widetilde{\boldsymbol{\sigma}}}_k-\frac{2 K_2}{N}\sum_{k = 1}^N \left(\vec{\widetilde{\boldsymbol{\sigma}}}_j \cdot \vec{\widetilde{\boldsymbol{\sigma}}}_k\right)^2\vec{\widetilde{\boldsymbol{\sigma}}}_k\nonumber \\
    &+\sqrt{2T} ~\mathbf{N}_j(t) \vec{\widetilde{\boldsymbol{\sigma}}}_j ,\label{eq: 2D vectorisation model1 2 final rotated}
\end{align}
which is identical to Eq.~\eqref{eq: 2D vectorisation model1 2 final}. Hence, the evolution equations of the rotated vectors $\vec{\widetilde{\boldsymbol{\sigma}}}_j \equiv \mathbf{R}\vec{\boldsymbol{\sigma}}_j~\forall~j$ are identical to that of $\vec{\boldsymbol{\sigma}}_j~\forall~j$. Thus, we conclude that the action of any constant rotation matrix on all the vectors denoting the position of the oscillators keeps the evolution equation invariant, making it a symmetry of the problem. This knowledge about symmetry will be useful in doing the center manifold analysis in Sec.~\ref{sec: CM}.

 In the following section, we consider the limit $N\to \infty$ of our system and discuss how one may analyze it using a Fokker-Planck equation approach.

\section{Fokker-Planck Equation\label{sec: FP model 1}}

Our aim in this section is to study the system~\eqref{eq: 2D vectorisation model1 2 final} in the thermodynamic limit $N\to\infty$. To this end, we may introduce a density function $F\left(\vec{\boldsymbol{\sigma}},t\right)$, denoting the density of oscillators at ``position" $\vec{\boldsymbol{\sigma}}$ at time $t$. The normalization condition reads as
\begin{eqnarray}
    \int_{|\vec{\boldsymbol{\sigma}}|=1} F\left(\vec{\boldsymbol{\sigma}},t\right) d\vec{\boldsymbol{\sigma}} = 1, \label{eq: Normalization}
\end{eqnarray}
where the integration is done over the $(D-1)$-dimensional surface of the $D$-dimensional unit sphere, represented as $\mathcal{S} \equiv S^{D-1} \subset \mathbb{R}^D$. We may now define a $ D$-dimensional spherical polar coordinate system (see Appendix~\ref{app: 0}), with $r\in[0,\infty),~\theta_1,\theta_2,\ldots,\theta_{D-2}\in[0,\pi]$ and $\phi\in[0,2\pi)$. Then, a differential surface element on the unit sphere $\mathcal{S}$ may be expressed as $d\Omega = d\phi\prod_{j=1}^{D-2}\sin^{D-1-j}{\theta_j}d\theta_j$, with the total area of the $(D-1)$-dimensional surface given by
\begin{eqnarray}
    \Omega_D &=& \int d\Omega = \int_0^{2\pi}d\phi \prod_{j=1}^{D-2}\int_0^{\pi} d \theta_j \sin^{D-1-j}{\theta_j}d\theta_j \nonumber \\
    &=& \frac{2 \pi^{D/2}}{\Gamma(D/2)},
\end{eqnarray}
and the differential volume element in $\mathbb{R}^D$ given by $dV_D = r^{D-1}drd\Omega$. In their terms, Eq.~\eqref{eq: Normalization} may be expressed as
\begin{eqnarray}
    \int  F\left(\vec{\boldsymbol{\sigma}},t\right)~d\Omega = 1. \label{eq: norm surface}
\end{eqnarray}

To obtain the corresponding Fokker-Planck equation, following~\cite{Sarthak2019}, we first extend the velocity flow field given by Eq.~\eqref{eq: 2D vectorisation model1 2 final} to the entire space $\mathbb{R}^D$, by defining $\vec{\boldsymbol{\sigma}} = \vec{\mathbf{r}}/r =  \hat{\mathbf{r}}$. Using this, we may write the velocity flow field from Eq.~\eqref{eq: 2D vectorisation model1 2 final} as
\begin{equation}
    \frac{d \vec{\mathbf{r}}}{dt} = K_1  \left[\vec{\mathbb{P}}-\frac{1}{r^2}\left(\vec{\mathbf{r}}\cdot \vec{\mathbb{P}}\right)\vec{\mathbf{r}}\right]+2 K_2\Bigg[\frac{1}{r}\mathbb{M}-\frac{1}{r^3}\left(\vec{\mathbf{r}} \cdot \mathbb{M}\vec{\mathbf{r}}\right)\mathbf{I}_D\Bigg]\vec{\mathbf{r}}.\label{eq: dr/dt model1 main text}
\end{equation}
Since $\hat{\mathbf{r}}\cdot d\vec{\mathbf{r}}/dt = 0$, these flow fields result in  surfaces of spheres with varying radius $r$ and centered at
$r=0$ as invariant manifolds under the dynamical evolution.

We may then extend the density function $F\left(\vec{\boldsymbol{\sigma}},t\right)$, which is defined on the surface $\mathcal{S}$, to the entire $\mathbb{R}^D$ space, by defining the density $\mathcal{F}\left(\vec{\mathbf{r}},t\right)$ as
\begin{eqnarray}
\mathcal{F}\left(\vec{\mathbf{r}},t\right) = F\left(\vec{\boldsymbol{\sigma}},t\right) \delta(r-1). \label{eq: F and curly F main text}
\end{eqnarray}
Hence, the normalization condition~\eqref{eq: norm surface} may be extended as
\begin{eqnarray}
    \int  \mathcal{F}\left(\vec{\mathbf{r}},t\right)~dV_D = 1. \label{eq: norm volume main text}
\end{eqnarray}

To obtain the time evolution of $\mathcal{F}\left(\vec{\mathbf{r}},t\right)$, we start by considering a continuous and differentiable test function $h_1(\vec{\boldsymbol{\sigma}}(t))$ defined on the surface $\mathcal{S}$ and using the standard It$\hat{\mathrm{o}}$-calculus~\cite{Denisov2009}, we obtain (see Appendix~\ref{app: FP model 1})

\begin{eqnarray}
    \frac{\partial\mathcal{F}}{\partial t} &=&-\vec{\nabla}  \cdot \left[\mathcal{F}\vec{\mathbf{v}}\right]+\frac{T}{r^2}\nabla^2_\mathcal{S}\mathcal{F},\label{eq: FP first form model 2}
\end{eqnarray}
where the velocity field $\vec{\mathbf{v}} = d\vec{\mathbf{r}}/dt$ is given in Eq.~\eqref{eq: dr/dt model1 main text}. Using the expression of $\vec{\nabla}$ from Eq.~\eqref{eq: nabla} and integrating Eq.~\eqref{eq: FP first form model 2} over $r$ from $1-\epsilon$ to $1+\epsilon$ for small $\epsilon$ simplifies Eq.~\eqref{eq: FP first form model 2} into the following form (see Appendix~\ref{app: 11}):
\begin{align}
    \frac{\partial F}{\partial t}=&-K_1 \left[ \vec{\nabla}_\mathcal{S}F-(D-1)F \vec{\boldsymbol{\sigma}}\right]\cdot \vec{\mathbb{P}}-2K_2 F \nonumber \\
    & -2K_2 \left[\vec{\nabla}_\mathcal{S}F -D F \vec{\boldsymbol{\sigma}}\right]\cdot \mathbb{M} \vec{\boldsymbol{\sigma}} +T~\nabla^2_\mathcal{S}F. \label{eq: continuity final model 3}
\end{align}
 The order parameters defined in Eqs.~\eqref{eq: order parameter 1}~and~\eqref{eq: order parameter 2} take the form
\begin{align}
    \vec{\mathbb{P}} &= \int \vec{\boldsymbol{\sigma}}~F\left(\vec{\boldsymbol{\sigma}},t\right)~d\Omega, \label{eq: order parameter 1 final model 2}\\
    \mathbb{M} &= \int \vec{\boldsymbol{\sigma}} \otimes \vec{\boldsymbol{\sigma}}~ F\left(\vec{\boldsymbol{\sigma}},t\right)~d\Omega. \label{eq: order parameter 2 final model 2}
\end{align}

Since $\vec{\boldsymbol{\sigma}}\cdot \vec{\nabla}_\mathcal{S}F = 0 $, we may rearrange Eq.~\eqref{eq: continuity final model 3} into the following compact form:
\begin{align}
    \frac{\partial F}{\partial t}=&-K_1 \left[ \vec{\nabla}_\mathcal{S}F-(D-1)F \vec{\boldsymbol{\sigma}}\right]\cdot \vec{\mathbb{P}}\nonumber \\
    &  -2K_2 \left[\vec{\nabla}_\mathcal{S}F -D F \vec{\boldsymbol{\sigma}}\right]\cdot \left[\mathbb{M} -\frac{1}{D}\mathbf{I}\right]\vec{\boldsymbol{\sigma}} +T~\nabla^2_\mathcal{S}F. \label{eq: continuity final model 2}
\end{align}
Equation~\eqref{eq: continuity final model 2} provides a closed macroscopic description of the dynamics in terms of the probability density on the unit sphere. This formulation allows us to systematically analyze stability and nonlinear dynamics in the thermodynamic limit.

Now that we have derived the Fokker-Planck equation, we show below that it admits an incoherent stationary state, and our objective will be to perform its nonlinear stability via a center manifold analysis.

\section{\label{sec: CM}Center Manifold Analysis}

The Fokker-Planck equation~\eqref{eq: continuity final model 2} along with Eqs.~\eqref{eq: order parameter 1 final model 2}~and~\eqref{eq: order parameter 2 final model 2} provides a closed-form description of the time evolution of the system in the thermodynamic limit. Consider the incoherent stationary state, when all the oscillators are uniformly spread out on the surface of the $D$-dimensional unit sphere. We may express the corresponding stationary-state density in the limit $N\to\infty$ as
\begin{eqnarray}
    F_0\left(\vec{\boldsymbol{\sigma}}\right)  = \frac{1}{\Omega_D},~\mathrm{with}~\Omega_D = \frac{2\pi^{\frac{D}{2}}}{\Gamma\left(\frac{D}{2}\right)}. \label{eq: F0 Incoherent}
\end{eqnarray}
Using this in Eqs.~\eqref{eq: order parameter 1 final model 2}~and~\eqref{eq: order parameter 2 final model 2},  we obtain
\begin{align}
    \vec{\mathbb{P}}_{F = F_0} &= \frac{1}{S_D}\int \vec{\boldsymbol{\sigma}} ~d\Omega= \vec{0}_D, \label{eq: order parameter 1 Incoherent model 2}\\
    \mathbb{M}_{F=F_0} &= \frac{1}{S_D}\int\vec{\boldsymbol{\sigma}} \otimes \vec{\boldsymbol{\sigma}}~d\Omega = \frac{1}{D} \mathbf{I}_D. \label{eq: order parameter 2 Incoherent}
\end{align}
where $\vec{0}_D$ is a $D\times1$ vector with all elements zero. Using Eq.~\eqref{eq: F0 Incoherent} along with Eqs.~\eqref{eq: order parameter 1 Incoherent model 2}~and~\eqref{eq: order parameter 2 Incoherent} in Eq.~\eqref{eq: continuity final model 2}, we obtain that the uniform distribution $F_0$ satisfies the Fokker-Planck equation in the stationary state.

Note that the incoherent state, characterised by $\vec{\mathbb{P}}=\vec{0}_D$, is satisfied by many possible incoherent state densities. Among all these incoherent state densities, we are interested in the stability of the uniformly-incoherent state given by Eq.~\eqref{eq: F0 Incoherent}. To study this, we initiate the dynamics in a state with small perturbation around the state $F_0\left(\vec{\boldsymbol{\sigma}}\right)$ and study the evolution of this perturbation under the dynamics~\eqref{eq: continuity final model 2}. For fixed $K_1$, $K_2$ and $D$, if the perturbation decreases to zero with time, then the uniformly-incoherent state $F_0\left(\vec{\boldsymbol{\sigma}}\right)$ is a stable solution of Eq.~\eqref{eq: continuity final model 2} and the system is in the incoherent phase. On the other hand, if the perturbation keeps on growing in time, then $F_0\left(\vec{\boldsymbol{\sigma}}\right)$ is an unstable solution of Eq.~\eqref{eq: continuity final model 2} and the system is in the synchronized phase.

In order to study the behavior of the aforementioned perturbation in time, we start by considering the probability density $F\left(\vec{\boldsymbol{\sigma}},t\right)$ to be of the form
\begin{equation}
    F\left(\vec{\boldsymbol{\sigma}},t \right) = \frac{1}{\Omega_D}+\eta\left(\vec{\boldsymbol{\sigma}},t \right), \label{eq: F and eta} 
\end{equation}
where $|\eta\left(\vec{\boldsymbol{\sigma}},t \right)|\ll 1$ denotes the perturbation around the state $F_0\left(\vec{\boldsymbol{\sigma}}\right) = 1/\Omega_D$. Normalization of $F\left(\vec{\boldsymbol{\sigma}},t \right)$ given in Eq.~\eqref{eq: norm surface} puts the following condition:
\begin{eqnarray}
    \int\eta\left(\vec{\boldsymbol{\sigma}},t \right) d\Omega = 0. \label{eq: eta norm surface}
\end{eqnarray}

We now analyze the stability of the incoherent state by projecting the dynamics onto the slow modes near criticality. This will allow us to derive reduced evolution equations for the order parameters and determine the nature of the phase transitions.

Using Eq.~\eqref{eq: F and eta} in Eq.~\eqref{eq: continuity final model 2}, we obtain the evolution equation of $\eta\left(\vec{\boldsymbol{\sigma}},t\right)$, which reads as
\begin{equation}
    \frac{\partial \eta}{\partial t} = \mathcal{L}\eta+\mathcal{N}[\eta], \label{eq: del eta del t}
\end{equation}
where the linear operator $\mathcal{L}$ has the expression
\begin{align}
    \mathcal{L}\eta &= T~\nabla^2_\mathcal{S}\eta + K_1 \bigg(\frac{D-1}{\Omega_D}\bigg) \vec{\boldsymbol{\sigma}} \cdot \int \vec{\boldsymbol{\sigma}}' \eta\left(\vec{\boldsymbol{\sigma}}',t\right)~d\Omega' \nonumber\\
    &+ 2K_2 \bigg(\frac{D}{\Omega_D}\bigg)\vec{\boldsymbol{\sigma}} \cdot \bigg[\int \vec{\boldsymbol{\sigma}}'\otimes \vec{\boldsymbol{\sigma}}'\eta\left(\vec{\boldsymbol{\sigma}}',t\right)~d\Omega'\bigg]\vec{\boldsymbol{\sigma}},\label{eq: L eta model 2}
\end{align}
and the non-linear operator $\mathcal{N}[\cdot]$ has the expression
\begin{align}
    \mathcal{N}[\eta] &= (D-1)K_1\eta \vec{\boldsymbol{\sigma}}\cdot \int \vec{\boldsymbol{\sigma}}'\eta\left(\vec{\boldsymbol{\sigma}}',t\right)~d\Omega' \nonumber\\
    &-K_1  \vec{\nabla}_\mathcal{S}\eta\cdot \int \vec{\boldsymbol{\sigma}}'\eta\left(\vec{\boldsymbol{\sigma}}',t\right)~d\Omega'
    \nonumber\\
    &+ 2DK_2 \eta \vec{\boldsymbol{\sigma}}\cdot \bigg[\int \vec{\boldsymbol{\sigma}}'\otimes \vec{\boldsymbol{\sigma}}'\eta\left(\vec{\boldsymbol{\sigma}}',t\right)~d\Omega'\bigg] \vec{\boldsymbol{\sigma}}\nonumber\\
    &-2K_2\vec{\nabla}_\mathcal{S}\eta\cdot \bigg[\int \vec{\boldsymbol{\sigma}}'\otimes \vec{\boldsymbol{\sigma}}'\eta\left(\vec{\boldsymbol{\sigma}}',t\right)~d\Omega'\bigg] \vec{\boldsymbol{\sigma}}. \label{eq: NN eta model 2}
\end{align}
Since $\eta$ is a small perturbation, $\mathcal{L}\eta$ dominates over $\mathcal{N}[\eta]$. Hence, in the leading order, the dynamics of $\eta$ is determined by the eigenspectrum of the operator $\mathcal{L}$. In passing, we define a new order parameter
\begin{equation}
    \widetilde{\mathbb{M}} \equiv \int \vec{\boldsymbol{\sigma}} \otimes \vec{\boldsymbol{\sigma}}~ \eta\left(\vec{\boldsymbol{\sigma}},t\right)~d\Omega = \mathbb{M}-\frac{1}{D}\mathbf{I}_D, \label{eq: M TILDE}
\end{equation}
where $\mathbb{M}$ is defined in Eq.~\eqref{eq: order parameter 2 final model 2}. From now on, we will use $\widetilde{\mathbb{M}}$ instead of $\mathbb{M}$ to characterize the corresponding transition.

Before solving the problem for general $D$ dimensions, let us build some intuition towards addressing the problem by first working out explicitly the case of $D=3$. We will then apply the insights so obtained for arbitrary dimensions.

\subsection{\texorpdfstring{$D=3$ Case}{D=3 Case}}

For illustrative purposes, let us assume $K_2 = 0$ in this section. In $D=3$, the probability density $\eta\left(\vec{\boldsymbol{\sigma}},t\right)$ is a function of $\theta_1,\phi$ and $t$. Hence, quite generally, we may express it as
\begin{equation}
    \eta(\theta_1,\phi,t) = \sum_{l=1}^\infty\sum_{m=-l}^la_{lm}(t) ~Y_l^m(\theta_1,\phi), \label{eq: eta expansion Ylm}
\end{equation}
where $Y_l^m(\theta_1,\phi)$'s are the standard spherical harmonic functions. Here, the summation over $l$ starts from $l=1$ instead of $l=0$ to satisfy Eq.~\eqref{eq: eta norm surface}.

By definition, spherical harmonics are eigenfunctions of $\nabla^2_\mathcal{S}$ with the eigenvalue equation
\begin{equation}
    \nabla^2_\mathcal{S}Y_l^m(\theta_1,\phi) = -l(l+1)Y_l^m(\theta_1,\phi). \label{eq: eig nab}
\end{equation}
From now on, for brevity of notation, we will use $Y_l^m$ to denote $Y_l^m(\theta_1,\phi)$. In Cartesian coordinates, any arbitrary unit vector may be written as (see Appendix~\ref{app: 0})
\begin{eqnarray}
    \vec{\boldsymbol{\sigma}} = 
    \begin{pmatrix}
        \cos{\theta_1}\\
        \sin{\theta_1}\cos{\phi}\\
        \sin{\theta_1}\sin{\phi}
    \end{pmatrix}
    =  
    \sqrt{\frac{2\pi}{3}}
    \begin{pmatrix}
        \sqrt{2}Y_1^0\\
        Y_1^{-1}-Y_1^1\\
        iY_1^{-1}+iY_1^1\\
    \end{pmatrix}. \label{eq: sigma in Y}
\end{eqnarray}
Note that the convention of naming the Cartesian axes is as given in Appendix~\ref{app: 0}, which differs from the usual convention.

Using Eq.~\eqref{eq: sigma in Y} along with the orthonormality properties of spherical harmonics, we may write
\begin{equation}
    \vec{\mathbb{P}} = \int \vec{\boldsymbol{\sigma}}' \eta\left(\vec{\boldsymbol{\sigma}}',t\right)~d\Omega' = 
    \sqrt{\frac{2\pi}{3}}
    \begin{pmatrix}
        \sqrt{2}a_{1,0}\\
        a_{1,-1}-a_{1,1}\\
        -ia_{1,1}-ia_{1,-1}
    \end{pmatrix},
\end{equation}
using which we obtain
\begin{equation}
     \frac{K_1}{2\pi}\vec{\boldsymbol{\sigma}}\cdot \int \vec{\boldsymbol{\sigma}}'\eta\left(\vec{\boldsymbol{\sigma}}',t\right)~d\Omega' = \frac{2K_1}{3}\sum_{m=-1}^1a_{1m}Y_1^m.\label{eq: P IN Y}
\end{equation}

Combining Eq.~\eqref{eq: P IN Y}~along with~Eq.~\eqref{eq: eig nab}, we obtain that $Y_l^m$ is also an eigenfunction of the linear operator $\mathcal{L}$ (with $K_2=0$) with the eigenvalue equation 
\begin{equation}
    \mathcal{L}Y_l^m = \lambda_lY_l^m,
\end{equation}
and with the eigenvalue
\begin{equation}
      \lambda_l = -Tl(l+1) + \frac{2K_1}{3}\delta_{l,1}. \label{eq: lambda l}
\end{equation}
Note that since the eigenvalues $\lambda_l$ are independent of $m$, each eigenvalue has a degeneracy $(2l+1)$. This is because for a fixed $l$, there are $(2l+1)$ number of $Y_l^m$'s possible.

On the basis of the above, we here conclude that in Eq.~\eqref{eq: eta expansion Ylm}, $\eta$ is expanded in the eigenbasis of the operator $\mathcal{L}$. Furthermore, since for a fixed $l$, the functions $Y_l^m$ for all $m$ have identical eigenvalue $\lambda_l$, any general linear combination $\sum_{m=-l}^lc_mY_l^m$ will also be an eigenfunction of $\mathcal{L}$ with eigenvalue $\lambda_l$. Hence, for a fixed $l$, instead of using $Y_l^m$ as our expansion basis for $\eta$, we may use a different basis that is simultaneously an eigenbasis of $\mathcal{L}$ and also suitable for invoking the symmetry of the problem as discussed in Sec.~\ref{app: sym}.

Another important observation is that any $Y_l^m$ with $l>1$ may be expressed as a linear combination of the product of $Y_1^{-1},~Y_1^0$, and~$Y_1^1$. For example, we have
\begin{eqnarray}
     Y_2^{-2}  &=& \sqrt{\frac{10 \pi}{3}} Y_1^{-1} Y_1^{-1},~~Y_2^{-1} =  \sqrt{\frac{20 \pi}{3}} Y_1^{-1} Y_1^0,\nonumber\\
     Y_2^{0} &=& \sqrt{\frac{20 \pi}{9}}\bigg[ Y_1^{-1} Y_1^{1}+Y_1^0Y_1^0\bigg],\nonumber\\
      Y_2^1 &=& \sqrt{\frac{20 \pi}{3}} Y_1^{1} Y_1^0,~~Y_2^2 = \sqrt{\frac{10 \pi}{3}} Y_1^{1} Y_1^1. \nonumber\\ \label{eq: Y2 in Y1}
\end{eqnarray}
Note that this decomposition is similar to addition of angular momentum, whereby eigenstates of total angular momentum are expressed in a product basis. Hence, the coefficents in the expansion in Eq.~\eqref{eq: Y2 in Y1} are simply the well-known Clebsch–Gordan coefficients. More generally, any $Y_l^m$ for general $(l,m)$ may be written as~\cite{Varshalovich1988}
\begin{align}
    Y_l^m(\theta_1,\phi) =& \sqrt{\frac{(4\pi)^{n-1}(2n+1)!!}{3^nn!}}\nonumber\\
    & \times\{\ldots\{\{\vec{\mathbb{Y}}_1\otimes\vec{\mathbb{Y}}_1\}_2\otimes\vec{\mathbb{Y}}_1\}_3\ldots\otimes\vec{\mathbb{Y}}_1\}_{lm}, \label{eq: Ylm in ten Y1}
\end{align}
where we have $\vec{\mathbb{Y}}_1 = \big(Y_1^{-1}(\theta_1,\phi),Y_1^{0}(\theta_1,\phi),Y_1^{1}(\theta_1,\phi)\big)$. Here $\{ A_{\ell_1} \otimes B_{\ell_2} \}_{\ell}$ denotes the irreducible spherical tensor of rank $\ell$ obtained by coupling two spherical tensors of ranks $\ell_1$ and $\ell_2$.  
The projection onto angular momentum $\ell$ is carried out via the standard Clebsch-Gordan decomposition of the tensor product.

With this insight, let us now focus on the expansion in Eq.~\eqref{eq: eta expansion Ylm}. Using Eq.~\eqref{eq: Y2 in Y1}, clearly, we may write
\begin{align}
    \sum_{m=-1}^1 a_{1m}(t) ~Y_1^m(\theta_1,\phi)&= \vec{\mathbf{a}} \cdot \vec{\mathbb{Y}}_1,\\
    \sum_{m=-2}^2a_{2m}(t) ~Y_2^m(\theta_1,\phi)&=\vec{\mathbb{Y}}_1 \cdot \mathbf{b}\vec{\mathbb{Y}}_1,
\end{align}
where $\vec{\mathbf{a}} = \big(a_{1,-1}(t),a_{1,0}(t),a_{1,1}(t)\big)$ and
\begin{equation}
    \mathbf{b} = 
     \sqrt{\frac{5 \pi}{9}}
    \begin{pmatrix}
        \sqrt{6}a_{2,-2}(t) & \sqrt{3}a_{2,-1}(t)& a_{2,0}(t)\\
        \sqrt{3}a_{2,-1}(t) & \sqrt{4}a_{2,0}(t)& \sqrt{3}a_{2,1}(t)\\
       a_{2,0}(t) & \sqrt{3}a_{2,1}(t)&\sqrt{6}a_{2,2}(t)
    \end{pmatrix}.
\end{equation}
Hence, we may rewrite Eq.~\eqref{eq: eta expansion Ylm} as
\begin{equation}
    \eta(\theta_1,\phi,t) = \vec{\mathbf{a}} \cdot \vec{\mathbb{Y}}_1+\vec{\mathbb{Y}}_1 \cdot \mathbf{b}\vec{\mathbb{Y}}_1 + \ldots. \label{eq: eta in Yy}
\end{equation}

We now express $\vec{\mathbb{Y}}_1$ in terms of $\vec{\boldsymbol{\sigma}}$. From Eq.~\eqref{eq: sigma in Y}, we have
\begin{equation}
    \vec{\boldsymbol{\sigma}} = \Xi^{-1} \vec{\mathbb{Y}}_1,~~~
    \Xi^{-1} = 
    \sqrt{\frac{2\pi}{3}}
    \begin{pmatrix}
    0 & \sqrt{2} & 0\\
        1 & 0 & -1\\
        i & 0 & i
    \end{pmatrix},
\end{equation}
inverting which, we may write
\begin{equation}
    \vec{\mathbb{Y}}_1 = \Xi~\vec{\boldsymbol{\sigma}} ,~~~
    \Xi = 
    \sqrt{\frac{3}{8\pi}}
    \begin{pmatrix}
      0&  1 & -i \\
     \sqrt{2} &   0 & 0 \\
       0& -1 & -i \\
    \end{pmatrix}. \label{eq: Y to sigma}
\end{equation}
Using Eq.~\eqref{eq: Y to sigma} in Eq.~\eqref{eq: eta in Yy}, we may express $\eta$ as
\begin{equation}
    \eta(\vec{\boldsymbol{\sigma}},t) = \vec{\mathbf{A}}\cdot  \vec{\boldsymbol{\sigma}}  + \vec{\boldsymbol{\sigma}} \cdot \mathbf{B} \vec{\boldsymbol{\sigma}} +\ldots, \label{eq: eta A B}
\end{equation}
where we have $\vec{\mathbf{A}} = \Xi^\top \vec{\boldsymbol{\sigma}} $ and $\mathbf{B} = \Xi^\top \mathbf{b} \Xi$. Direct computation gives us $\mathbf{B}^\top = \mathbf{B}$ and $\mathrm{Tr}[\mathbf{B}]=0$. Representing $\vec{\boldsymbol{\sigma}} = (\sigma_1,\sigma_2,\sigma_3),~\vec{\mathbf{A}}=(A_1,A_2,A_3)$ and $ij$-th element of $\mathbf{B}$ as $B_{ij}$ in the Cartesian coordinates, we may rewrite Eq.~\eqref{eq: eta A B} as
\begin{equation}
    \eta(\vec{\boldsymbol{\sigma}},t) = \sum_{i_1=1}^3A_{i_1}\sigma_{i_1}+\sum_{i_1=1}^3\sum_{i_2=1}^3 B_{i_1i_2} \sigma_{i_1}\sigma_{i_2}+\ldots, \label{eq: eta A B ten}
\end{equation}
with $\sum_{i_1=1}^3 B_{i_1i_1}=0$. More generally, combining Eq.~\eqref{eq: Ylm in ten Y1} along with Eq.~\eqref{eq: Y to sigma}, we may write the general expansion of $ \eta(\vec{\boldsymbol{\sigma}},t)$ in the following way: If $\mathbb{T}^{(n)}$ is a symmetric traceless tensor of order $n$ with $T^{(n)}_{i_1,i_2,\ldots,i_n}$ its $i_1i_2\ldots i_n$-th element, then we may express $\eta$ as~\cite{PhysRevResearch.2.043061, andrews1999special}
\begin{equation}
    \eta(\vec{\boldsymbol{\sigma}},t) = \sum_{n=1}^\infty \sum_{i_1=1}^3\sum_{i_2=1}^3 \cdots \sum_{i_n=1}^3T^{(n)}_{i_1,i_2,\ldots,i_n} \sigma_{i_1}\sigma_{i_2} \ldots\sigma_{i_n}, \label{eq: eta gen T exp}
\end{equation}
where $\mathbb{T}^{(n)}$ being symmetric implies that swapping of any indices does not change it, i.e.,
\begin{equation}
    T^{(n)}_{i_1,i_2,\ldots,i_n} = T^{(n)}_{i_{P(1)},i_{P(2)},\ldots,i_{P(n)}},
\end{equation}
with $P$ being any arbitrary permutation of the indices, and traceless means contracting any two indices gives zero, i.e.,
\begin{equation}
    \sum_{i_1=1}^3\sum_{i_2=1}^3 \cdots \sum_{i_n=1}^3 \delta_{i_pi_q}T^{(n)}_{i_1,i_2,\ldots,i_p,\ldots,i_q,\ldots,i_n} =0,
\end{equation}
for arbitrary $p,q$. Comparing Eq.~\eqref{eq: eta gen T exp} with Eq.~\eqref{eq: eta A B ten} we observe that $\mathbb{T}^{(1)}\equiv \vec{\mathbf{A}}$ and $\mathbb{T}^{(2)}\equiv \mathbf{B}$. We may introduce a compact notation
\begin{equation}
    \eta(\vec{\boldsymbol{\sigma}},t) = \sum_{n=1}^\infty \Big( \mathbb{T}^{(n)},\vec{\boldsymbol{\sigma}}^{\otimes n}\Big), \label{eq: eta compact tensor}
\end{equation}
where we have
\begin{equation}
    \Big( \mathbb{T}^{(n)},\vec{\boldsymbol{\sigma}}^{\otimes n}\Big) \equiv \sum_{i_1=1}^3\sum_{i_2=1}^3 \cdots \sum_{i_n=1}^3T^{(n)}_{i_1,i_2,\ldots,i_n} \sigma_{i_1}\sigma_{i_2} \ldots\sigma_{i_n}.
\end{equation}

Having discussed in detail the $D=3$ case, we now extend our analysis to the general $D$-dimensional case.

\subsection{\texorpdfstring{General $D$-Dimensional Case}{General D-Dimensional Case}}
\label{sec: Gen D}

Having now an explicit understanding of the structure of $\eta(\vec{\boldsymbol{\sigma}},t)$ in $D=3$, here we generalize our discussions to any general dimension $D$; here, we will consider the general case with $K_1$ and $K_2$ both non-zero. Note that Eq.~\eqref{eq: eta A B} as well as Eq.~\eqref{eq: eta compact tensor} can be generalized to $D$ dimensions with the extension that both $\vec{\boldsymbol{\sigma}} = (\sigma_1,\sigma_2,\ldots, \sigma_D)$ and $\vec{\mathbf{A}}=(A_1,A_2,\ldots,A_D)$ are $D$-dimensional vectors, $\mathbf{B}$ is a $D\times D$ matrix,   
\begin{equation}
    \Big( \mathbb{T}^{(n)},\vec{\boldsymbol{\sigma}}^{\otimes n}\Big) = \sum_{i_1=1}^D\sum_{i_2=1}^D \cdots \sum_{i_n=1}^DT^{(n)}_{i_1,i_2,\ldots,i_n} \sigma_{i_1}\sigma_{i_2} \ldots\sigma_{i_n}, \label{eq: T sigma expan D DIM}
\end{equation}
and the traceless property reads as
\begin{equation}
    \sum_{i_1=1}^D\sum_{i_2=1}^D \cdots \sum_{i_n=1}^D \delta_{i_pi_q}T^{(n)}_{i_1,i_2,\ldots,i_p,\ldots,i_q,\ldots,i_n} =0. \label{eq: traceless D}
\end{equation}

Before moving on, we have to first prove that Eq.~\eqref{eq: eta compact tensor} is still an useful expansion in general $D$ dimensions. In Appendix~\ref{app: 7}, we prove that $\Big( \mathbb{T}^{(n)},\vec{\boldsymbol{\sigma}}^{\otimes n}\Big)$ is an eigenfunction of $\nabla^2_\mathcal{S}$ with the eigenvalue equation
\begin{equation}
    \nabla^2_\mathcal{S}\Big( \mathbb{T}^{(n)},\vec{\boldsymbol{\sigma}}^{\otimes n}\Big) = -n(n+D-2)\Big( \mathbb{T}^{(n)},\vec{\boldsymbol{\sigma}}^{\otimes n}\Big). \label{eq: eig eq tensor n}
\end{equation}
Moreover, in Appendix~\ref{app: 8}, we derive that
\begin{equation}
    \vec{\boldsymbol{\sigma}}\cdot \int \vec{\boldsymbol{\sigma}}'\Big( \mathbb{T}^{(n)},\vec{\boldsymbol{\sigma}}'^{\otimes n}\Big) d\Omega'= \delta_{n,1}\frac{\Omega_D}{D}\Big( \mathbb{T}^{(n)},\vec{\boldsymbol{\sigma}}^{\otimes n}\Big), \label{eq: eig eq ten K1 INT}
\end{equation}
and in Appendix~\ref{app: 12}, we derive that
\begin{equation}
    \vec{\boldsymbol{\sigma}} \cdot \bigg[\int \vec{\boldsymbol{\sigma}}'\otimes \vec{\boldsymbol{\sigma}}'\Big( \mathbb{T}^{(n)},\vec{\boldsymbol{\sigma}}'^{\otimes n}\Big) d\Omega'\bigg]\vec{\boldsymbol{\sigma}}= \frac{\delta_{n,2}2\Omega_D}{D(D+2)}\Big( \mathbb{T}^{(n)},\vec{\boldsymbol{\sigma}}^{\otimes n}\Big). \label{eq: eig eq ten K2 INT}
\end{equation}

Putting Eqs.~\eqref{eq: eig eq tensor n},~\eqref{eq: eig eq ten K1 INT},~and~\eqref{eq: eig eq ten K2 INT} back into Eq.~\eqref{eq: L eta model 2}, we observe that $\Big( \mathbb{T}^{(n)},\vec{\boldsymbol{\sigma}}^{\otimes n}\Big)$ is an eigenfunction of $\mathcal{L}$ with the eigenvalue equation
\begin{eqnarray}
    \mathcal{L}\Big( \mathbb{T}^{(n)},\vec{\boldsymbol{\sigma}}^{\otimes n}\Big) = \lambda_n\Big( \mathbb{T}^{(n)},\vec{\boldsymbol{\sigma}}^{\otimes n}\Big), \label{eq: lambda n}
\end{eqnarray}
where the corresponding eigenvalues are
\begin{equation}
    \lambda_n = -n(n+D-2)T + \frac{(D-1)K_1}{D}\delta_{n,1} + \frac{4K_2}{(D+2)}\delta_{n,2}. \label{eq: eigen gen D}
\end{equation}
Note that putting $D = 3$ and $K_2=0$ in Eq.~\eqref{eq: lambda n}, we get back Eq.~\eqref{eq: lambda l}.  The modes $n=1$ and $n=2$ play a distinguished role: the former governs the onset of global synchronization, while the latter captures correlation-driven ordering. Higher modes are always stable and do not contribute to the critical dynamics.

In a symmetric traceless tensor $\mathbb{T}^{(n)}$ of order $n$, the total number of independent elements is given by
\begin{equation}
    \mathrm{dim}\Big[\mathbb{T}^{(n)}\Big] = 
    \begin{pmatrix}
        D+n-1\\
        n
    \end{pmatrix}
    -
    \begin{pmatrix}
        D+n-3\\
        n-2
    \end{pmatrix}.
\end{equation}
Hence, each of the eigenvalues $\lambda_n$ has degeneracy $\mathrm{dim}\big[\mathbb{T}^{(n)}\big]$. Clearly, eigenvalue $\lambda_1$ has degeneracy $D$, eigenvalue $\lambda_2$ has degeneracy $ D(D+1)/2-1$, and so on.

In the light of the above discussion, we conclude that the expansion of $\eta(\vec{\boldsymbol{\sigma}},t)$ in Eq.~\eqref{eq: eta compact tensor} is done in the eigenbasis of the operator $\mathcal{L}$.

Further analysis necessitates taking into account effects of symmetry, which is done in the following section.

\subsection{Effect of Symmetry}

Let us consider arbitrary $D$ dimensions and go back to the expansion in Eq.~\eqref{eq: eta compact tensor}, which provides the general form of $\eta(\vec{\boldsymbol{\sigma}},t)$. Writing the first few terms, we get
\begin{align}
    \eta(\vec{\boldsymbol{\sigma}},t) &= \Big( \mathbb{T}^{(1)},\vec{\boldsymbol{\sigma}}\Big) + \Big( \mathbb{T}^{(2)},\vec{\boldsymbol{\sigma}}\otimes \vec{\boldsymbol{\sigma}}\Big) + \cdots, \nonumber\\
    &= \vec{\mathbf{A}} \cdot \vec{\boldsymbol{\sigma}} +\vec{\boldsymbol{\sigma}}\cdot\mathbf{B}\vec{\boldsymbol{\sigma}}+\cdots, \label{eq: eta in a b in gen d}
\end{align}
where $\vec{\mathbf{A}}$ and $\vec{\boldsymbol{\sigma}} $ are $D$-dimensional vectors and $\mathbf{B}$ is a $D\times D$ symmetric traceless matrix.

Consider all the vectors denoting the position of the oscillators being rotated by a constant rotation matrix $\mathbf{R}$. Hence, the position vector $\vec{\boldsymbol{\sigma}}$ is transformed to $\vec{\boldsymbol{\widetilde{\sigma}}} \equiv \mathbf{R}\vec{\boldsymbol{\sigma}} $ and the probability density $\eta(\vec{\boldsymbol{\sigma}},t)$ is transformed to $\widetilde{\eta}(\vec{\boldsymbol{\widetilde{\sigma}}},t)$. Since
 Eq.~\eqref{eq: eta in a b in gen d} provides the general form of probability density on $\mathcal{S}$, the probability density $\widetilde{\eta}(\vec{\boldsymbol{\widetilde{\sigma}}},t)$ can also be expressed as
\begin{equation}
    \widetilde{\eta}(\vec{\boldsymbol{\widetilde{\sigma}}},t) = \vec{\mathbf{\widetilde{A}}} \cdot \vec{\boldsymbol{\widetilde{\sigma}}} +\vec{\boldsymbol{\widetilde{\sigma}}}\cdot\mathbf{\widetilde{B}}\vec{\boldsymbol{\widetilde{\sigma}}}+\cdots. \label{eq: eta tilde sigma tilde}
\end{equation}
Using $\vec{\boldsymbol{\widetilde{\sigma}}} \equiv \mathbf{R}\vec{\boldsymbol{\sigma}} $ in Eq.~\eqref{eq: eta tilde sigma tilde}, we obtain 
\begin{equation}
    \widetilde{\eta}(\vec{\boldsymbol{\widetilde{\sigma}}},t) = \Big(\mathbf{R}^\top\vec{\mathbf{\widetilde{A}}} \Big)\cdot \vec{\boldsymbol{\sigma}} +\vec{\boldsymbol{\sigma}}\cdot \Big(\mathbf{R}^\top\mathbf{\widetilde{B}} \mathbf{R}\Big)\vec{\boldsymbol{\sigma}}+\cdots. \label{eq: eta tilde sigma}
\end{equation}

From the discussion in Sec.~\ref{app: sym}, the system remains invariant under rotation by any constant $\mathbf{R}$. Probability conservation implies
\begin{equation}
   \eta(\vec{\boldsymbol{\sigma}},t) =  \widetilde{\eta}(\vec{\boldsymbol{\widetilde{\sigma}}},t)~\mathrm{det}\big[\mathbf{R}\big],\label{eq: symmetry condition}
\end{equation}
where $\mathrm{det}\big[\mathbf{R}\big]$ is the determinant of the Jacobian. From the properties of the rotation matrix, we have $\mathrm{det}\big[\mathbf{R}\big] =1$. Using this along with Eqs.~\eqref{eq: eta in a b in gen d}~and~\eqref{eq: eta tilde sigma} in Eq.~\eqref{eq: symmetry condition} and comparing both sides, we obtain
\begin{equation}
    \vec{\mathbf{\widetilde{A}}} = \mathbf{R}\vec{\mathbf{A}},~~~\widetilde{\mathbf{B}} = \mathbf{R} \mathbf{B}\mathbf{R}^\top. \label{eq: A B Rule}
\end{equation}
Hence, the rotational symmetry of the problem restricts $\vec{\mathbf{A}}$ and $\mathbf{B}$ to transform following Eq.~\eqref{eq: A B Rule}. Going in this way, we may say, for any general traceless symmetric tensor $\mathbb{T}^{(n)}$, it has to transform following the rule
\begin{equation}
    \widetilde{T}^{(n)}_{i_1,i_2,\ldots,i_n} = \sum_{j_1=1}^D\cdots\sum_{j_n=1}^D T^{(n)}_{j_1,j_2,\ldots,j_n}R_{i_1j_1}R_{i_2j_2}\cdots R_{i_nj_n}, \label{eq: A B Rule general}
\end{equation}
where $R_{ab}$ is $ab$-th component of $\mathbf{R}$ and $\widetilde{T}^{(n)}_{i_1,i_2,\ldots,i_n}$ are the components of the rotated tensor $\widetilde{\mathbb{T}}^{(n)}$. Note that, the transformation rule \eqref{eq: A B Rule general} preserves the symmetric and traceless properties of the tensor $\mathbb{T}^{(n)}$.

\subsubsection{Critical Point}
\label{sec: model 1 critical point}

 We now focus on the eigenvalues $\lambda_{n}$, given in Eq.~\eqref{eq: eigen gen D}. For $n\geq3$, the eigenvalues $\lambda_n = -n(n+D-2)T$ are independent of the interaction strengths $K_1,K_2$ and always negative. Hence, any perturbation given by the eigenfunctions corresponding to these eigenvalues or linear combinations thereof will quickly die down in time and go to zero under evolution solely due to the linear operator in Eq.~\eqref{eq: del eta del t}. On the other hand, considering the eigenvalue $\lambda_1$, upon changing $K_1$, it changes sign at $K_1 = K_1^\mathrm{c} \equiv DT$. Similarly, upon changing $K_2$, eigenvalue $\lambda_2$ changes sign at $K_2 = K_2^\mathrm{c} \equiv D(D+2)T/2$. When $K_1 < K_1^\mathrm{c} \equiv DT$ and $K_2 < K_2^\mathrm{c} \equiv D(D+2)T/2$, the eigenvalues $\lambda_1$ and $\lambda_2$ are negative. Since the rest of the eigenvalues are also negative, any perturbation will go to zero in time in this regime, making the uniformly incoherent state stable in the region $K_1 < K_1^\mathrm{c}$ and $K_2 < K_2^\mathrm{c}$. 
 
 For $K_2 \leq K_2^\mathrm{c}$ and $K_1>K_1^\mathrm{c}$, we have $\lambda_1 >0$ and rest of the eigenvalues negative. Hence,  any perturbation of the form $\big(\mathbb{T}^{(1)},\vec{\boldsymbol{\sigma}}\big) \equiv \vec{\mathbf{A}}\cdot \vec{\boldsymbol{\sigma}}$, which is an eigenfunction of $\mathcal{L}$ with eigenvalue $\lambda_1$, will become linearly unstable. As a result, when $\lambda_1 >0$, any perturbation in its eigenspace grows with time. Hence, in the leading order, $\eta\left(\vec{\boldsymbol{\sigma}},t \right)$ will be of the form $\vec{\mathbf{A}}\cdot \vec{\boldsymbol{\sigma}}$. From the results in Appendix~\ref{app: 10}, we have $\vec{\mathbb{P}} = \int \vec{\boldsymbol{\sigma}}\eta\left(\vec{\boldsymbol{\sigma}},t \right)~d\Omega \approx \int \vec{\boldsymbol{\sigma}}\big(\vec{\mathbf{A}}\cdot \vec{\boldsymbol{\sigma}}\big)d\Omega = \big(\Omega_D/D\big)\vec{\mathbf{A}}$.  Since the perturbations having the form $\vec{\mathbf{A}}\cdot \vec{\boldsymbol{\sigma}}$ grows in time, the quantity $\vec{\mathbf{A}}$ also grows in time, indicating $\vec{\mathbb{P}}$ becomes nonzero in this range. Hence, the order parameter $\vec{\mathbb{P}}$ shows a phase transition at
\begin{equation}
    K_1^\mathrm{c} = DT.
\end{equation}
This shows that the critical point of $\vec{\mathbb{P}}$ is independent of the interaction strength $K_2$ as long as $K_2 \leq K_2^\mathrm{c}$. On the other hand, the nature of the criticality depends on $K_2$, which we will discuss in the next section. Note that when the perturbation grows sufficiently large, $\mathcal{N}[\eta]$ starts contributing in the dynamics of  $\eta(\vec{\boldsymbol{\sigma}},t)$ in the next leading order. From the discussion in Sec.~\ref{sec: Onset model 1}, this contribution is of the form $ \vec{\boldsymbol{\sigma}}\cdot\mathbf{B}\vec{\boldsymbol{\sigma}}$. Hence, up to the second leading order, we have $ \eta(\vec{\boldsymbol{\sigma}},t) \approx \vec{\mathbf{A}} \cdot \vec{\boldsymbol{\sigma}} +\vec{\boldsymbol{\sigma}}\cdot\mathbf{B}\vec{\boldsymbol{\sigma}}$. From the results in Appendix~\ref{app: 10}, we further have $\widetilde{\mathbb{M}}= \int \vec{\boldsymbol{\sigma}}\otimes \vec{\boldsymbol{\sigma}}\eta\left(\vec{\boldsymbol{\sigma}},t \right)~d\Omega  = \big[2\Omega_D/D(D+2)\big]\mathbf{B}$. This indicates that order parameter $\widetilde{\mathbb{M}}$ also shows a transition at $K_1=K_1^\mathrm{c}$.

For $K_1\leq K_2^\mathrm{c}$ and $K_2 > K_2^\mathrm{c}$, we have $\lambda_2 >0$ and rest of the eigenvalues negative. Hence,  any perturbation of the form $\big(\mathbb{T}^{(2)},\vec{\boldsymbol{\sigma}}\otimes \vec{\boldsymbol{\sigma}}\big) \equiv \vec{\boldsymbol{\sigma}}\cdot \mathbf{B}\vec{\boldsymbol{\sigma}}$, which is an eigenfunction of $\mathcal{L}$ with eigenvalue $\lambda_2$, will become linearly unstable. When $\lambda_2 >0$, any perturbation in its eigenspace grows with time. Hence, in the leading order, $\eta\left(\vec{\boldsymbol{\sigma}},t \right)$ will be of the form $\vec{\boldsymbol{\sigma}}\cdot \mathbf{B}\vec{\boldsymbol{\sigma}}$. From the results in Appendix~\ref{app: 10}, we have $\vec{\mathbb{P}} = \int \vec{\boldsymbol{\sigma}}\eta\left(\vec{\boldsymbol{\sigma}},t \right)~d\Omega \approx \int \vec{\boldsymbol{\sigma}}\big(\vec{\boldsymbol{\sigma}}\cdot \mathbf{B}\vec{\boldsymbol{\sigma}}\big)d\Omega = \vec{0}_D$ and $\widetilde{\mathbb{M}} = \int \vec{\boldsymbol{\sigma}} \otimes \vec{\boldsymbol{\sigma}}\eta\left(\vec{\boldsymbol{\sigma}},t \right)~d\Omega \approx \int \vec{\boldsymbol{\sigma}} \otimes \vec{\boldsymbol{\sigma}}\big(\vec{\boldsymbol{\sigma}}\cdot \mathbf{B}\vec{\boldsymbol{\sigma}}\big)d\Omega = \big\{2\Omega_D/[D(D+2)]\}\mathbf{B}$.  Since the perturbations having the form $\vec{\boldsymbol{\sigma}}\cdot \mathbf{B}\vec{\boldsymbol{\sigma}}$ grows in time, the quantity $\mathbf{B}$ also grows in time, indicating $\widetilde{\mathbb{M}}$ becomes nonzero in this range. Hence, the order parameter $\widetilde{\mathbb{M}}$ shows a phase transition at
\begin{equation}
    K_2^\mathrm{c} = \frac{D(D+2)T}{2},
\end{equation}
while the order-parameter $\vec{\mathbb{P}}$ does not show any transition. Note that the critical point of $\widetilde{\mathbb{M}}$ is independent of the interaction strength $K_1$ as long as $K_1 \leq K_1^\mathrm{c}$. On the other hand, the nature of the criticality depends on the dimension of the system, which we will discuss in the next section. As we will discuss in Sec.~\ref{sec: tran in M}, to probe the transition in the order parameter $\widetilde{\mathbb{M}}$, we will use the quantity $\mathbb{B} \equiv \mathrm{Tr}\big[\widetilde{\mathbb{M}}^2\big]$.  Let us first discuss in the following section the issue of transitions exhibited by the quantity $\vec{\mathbb{P}}$.

\subsection{\texorpdfstring{Transition in $\vec{\mathbb{P}}$}{Transition in P}}
\label{sec: Onset model 1}

We now focus on obtaining the evolution equation of $\mathbb{P}$ up to second leading order to understand the nature of the transition. We consider the parameter region $K_2 \leq K_2^\mathrm{c}$ and $K_1>K_1^\mathrm{c}$ but close to the critical point. If the initial perturbation at $t=0$ is small, then the dynamics of $\eta$ at initial times will be dominated by $\mathcal{L}$. In this parameter region, the eigenfunctions corresponding to the eigenvalue $\lambda_1$ is linearly unstable and all other eigenfunctions are linearly stable. Hence, after some time $t$, the leading order contribution to $\eta(\vec{\boldsymbol{\sigma}},t)$ will come from the eigenfunctions of $\lambda_1$, whereas contribution from the eigenfunctions corresponding to any other eigenvalue $\lambda_n,n\geq2$ will be sub-leading. During the initial times of the dynamics, we may write
\begin{eqnarray}
    \eta(\vec{\boldsymbol{\sigma}},t) \approx \vec{\mathbf{A}}\cdot \vec{\boldsymbol{\sigma}},
\end{eqnarray}
where the components of $\vec{\mathbf{A}}$ denote the amplitude along the eigenfunctions of $\lambda_1$. At late times, when the components of $\vec{\mathbf{A}}$ are sufficiently large, $\mathcal{N}[\eta]$ will start contributing to the dynamics of $\eta$. The eigenfunctions with the least negative eigenvalue will contribute in the next leading order. In our case, these eigenfunctions are $\vec{\boldsymbol{\sigma}}\cdot\mathbf{B}\vec{\boldsymbol{\sigma}}$ with the eigenvalue $\lambda_2 = -2DT$. Hence, in the next leading order, we may write
\begin{equation}
    \eta(\vec{\boldsymbol{\sigma}},t) \approx \vec{\mathbf{A}} \cdot \vec{\boldsymbol{\sigma}} +\vec{\boldsymbol{\sigma}}\cdot\mathbf{B}\vec{\boldsymbol{\sigma}}. \label{eq: eta A B Approx}
\end{equation}
Since the contribution from $\mathcal{N}[\eta]$ originates as an effect of the leading-order contribution from the most unstable eigenmode, we may assume that $\mathbf{B}$ is a function of $\vec{\mathbf{A}}$. Hence, to satisfy the transformation rule given in Eq.~\eqref{eq: A B Rule}, the matrix $\mathbf{B}$ is restricted to have the following form:
\begin{equation}
    \mathbf{B} = b_0\Big(\vec{\mathbf{A}}\cdot\vec{\mathbf{A}}\Big) \mathbf{I}_D+b_1\Big(\vec{\mathbf{A}}\cdot\vec{\mathbf{A}}\Big) \vec{\mathbf{A}}\otimes\vec{\mathbf{A}}, \label{eq: B in A A 1}
\end{equation}
where $b_0$ and $b_1$ are scalar functions of $\big|\vec{\mathbf{A}}\big|^2$. To preserve the traceless property $\mathrm{Tr}\big[\mathbf{B}\big] = 0$, we must have $b_0\big(\vec{\mathbf{A}}\cdot\vec{\mathbf{A}}\big) = -D^{-1}\big|\vec{\mathbf{A}}\big|^2b_1\big(\vec{\mathbf{A}}\cdot\vec{\mathbf{A}}\big)$. 
Further assuming $b_1\Big(\vec{\mathbf{A}}\cdot\vec{\mathbf{A}}\Big)$ has a Taylor expansion $b_1\Big(\vec{\mathbf{A}}\cdot\vec{\mathbf{A}}\Big) = b_{10} + b_{12}\big|\vec{\mathbf{A}}\big|^2+b_{14} \big|\vec{\mathbf{A}}\big|^4 +\ldots$, we obtain in the leading order that $\mathbf{B}$ can be written as
\begin{equation}
    \mathbf{B} =b_{10}\bigg[\vec{\mathbf{A}}\otimes\vec{\mathbf{A}}-\frac{\big|\vec{\mathbf{A}}\big|^2}{D}\mathbf{I}_D\bigg]+\ldots, \label{eq: B leading exp}
\end{equation}
where $b_{10}$ is to be determined later. Using this in Eq.~\eqref{eq: eta A B Approx}, we can express $\eta(\vec{\boldsymbol{\sigma}},t)$ entirely in terms of $\vec{\mathbf{A}}$. Our goal now is to obtain an evolution equation for $\vec{\mathbf{A}}$ from the evolution equation of $\eta$ as given in Eq.~\eqref{eq: del eta del t}.

Putting Eq.~\eqref{eq: eta A B Approx} back into Eqs.~\eqref{eq: L eta model 2}~and~\eqref{eq: NN eta model 2} along with the results from Appendices~\ref{app: 6}~and~\ref{app: 10}, we obtain
\begin{equation}
    \mathcal{L}\eta = \lambda_1 \vec{\mathbf{A}} \cdot\vec{\boldsymbol{\sigma}} +\lambda_2~\vec{\boldsymbol{\sigma}}\cdot \mathbf{B}\vec{\boldsymbol{\sigma}}, \label{eq: L ETA ANSATZ}
\end{equation}
and
\begin{align}
    \mathcal{N}[\eta] &= K_1 \Omega_D \Big(\vec{\mathbf{A}}\cdot  \vec{\boldsymbol{\sigma}}\Big)^2 -\frac{2\Omega_D}{D}\bigg[K_1+\frac{2K_2}{(D+2)}\bigg]\vec{\mathbf{A}} \cdot \mathbf{B}\vec{\boldsymbol{\sigma}}\nonumber\\
    &-\frac{K_1 \Omega_D}{D}\vec{\mathbf{A}}\cdot \vec{\mathbf{A}} + \frac{4(D+1)K_2 \Omega_D}{D(D+2)}\Big( \vec{\boldsymbol{\sigma}}\cdot \mathbf{B}\vec{\boldsymbol{\sigma}} \Big)^2\nonumber\\
    &+ \frac{(D+1)\Omega_D}{D}\bigg[K_1+\frac{4K_2}{(D+2)}\bigg]
    \Big(\vec{\mathbf{A}}\cdot  \vec{\boldsymbol{\sigma}}\Big)\Big(\vec{\boldsymbol{\sigma}} \cdot \mathbf{B} \vec{\boldsymbol{\sigma}}\Big)\nonumber\\
    & -\frac{8K_2\Omega_D}{D(D+2)}\vec{\boldsymbol{\sigma}}\cdot \mathbf{B}^2\vec{\boldsymbol{\sigma}}. \label{eq: N ETA Ansatz model 2}
\end{align}
The eigenvalues $\lambda_1$ and $\lambda_2$ are given in Eq.~\eqref{eq: eigen gen D}.

Taking a derivative with respect to time on both sides of Eq.~\eqref{eq: eta A B Approx}, we get
\begin{equation}
    \frac{\partial \eta}{\partial t} = \frac{d\vec{\mathbf{A}}}{dt} \cdot \vec{\boldsymbol{\sigma}} +\vec{\boldsymbol{\sigma}}\cdot\frac{d\mathbf{B}}{dt}\vec{\boldsymbol{\sigma}}. \label{eq: deta da}
\end{equation}
Multiplying both sides of Eq.~\eqref{eq: deta da} with $\vec{\boldsymbol{\sigma}}$ and integrating over $\mathcal{S}$ and using the result from Appendix~\ref{app: 10}, we obtain
\begin{equation}
    \int \vec{\boldsymbol{\sigma}}\frac{\partial \eta}{\partial t}~d\Omega = \frac{\Omega_D}{D}\frac{d\vec{\mathbf{A}}}{dt},
\end{equation}
which, along with Eq.~\eqref{eq: del eta del t}, gives
\begin{equation}
    \frac{d\vec{\mathbf{A}}}{dt}=\frac{D}{\Omega_D}\int \vec{\boldsymbol{\sigma}}\mathcal{L}\eta~d\Omega+\frac{D}{\Omega_D}\int \vec{\boldsymbol{\sigma}}\mathcal{N}[\eta]~d\Omega. \label{eq: da int l n}
\end{equation}
Computing with the expressions of $\mathcal{L}\eta$ and $\mathcal{N}[\eta]$ from Eqs.~\eqref{eq: L ETA ANSATZ}~and~\eqref{eq: N ETA Ansatz model 2} and using the results from Appendix~\ref{app: 10}, we obtain 
\begin{equation}
    \frac{D}{\Omega_D}\int \vec{\boldsymbol{\sigma}}\mathcal{L}\eta~d\Omega = \lambda_1\vec{\mathbf{A}},\label{eq: Leta int K1}
\end{equation}
and
\begin{align}
    &\frac{D}{\Omega_D}\int \vec{\boldsymbol{\sigma}}\mathcal{N}[\eta]~d\Omega  = -\frac{2\Omega_D}{D(D+2)}\Bigg[K_1-    \frac{2D}{(D+2)}K_2\Bigg]\mathbf{B}\vec{\mathbf{A}} \nonumber\\
    &= -\frac{2(D-1)\Omega_D}{D^2(D+2)}\Bigg[K_1-    \frac{2D}{(D+2)}K_2\Bigg]b_{10}\big|\vec{\mathbf{A}}\big|^2\vec{\mathbf{A}}, \label{eq: Neta int K1 K2}
\end{align}
where in obtaining the last equality, we have used the leading order expression of $\mathbf{B}$ as given in Eq.~\eqref{eq: B leading exp}. Putting Eqs.~\eqref{eq: Leta int K1}~and~\eqref{eq: Neta int K1 K2} back into Eq.~\eqref{eq: da int l n}, we obtain
\begin{equation}
    \frac{d\vec{\mathbf{A}}}{dt} = \lambda_1\vec{\mathbf{A}}  -\frac{2(D-1)\Omega_D}{D^2(D+2)}\Bigg[K_1-    \frac{2D}{(D+2)}K_2\Bigg]b_{10}\big|\vec{\mathbf{A}}\big|^2\vec{\mathbf{A}}. \label{eq: da/dt almost model 2}
\end{equation}

Thus, we have obtained the evolution equation of $\vec{\mathbf{A}}$ upto the unknown constant $b_{10}$. We now have to compute the coefficient $b_{10}$. Rearranging Eq.~\eqref{eq: deta da} together with Eq.~\eqref{eq: del eta del t}, we obtain
\begin{equation}
\vec{\boldsymbol{\sigma}}\cdot\frac{d\mathbf{B}}{dt}\vec{\boldsymbol{\sigma}} = \mathcal{L}\eta +\mathcal{N}[\eta]-\frac{d\vec{\mathbf{A}}}{dt} \cdot \vec{\boldsymbol{\sigma}} . \label{eq: to b10 1}
\end{equation}
Using Eqs.~\eqref{eq: L ETA ANSATZ},~\eqref{eq: N ETA Ansatz model 2}~and~\eqref{eq: da/dt almost model 2} in Eq.~\eqref{eq: to b10 1} and keeping terms up to $\mathcal{O}\big(\big|\vec{\mathbf{A}}\big|^2\big)$, we obtain
\begin{equation}
    \vec{\boldsymbol{\sigma}}\cdot\frac{d\mathbf{B}}{dt}\vec{\boldsymbol{\sigma}} =  \lambda_2\vec{\boldsymbol{\sigma}}\cdot \mathbf{B}\vec{\boldsymbol{\sigma}}+ \Omega_D K_1\Big(\vec{\mathbf{A}}\cdot  \vec{\boldsymbol{\sigma}}\Big)^2 -\frac{ \Omega_DK_1}{D}\vec{\mathbf{A}}\cdot \vec{\mathbf{A}}, \label{eq: to b10 2 model 2}
\end{equation}
where we have used the fact that in leading order, $\mathbf{B}$ is $\mathcal{O}\big(\big|\vec{\mathbf{A}}\big|^2\big)$, which is evident from Eq.~\eqref{eq: B leading exp}. Again, using Eq.~\eqref{eq: B leading exp}, we further obtain, up to leading order, that
\begin{equation}
    \vec{\boldsymbol{\sigma}}\cdot \mathbf{B}\vec{\boldsymbol{\sigma}}=b_{10}\Big[ \Big(\vec{\mathbf{A}}\cdot  \vec{\boldsymbol{\sigma}}\Big)^2 -\frac{ 1}{D}\vec{\mathbf{A}}\cdot \vec{\mathbf{A}}\Big] + \ldots. \label{eq: sigma B sigma leading}
\end{equation}
Putting this back into Eq.~\eqref{eq: to b10 2 model 2} we obtain
\begin{equation}
\vec{\boldsymbol{\sigma}}\cdot\frac{d\mathbf{B}}{dt}\vec{\boldsymbol{\sigma}} =  \Big[\lambda_2b_{10}+ \Omega_D K_1\Big]\Big[\Big(\vec{\mathbf{A}}\cdot  \vec{\boldsymbol{\sigma}}\Big)^2 -\frac{ 1}{D}\vec{\mathbf{A}}\cdot \vec{\mathbf{A}}\Big]. \label{eq: to b10 3}
\end{equation}

\begin{figure}
\includegraphics[width=1.0\linewidth]{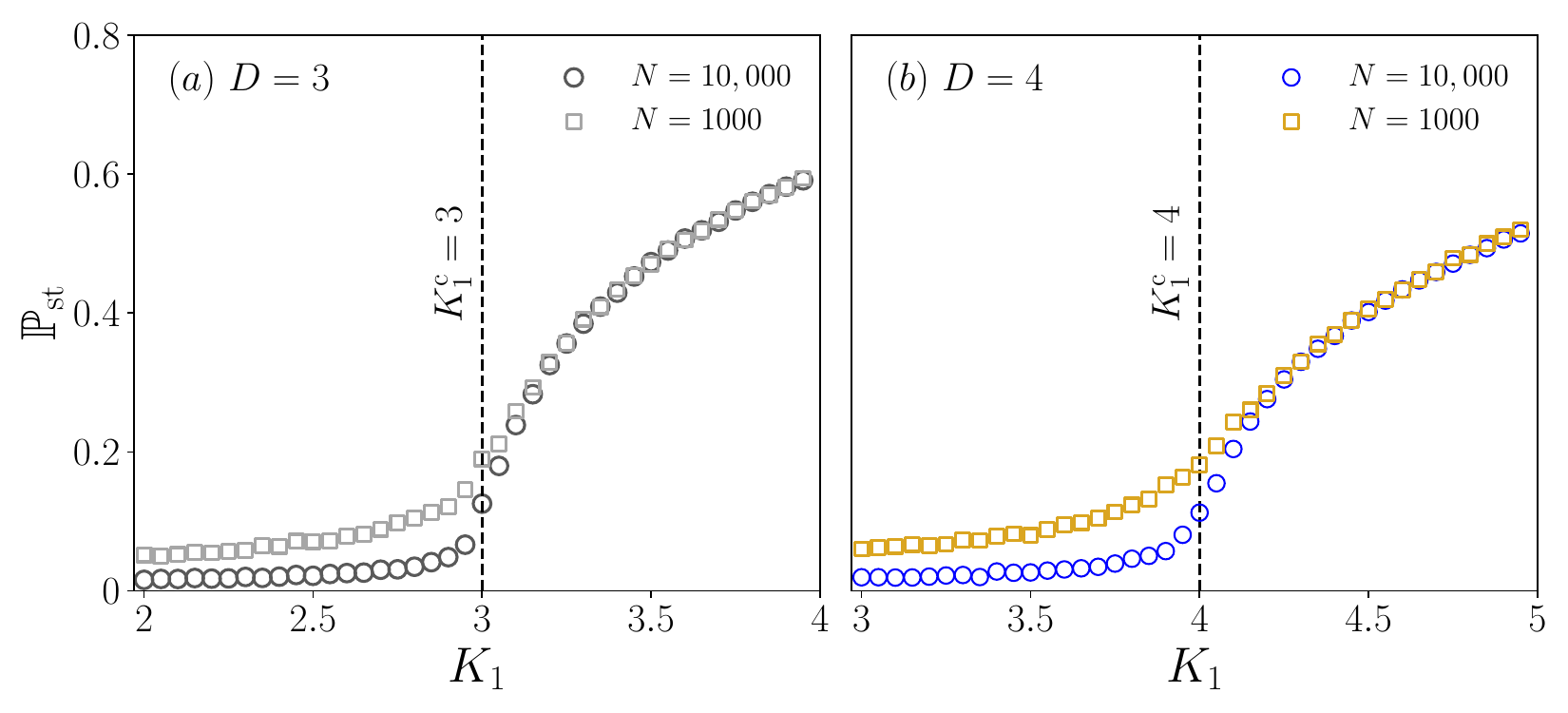}
    \caption{The figure shows numerical-integration results for the stationary state $|\vec{\mathbb{P}}|$, i.e., $\mathbb{P}_\mathrm{st}$ versus $K_1$ for $D = 3$ and $4$, respectively for the case with the noise strength $T = 1$ and $K_2=0$. The black dashed line represents the theoretical prediction of the critical points. All simulation data presented in this paper were generated by integrating the equations of motion using a combination of standard fourth-order Runge-Kutta and Euler algorithms, with integration time step equal to $0.001$.}
    \label{fig:2}
\end{figure}

We now focus on computing the left hand side of Eq.~\eqref{eq: to b10 3} directly from Eq.~\eqref{eq: B leading exp}. Taking the time derivative of Eq.~\eqref{eq: B leading exp}, we obtain
\begin{equation}
    \frac{d\mathbf{B} }{dt}=b_{10}\bigg[\frac{d\vec{\mathbf{A}}}{dt}\otimes\vec{\mathbf{A}}+\vec{\mathbf{A}}\otimes\frac{d\vec{\mathbf{A}}}{dt}-\frac{2}{D}\bigg(\vec{\mathbf{A}}\cdot\frac{d\vec{\mathbf{A}}}{dt}\bigg)\mathbf{I}_D\bigg]+\ldots. \label{eq: db/dt from exp}
\end{equation}
Using Eq.~\eqref{eq: da/dt almost model 2} up to leading order, we obtain
\begin{equation}
    \frac{d\mathbf{B} }{dt}=2\lambda_1b_{10}\bigg[\vec{\mathbf{A}}\otimes\vec{\mathbf{A}}-\frac{\big|\vec{\mathbf{A}}\big|^2}{D}\mathbf{I}_D\bigg]+\ldots,
\end{equation}
which immediately gives
\begin{equation}
    \vec{\boldsymbol{\sigma}}\cdot\frac{d\mathbf{B}}{dt}\vec{\boldsymbol{\sigma}} =  2\lambda_1b_{10}\bigg[\Big(\vec{\mathbf{A}}\cdot  \vec{\boldsymbol{\sigma}}\Big)^2 -\frac{ 1}{D}\vec{\mathbf{A}}\cdot \vec{\mathbf{A}}\bigg]. \label{eq: to b10 4}
\end{equation}
Comparing Eqs.~\eqref{eq: to b10 3}~and~\eqref{eq: to b10 4}, we finally get
\begin{equation}
     b_{10} = \frac{\Omega_DK_1}{2\lambda_1-\lambda_2}.\label{eq: b10}
\end{equation}

We have thus completely characterized Eq.~\eqref{eq: da/dt almost model 2}. Let us now convert the evolution equation of $\vec{\mathbf{A}}$ into the evolution equation of the order parameter $\mathbb{P} = |\vec{\mathbb{P}}|$ as defined in Eq.~\eqref{eq: order parameter 1 final model 2}. Putting Eq.~\eqref{eq: F and eta}~along with Eq.~\eqref{eq: eta A B Approx} into Eq.~\eqref{eq: order parameter 1 final model 2}, we obtain (see Appendix~\ref{app: 10})
\begin{equation}
    \vec{\mathbb{P}} = \int \vec{\boldsymbol{\sigma}}~\eta\left(\vec{\boldsymbol{\sigma}},t\right)d\Omega = \int \vec{\boldsymbol{\sigma}}\Big(\vec{\mathbf{A}} \cdot \vec{\boldsymbol{\sigma}} \Big)d\Omega = \frac{\Omega_D}{D}\vec{\mathbf{A}}. \label{eq: P to A}
\end{equation}
Substituting this into Eq.~\eqref{eq: da/dt almost model 2} and taking the inner product with $\vec{\mathbb{P}}$ on both sides and noting that $\mathbb{P}^2 = \vec{\mathbb{P}}\cdot\vec{\mathbb{P}}$, we finally obtain that near criticality, the evolution of the order parameter $\mathbb{P}$ satisfies, up to second order, the following equation:
\begin{equation}
    \frac{d\mathbb{P}}{dt} = \lambda_1\mathbb{P}-c_3\mathbb{P}^3, \label{eq: dp/dt final}
\end{equation}
where we have
\begin{align}
    c_3&=\frac{D(D-1)K_1}{(D+2)}\nonumber\\
    &\times\left[\frac{(D+2)K_1-2DK_2}{(D-1)(D+2)K_1-2DK_2+D(D+2)T}\right]\label{eq: c3 k1 k2}.
\end{align}
Note that Eq.~\eqref{eq: dp/dt final} has the usual normal form of a pitchfork bifurcation. 

At criticality, we have $K_1 = K_1^\mathrm{c} = DT$. Putting this into the expression of $c_3$, we obtain
\begin{equation}
    c_3\big|_\mathrm{critical} = \frac{D(D-1)K_1}{(D+2)}\left[\frac{(D+2)T-2K_2}{D(D+2)T-2K_2}\right]. \label{eq: c3 critical model 2}
\end{equation}
The parameter region we are interested in here is $K_2\leq K_2^\mathrm{c} = D(D+2)T/2$. From Eq.~\eqref{eq: c3 critical model 2}, it is clear that in this parameter region of our interest, the denominator of $c_3\big|_\mathrm{critical}$ is always positive. Hence, the sign of $c_3\big|_\mathrm{critical}$ is entirely dependent on its numerator. Explicitly, we have
\begin{subequations}\label{eq:c3critical}
\begin{align}
    c_3\big|_\mathrm{critical} &> 0, 
        &&\text{when } K_2 < (D+2)T/2, \label{eq:c3critical:a}\\
     &= 0, 
        &&\text{when } K_2 = (D+2)T/2, \label{eq:c3critical:b}\\
    &< 0, 
        &&\text{when } K_2 > (D+2)T/2. \label{eq:c3critical:c}
\end{align}
\end{subequations}
Hence, the order parameter $\mathbb{P}$ shows at $K_1 = K_1^\mathrm{c}\equiv DT$ a continuous transition for $0 \leq K_2 < (D+2)T/2$ and a discontinuous transition for $(D+2)T/2 \leq K_2 < D(D+2)T/2$. Hence, $K_2 = K_2^\mathrm{tc} \equiv  (D+2)/2$ is similar to a tri-critical point, around which the system changes its nature of phase transition. The corresponding phase diagram in shown in Fig.~\ref{fig:schematic}.

To close, let us compare our expressions with respect to standard results. For $D=2$, Eq.~\eqref{eq: dp/dt final} reduces to
\begin{equation}
    \frac{d\mathbb{P}}{dt} = \bigg[\frac{K_1}{2}-T\bigg]\mathbb{P}-\frac{K_1}{2}\left[\frac{K_1-K_2}{K_1-K_2+2T}\right]\mathbb{P}^3, 
\end{equation}
a result reported recently in Ref.~\cite{rupakfinite2025}, see \textit{Application 1} (note that $\mathbb{P}$ is identical to $R_1 = 2\pi |A|$ used therein; also, in this reference, the symbol $D$ was used to denote the noise strength).

When $K_2=0$, at criticality, we have $K_1 = K_1^\mathrm{c} = DT$. Putting this into the expression of $c_3$, we obtain
\begin{equation}
    c_3\big|_\mathrm{critical} = \frac{D(D-1)T}{(D+2)}.
\end{equation}
Since $ c_3\big|_\mathrm{critical}>0$ for any general dimension $D\geq2$, we conclude that in the presence of noise and without any quenched disorder, the transition becomes continuous in arbitrary dimensions. This is unlike the noiseless case with quenched disorder as discussed in Ref.~\cite{Sarthak2019}, where even and odd dimensions behave differently.

\begin{figure*}
    \centering
    \includegraphics[width=\linewidth]{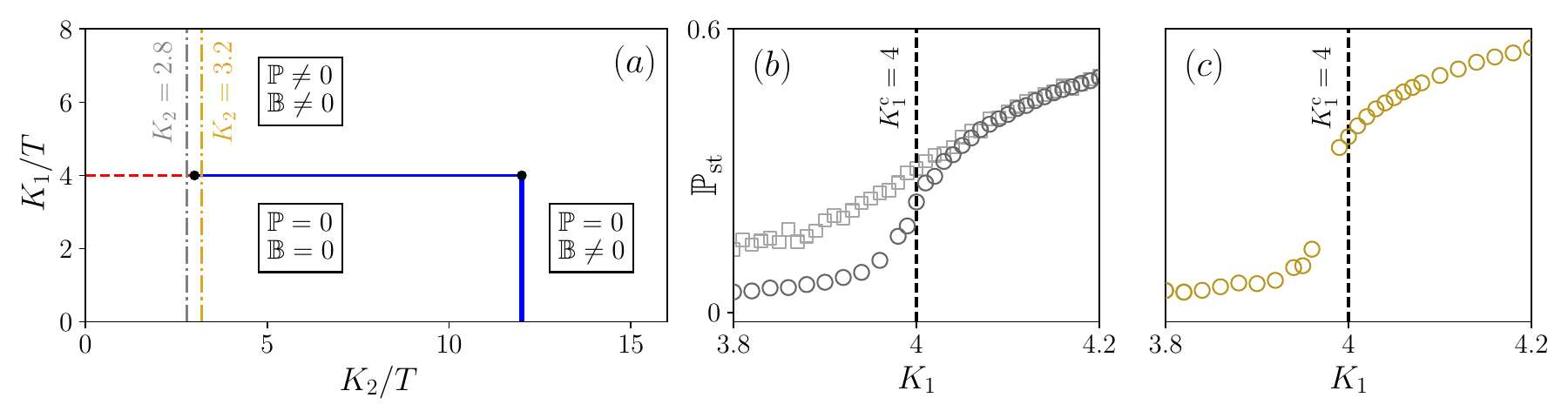}
    \caption{Figure (a) shows the $N\to \infty$ phase diagram of the model defined in Sec.~\ref{sec: model 1 formulation} for $D=4$. The red dashed line marks a continuous transition, while the blue solid line indicates a discontinuous transition; at both transitions, the synchronization order parameter $\mathbb{P}_\mathrm{st}$ become nonzero simultaneously. The thick blue vertical solid line denotes a discontinuous transition in $\mathbb{B}_\mathrm{st}$ alone, across which $\mathbb{P}_\mathrm{st}$ remains zero. Panels (b) and (c) show the behavior of numerically obtained $\mathbb{P}_\mathrm{st}$ as a function of $K_1$ for a continuous case $(K_2 = 2.8)$ and a discontinuous case $(K_2=3.2)$, respectively. The noise strength is $T=1.0$. The square and the circular markers correspond to $N=10^3$ and $10^4$ respectively.}
    \label{fig:placeholder}
\end{figure*}

In Figs.~\ref{fig:2} and \ref{fig:placeholder}, we demonstrate that our 
theoretical predictions for both the critical points (the one for $\mathbb{P}$ and the one for $\mathbb{B}$) and the nature of the 
associated phase transitions are consistent with simulations for both 
odd and even dimensions. Specifically, the $D=3$ case is shown in 
Fig.~\ref{fig:2}(a), while the $D=4$ results are shown in 
Fig.~\ref{fig:2}(b) and Fig.~\ref{fig:placeholder}. Figure~\ref{fig:2} 
presents simulation results in the absence of second-harmonic interaction for 
$D=3$ and $D=4$, whereas Fig.~\ref{fig:placeholder} illustrates the phase 
behavior in the presence of second-harmonic interaction for the representative 
case $D=4$. Figure~\ref{fig:placeholder}(a) shows the corresponding phase diagram, with 
the values of the order parameters in the different phases indicated. In 
panel (b), we plot the stationary-state value of the order parameter $\mathbb{P}_{\mathrm{st}}$ as 
a function of $K_1$ for a representative value $K_2=2.8$, where the system 
undergoes a continuous transition. In contrast, panel (c) shows 
$\mathbb{P}_{\mathrm{st}}$ versus $K_1$ for $K_2=3.2$, illustrating a 
discontinuous transition. Finally, the dependence of the
order parameter $\mathbb{B}_{\mathrm{st}}$ on $K_2$ at fixed $K_1$ for 
different dimensions is shown in Fig.~\ref{fig: 411}.

We may also compute higher order terms in Eq.~\eqref{eq: dp/dt final} by considering $\eta(\vec{\boldsymbol{\sigma}},t)$ in Eq.~\eqref{eq: eta A B Approx} up to third leading order as
\begin{equation}
    \eta(\vec{\boldsymbol{\sigma}},t) \approx \vec{\mathbf{A}} \cdot \vec{\boldsymbol{\sigma}} +\vec{\boldsymbol{\sigma}}\cdot\mathbf{B}\vec{\boldsymbol{\sigma}}+\Big(\mathbb{T}^{(3)},\vec{\boldsymbol{\sigma}}^{\otimes3}\Big), \label{eq: eta A B Approx model 2 3rd}
    \end{equation}
where the symmetric and traceless property, along with the symmetry condition~\eqref{eq: A B Rule general} requires
\begin{align}
    T^{(3)}_{ijk} &= h_0\Big(\vec{\mathbf{A}}\cdot\vec{\mathbf{A}}\Big) \nonumber\\
    &\times\bigg\{A_iA_jA_k-\frac{A^2}{(D+2)}\Big[\delta_{ij}A_k+\delta_{ik}A_j+\delta_{jk}A_i\Big]\bigg\},
\end{align}
where $h_0(\cdot)$ is a scalar function. Following the same procedure as before, we obtain the modification to  Eq.~\eqref{eq: dp/dt final}, which reads as

\begin{equation}
    \frac{d\mathbb{P}}{dt} = \lambda_1\mathbb{P}-c_3\mathbb{P}^3-c_5\mathbb{P}^5,\label{eq: dp/dt final model 2 3rd}
\end{equation}
where the coefficient $c_5$ reads as
\begin{align}
    c_5 &= \frac{2(D-1)\Omega_D}{D^2(D+2)^2}\Bigg\{\bigg[(D+2)K_1-2DK_2\bigg]b_{12}\nonumber\\
    &+\frac{12DK_2}{(D+2)(D+4)}b_{10}h_{10}\Bigg\},
\end{align}
where we have
\begin{align}
    h_{10} = \frac{(D+1)\Omega_D}{D(3\lambda_1-\lambda_3)}\bigg[K_1+\frac{4}{(D+2)}K_2\bigg]b_{10},
\end{align}
and 
\begin{align}
    b_{12} &= \frac{1}{(4\lambda_1-\lambda_2)}\bigg\{2c_3b_{10} + \frac{2\Omega_D}{D(D+2)(D+4)}\nonumber\\
    &\times\Big[4(D-2)K_2b_{10}^2-3DK_1h_{10}\Big]\bigg\}.
\end{align}
The coefficients $\lambda_1, c_3$, and $b_{10}$ are already given in Eqs.~\eqref{eq: eigen gen D},~\eqref{eq: c3 k1 k2}~and~\eqref{eq: b10}. 

The main summary here is that the synchronization transition remains continuous with mean-field scaling, independent of dimension.
 Having discussed the transition in $\vec{\mathbb{P}}$, we now discuss in the following section the transition exhibited by the quantity $\widetilde{\mathbb{M}}$.

\subsection{\texorpdfstring{Transition in $\widetilde{\mathbb{M}}$}{Transition in M}\label{sec: tran in M}}
In this section, the parameter region we are interested in is $K_1 \leq K_1^\mathrm{c}$ and $K_2>K_2^\mathrm{c}$. In this parameter range, eigenvalue $\lambda_2$ is positive and the rest of the eigenvalues are negative. Hence, the eigenfunctions corresponding to the eigenvalue $\lambda_2$ become linearly unstable, whereas the rest of the eigenfunctions remain linearly stable. As a result, any perturbation in the eigenspace of $\lambda_2$, which has the form $\big(\mathbb{T}^{(2)},\vec{\boldsymbol{\sigma}}\otimes \vec{\boldsymbol{\sigma}}\big) \equiv \vec{\boldsymbol{\sigma}}\cdot \mathbf{B}\vec{\boldsymbol{\sigma}}$, will keep on growing with time in the leading order, while perturbation along any other eigenspace will go to zero. If the initial perturbation at $t=0$ is small, then the dynamics of $\eta$ at initial times will be dominated by $\mathcal{L}$. Hence, after some time $t$, the leading order contribution to $\eta(\vec{\boldsymbol{\sigma}},t)$ will come from the eigenfunctions of $\lambda_2$, whereas contribution from the eigenfunctions corresponding to any other eigenvalue $\lambda_n,n\neq2$ will be sub-leading. During the initial times of the dynamics, $\eta(\vec{\boldsymbol{\sigma}},t)$ can be written up to leading order as
\begin{eqnarray}
    \eta(\vec{\boldsymbol{\sigma}},t) \approx \vec{\boldsymbol{\sigma}}\cdot \mathbf{B}\vec{\boldsymbol{\sigma}}.
\end{eqnarray}

At late times, when the elements of $\mathbf{B}$ are sufficiently large, $\mathcal{N}[\eta]$ will start contributing to the dynamics of $\eta$. Usually, the eigenfunctions with the least negative eigenvalue will contribute in the next leading order. In our case, these eigenfunctions are either $\vec{\mathbf{A}}\cdot\vec{\boldsymbol{\sigma}}$ with the eigenvalue $\lambda_1$ or $\Big(\mathbb{T}^{(3)},\vec{\boldsymbol{\sigma}}\otimes \vec{\boldsymbol{\sigma}}\otimes \vec{\boldsymbol{\sigma}}\Big)$ with the eigenvalue and $\lambda_3$. However, since the contribution from $\mathcal{N}[\eta]$ originates as an effect of the leading-order contribution from the most unstable eigenmode, we may assume that the tensor elements of the second leading-order term can be written in terms of the elements of the highest leading-order term, i.e., the elements of $\mathbf{B}$. Since $\vec{\mathbf{A}}$ is a vector and $\mathbb{T}^{(3)}$ is a rank-$3$ traceless symmetric tensor, its elements can never be written in terms of the elements of a rank-$2$ traceless symmetric tensor $\mathbf{B}$. The immediate next higher-order symmetric traceless tensor that can be constructed with the elements of $\mathbf{B}$ is rank-$4$. Hence, in the next leading order, we may write
\begin{equation}
    \eta(\vec{\boldsymbol{\sigma}},t) \approx \vec{\boldsymbol{\sigma}}\cdot\mathbf{B}\vec{\boldsymbol{\sigma}}+\Big(\mathbb{T}^{(4)},\vec{\boldsymbol{\sigma}}^{\otimes4}\Big),\label{eq: eta B T Approx model 2}
\end{equation}
where $\mathbb{T}^{(4)}$ is a traceless symmetric rank-$4$ tensor. The most general $\mathbb{T}^{(4)}$ one can construct is~\cite{JARIC20032123}
\begin{align}
    &T^{(4)}_{ijkl} = b_4\Big(\mathrm{Tr}\big[\mathbf{B}^2\big],\mathrm{Tr}\big[\mathbf{B}^3\big],\ldots,\mathrm{Tr}\big[\mathbf{B}^D\big]\Big)\Bigg[\Big(B_{ij}B_{kl} \nonumber\\
    &+ B_{ik}B_{jl}+B_{il}B_{jk}\Big)-\frac{2}{(D+4)}\Big[\delta_{ij}\big[\mathbf{B}^2\big]_{kl}+\delta_{ik}\big[\mathbf{B}^2\big]_{jl}\nonumber\\
    &+\delta_{il}\big[\mathbf{B}^2\big]_{jk}+\delta_{jk}\big[\mathbf{B}^2\big]_{il}+\delta_{jl}\big[\mathbf{B}^2\big]_{ik}+\delta_{kl}\big[\mathbf{B}^2\big]_{ij}\Big] \nonumber\\
    &+\frac{2\mathrm{Tr}\big[\mathbf{B}^2\big]}{(D+2)(D+4)} \Big(\delta_{ij}\delta_{kl} + \delta_{ik}\delta_{jl}+\delta_{il}\delta_{jk}\Big)\Bigg], \label{eq: TIJKL}
\end{align}
where $T^{(4)}_{ijkl}$ denotes the $ijkl$-th element of $\mathbb{T}^{(4)}$. In Appendix~\ref{app: 13}, we proved that the expression of the elements of $\mathbb{T}^{(4)}$ given in Eq.~\eqref{eq: TIJKL} is traceless and symmetric. Using this expression, we may express $\big(\mathbb{T}^{(4)},\vec{\boldsymbol{\sigma}}^{\otimes 4}\big)$ in a compact form as
\begin{align}
    &\Big(\mathbb{T}^{(4)},\vec{\boldsymbol{\sigma}}^{\otimes4}\Big) = b_4\Big(\mathrm{Tr}\big[\mathbf{B}^2\big],\ldots,\mathrm{Tr}\big[\mathbf{B}^D\big]\Big)\bigg[3\Big(\vec{\boldsymbol{\sigma}}\cdot\mathbf{B}\vec{\boldsymbol{\sigma}}\Big)^2\nonumber\\
    &-\frac{12}{(D+4)}\Big(\vec{\boldsymbol{\sigma}}\cdot\mathbf{B}^2\vec{\boldsymbol{\sigma}}\Big)+\frac{6}{(D+2)(D+4)}\mathrm{Tr}\big[\mathbf{B}^2\big]\bigg]. \label{eq: T4 B4 S4 FULL}
\end{align}
Performing a Taylor series expansion of the function $b_4$ as
\begin{equation}
    b_4\Big(\mathrm{Tr}\big[\mathbf{B}^2\big],\ldots,\mathrm{Tr}\big[\mathbf{B}^D\big]\Big) = b_{40} + \sum_{m=2}^D b_{4,m}\mathrm{Tr}\big[\mathbf{B}^m\big]+\cdots,
\end{equation}
using which, in the leading order, we may write
\begin{align}
    \Big(\mathbb{T}^{(4)},\vec{\boldsymbol{\sigma}}^{\otimes4}\Big) &\approx 3b_{40}\bigg[\Big(\vec{\boldsymbol{\sigma}}\cdot\mathbf{B}\vec{\boldsymbol{\sigma}}\Big)^2-\frac{4}{(D+4)}\Big(\vec{\boldsymbol{\sigma}}\cdot\mathbf{B}^2\vec{\boldsymbol{\sigma}}\Big)\nonumber\\
    &+\frac{2\mathrm{Tr}\big[\mathbf{B}^2\big]}{(D+2)(D+4)}\bigg]. \label{eq: T4 contraction exp}
\end{align}
Putting Eq.~\eqref{eq: eta B T Approx model 2} back into Eqs.~\eqref{eq: L eta model 2}~and~\eqref{eq: NN eta model 2} along with the results from Appendices~\ref{app: 6},~\ref{app: 10}, and~\ref{app20}, we obtain
\begin{equation}
    \mathcal{L}\eta = \lambda_2\vec{\boldsymbol{\sigma}}\cdot \mathbf{B}\vec{\boldsymbol{\sigma}}+\lambda_4\Big(\mathbb{T}^{(4)},\vec{\boldsymbol{\sigma}}^{\otimes 4} \Big), \label{eq: L ETA ANSATZ model 2 M}
\end{equation}
and
\begin{align}
    \mathcal{N}[\eta] &= \frac{4\Omega_DK_2}{D}\bigg[\Big(\vec{\boldsymbol{\sigma}}\cdot\mathbf{B}\vec{\boldsymbol{\sigma}}\Big)^2-\frac{2}{(D+2)}\Big(\vec{\boldsymbol{\sigma}}\cdot\mathbf{B}^2\vec{\boldsymbol{\sigma}}\Big)\bigg] \nonumber\\
    &+\frac{12\Omega_DK_2}{D(D+2)(D+4)}b_{40} \bigg[(D+4)^2\Big(\vec{\boldsymbol{\sigma}}\cdot\mathbf{B}\vec{\boldsymbol{\sigma}}\Big)^3 \nonumber\\
    &+8\Big(\vec{\boldsymbol{\sigma}}\cdot\mathbf{B}^3\vec{\boldsymbol{\sigma}}\Big)-8(D+3)\Big(\vec{\boldsymbol{\sigma}}\cdot\mathbf{B}\vec{\boldsymbol{\sigma}}\Big)\Big(\vec{\boldsymbol{\sigma}}\cdot\mathbf{B}^2\vec{\boldsymbol{\sigma}}\Big) \nonumber\\
    &+\frac{2D\mathrm{Tr}\big[\mathbf{B}^2\big]}{(D+2)}\Big(\vec{\boldsymbol{\sigma}}\cdot\mathbf{B}\vec{\boldsymbol{\sigma}}\Big)
    \bigg]. \label{eq: N ETA Ansatz model 2 M}
\end{align}

Taking a derivative with respect to time on both sides Eq.~\eqref{eq: eta B T Approx model 2}, we get
\begin{equation}
    \frac{\partial \eta}{\partial t} = \vec{\boldsymbol{\sigma}}\cdot\frac{d\mathbf{B}}{dt}\vec{\boldsymbol{\sigma}}+\left(\frac{d\mathbb{T}^{(4)}}{dt},\vec{\boldsymbol{\sigma}}^{\otimes4}\right). \label{eq: deta da model 2}
\end{equation}
Multiplying both sides of Eq.~\eqref{eq: deta da model 2} with $\vec{\boldsymbol{\sigma}}\otimes \vec{\boldsymbol{\sigma}}$ and integrating over $\mathcal{S}$ and using the result from Appendix~\ref{app: 10}, we obtain
\begin{equation}
    \int \vec{\boldsymbol{\sigma}}\otimes\vec{\boldsymbol{\sigma}}~\frac{\partial \eta}{\partial t}~d\Omega = \frac{2\Omega_D}{D(D+2)}\frac{d\mathbf{B}}{dt},
\end{equation}
which, along with Eq.~\eqref{eq: del eta del t}, gives
\begin{align}
    \frac{d\mathbf{B}}{dt}&=\frac{D(D+2)}{2\Omega_D}\int \vec{\boldsymbol{\sigma}}\otimes\vec{\boldsymbol{\sigma}}~\mathcal{L}\eta~d\Omega\nonumber\\
    &+\frac{D(D+2)}{2\Omega_D}\int \vec{\boldsymbol{\sigma}}\otimes\vec{\boldsymbol{\sigma}}~\mathcal{N}[\eta]~d\Omega. \label{eq: da int l n model 2}
\end{align}
Computing with the expressions of $\mathcal{L}\eta$ and $\mathcal{N}[\eta]$ from Eqs.~\eqref{eq: L ETA ANSATZ model 2 M}~and~\eqref{eq: N ETA Ansatz model 2 M} and using the results from Appendix~\ref{app: 10}, we obtain from the linear part that
\begin{equation}
    \frac{D(D+2)}{2\Omega_D}\int \vec{\boldsymbol{\sigma}}\otimes \vec{\boldsymbol{\sigma}}~\mathcal{L}\eta~d\Omega = \lambda_2\mathbf{B}.\label{eq: Leta int K1 K2 M}
\end{equation}
The integral of the non-linear part gives
\begin{align}
    &\frac{D(D+2)}{2\Omega_D}\int \vec{\boldsymbol{\sigma}}\otimes \vec{\boldsymbol{\sigma}}~\mathcal{N}[\eta]~d\Omega \nonumber\\
    &= \frac{8\Omega_DK_2}{(D+2)(D+4)}\Bigg[\mathbf{B}^2-\frac{\mathrm{Tr}\big[\mathbf{B}^2\big]}{D}\mathbf{I}_D\Bigg] \nonumber\\
    &+\alpha b_{40}\Bigg[2\mathrm{Tr}\big[\mathbf{B}^3\big]\mathbf{I}_D-\alpha_1 \mathrm{Tr}\big[\mathbf{B}^2\big]\mathbf{B}-2D\mathbf{B}^3\Bigg],\label{eq: Neta int K1 K2 M}
\end{align}
where we have $\alpha_1 = (D^2+6D+12)/(D+2)$ and $\alpha = 96 \Omega_DK_2/[D(D+2)(D+4)^2(D+6)]$. Renaming
\begin{align}
    \beta\equiv \frac{8\Omega_DK_2}{(D+2)(D+4)},
\end{align}
and combining Eqs.~\eqref{eq: da int l n model 2},~\eqref{eq: Leta int K1 K2 M} and ~\eqref{eq: Neta int K1 K2 M}, we obtain
\begin{align}
    \frac{d\mathbf{B}}{dt} &= \lambda_2\mathbf{B}+\beta\Bigg[\mathbf{B}^2-\frac{\mathrm{Tr}\big[\mathbf{B}^2\big]}{D}\mathbf{I}_D\Bigg] \nonumber\\
    &+\alpha b_{40}\Bigg[2\mathrm{Tr}\big[\mathbf{B}^3\big]\mathbf{I}_D-\alpha_1 \mathrm{Tr}\big[\mathbf{B}^2\big]\mathbf{B}-2D\mathbf{B}^3\Bigg]. \label{eq: DM DT almost}
\end{align}

We now have to compute the coefficient $b_{40}$. Rearranging Eq.~\eqref{eq: deta da model 2} together with Eq.~\eqref{eq: del eta del t}, we obtain
\begin{equation}
\left(\frac{d\mathbb{T}^{(4)}}{dt},\vec{\boldsymbol{\sigma}}^{\otimes4}\right) = \mathcal{L}\eta +\mathcal{N}[\eta]-\vec{\boldsymbol{\sigma}}\cdot\frac{d\mathbf{B}}{dt}\vec{\boldsymbol{\sigma}} . \label{eq: to b40 1}
\end{equation}
Using Eqs.~\eqref{eq: L ETA ANSATZ model 2 M},~\eqref{eq: N ETA Ansatz model 2 M}~and~\eqref{eq: DM DT almost} and keeping terms up to $\mathcal{O}\big(\mathbf{B}^2\big)$, we obtain
\begin{align}
    &\left(\frac{d\mathbb{T}^{(4)}}{dt},\vec{\boldsymbol{\sigma}}^{\otimes4}\right) = \lambda_4\Big(\mathbb{T}^{(4)},\vec{\boldsymbol{\sigma}}^{\otimes 4} \Big)\nonumber\\
&+\frac{4\Omega_DK_2}{D}\bigg[\Big(\vec{\boldsymbol{\sigma}}\cdot\mathbf{B}\vec{\boldsymbol{\sigma}}\Big)^2-\frac{2}{(D+2)}\Big(\vec{\boldsymbol{\sigma}}\cdot\mathbf{B}^2\vec{\boldsymbol{\sigma}}\Big)\bigg] 
. \label{eq: to b40 2}
\end{align}

We now focus on computing the left hand side of Eq.~\eqref{eq: to b40 2} directly from Eq.~\eqref{eq: T4 contraction exp}. Taking the time derivative of Eq.~\eqref{eq: T4 contraction exp} and using Eq.~\eqref{eq: DM DT almost} up to leading order, we obtain
\begin{align}
    \bigg(\frac{d\mathbb{T}^{(4)}}{dt},\vec{\boldsymbol{\sigma}}^{\otimes4}\bigg) =2\lambda_2\Big(\mathbb{T}^{(4)},\vec{\boldsymbol{\sigma}}^{\otimes 4} \Big). \label{eq: dTDt direct}
\end{align}
Comparing Eqs.~\eqref{eq: to b40 2}~and~\eqref{eq: dTDt direct}, we obtain
\begin{align}
    &\Big(\mathbb{T}^{(4)},\vec{\boldsymbol{\sigma}}^{\otimes 4} \Big) \nonumber\\
    &= \frac{4\Omega_DK_2}{D(2\lambda_2-\lambda_4)}\bigg[\Big(\vec{\boldsymbol{\sigma}}\cdot\mathbf{B}\vec{\boldsymbol{\sigma}}\Big)^2 -\frac{2}{(D+2)}\Big(\vec{\boldsymbol{\sigma}}\cdot\mathbf{B}^2\vec{\boldsymbol{\sigma}}\Big)\bigg]. \label{eq: eq: to b40 3}
\end{align}

In the standard language of the irreducible representation of a rank-$4$ symmetric tensor, the left hand side of Eq.~\eqref{eq: eq: to b40 3} contains only spin-$4$ terms. The right hand side of Eq.~\eqref{eq: eq: to b40 3} contains spin-$4$ and spin-$2$ terms~\cite{JARIC20032123}. Since each of these irreducible subspaces is orthogonal to the others, we should only compare the spin-$4$ part of the right hand side of Eq.~\eqref{eq: eq: to b40 3} with the left hand side. To extract the spin-$4$ part of the right hand side, we apply a projection operator $\mathcal{P}_4$, which projects onto the spin-$4$ subspace. We can construct this operator as
\begin{align}
    \mathcal{P}_4 \equiv \frac{1}{8(D+2)(D+4)}\nabla^2_\mathcal{S}\Big[\nabla^2_\mathcal{S}+2D\Big]. \label{eq: proj def}
\end{align}

Before applying $\mathcal{P}_4$ on both sides of Eq.~\eqref{eq: eq: to b40 3}, let us first rewrite its right hand side in a suitable form. Using Eq.~\eqref{eq: T4 contraction exp}, we may write
\begin{align}
    &\bigg[\Big(\vec{\boldsymbol{\sigma}}\cdot\mathbf{B}\vec{\boldsymbol{\sigma}}\Big)^2-\frac{2}{(D+2)}\Big(\vec{\boldsymbol{\sigma}}\cdot\mathbf{B}^2\vec{\boldsymbol{\sigma}}\Big)\bigg] = \frac{1}{3b_{40}}\Big(\mathbb{T}^{(4)},\vec{\boldsymbol{\sigma}}^{\otimes 4} \Big)\nonumber\\
    &+\frac{2D}{(D+2)(D+4)} \bigg[\Big(\vec{\boldsymbol{\sigma}}\cdot\mathbf{B}^2\vec{\boldsymbol{\sigma}}\Big)-\frac{1}{D}\mathrm{Tr}\big[\mathbf{B}^2\big]\bigg].
\end{align}

Putting it back into Eq.~\eqref{eq: eq: to b40 3} and rearranging, we may write
\begin{align}
    &\bigg[1-\frac{4\Omega_D K_2}{3D(2\lambda_2-\lambda_4)}\frac{1}{b_{40}}\bigg]\Big(\mathbb{T}^{(4)},\vec{\boldsymbol{\sigma}}^{\otimes 4} \Big)\nonumber\\
    &=\frac{8\Omega_DK_2}{(D+2)(D+4)(2\lambda_2-\lambda_4)} \bigg[\Big(\vec{\boldsymbol{\sigma}}\cdot\mathbf{B}^2\vec{\boldsymbol{\sigma}}\Big)-\frac{1}{D}\mathrm{Tr}\big[\mathbf{B}^2\big]\bigg]. \label{eq: eq: to b40 4}
\end{align}
In Appendix~\ref{app: 21}, we show that
\begin{align}
    &\mathcal{P}_4\bigg[\Big(\vec{\boldsymbol{\sigma}}\cdot\mathbf{B}^2\vec{\boldsymbol{\sigma}}\Big)-\frac{1}{D}\mathrm{Tr}\big[\mathbf{B}^2\big]\bigg] = 0, \label{eq: 145}\\
    &\mathcal{P}_4\Big(\mathbb{T}^{(4)},\vec{\boldsymbol{\sigma}}^{\otimes 4} \Big) = \Big(\mathbb{T}^{(4)},\vec{\boldsymbol{\sigma}}^{\otimes 4} \Big). \label{eq: 146}
\end{align}
Hence, applying $\mathcal{P}_4$ on both sides of Eq.~\eqref{eq: eq: to b40 4}, we immediately obtain
\begin{align}
    b_{40} = \frac{4\Omega_D K_2}{3D(2\lambda_2-\lambda_4)}.
\end{align}
Putting this back into Eq.~\eqref{eq: DM DT almost}, we have obtained the evolution equation of the matrix $\mathbf{B}$.

 Let us now convert the evolution equation of $\mathbf{B}$ into the evolution equation of the order parameter $\widetilde{\mathbb{M}}$ as defined in Eq.~\eqref{eq: M TILDE}. Putting Eq.~\eqref{eq: eta B T Approx model 2} into Eq.~\eqref{eq: M TILDE}, we obtain (see Appendix~\ref{app: 10})
\begin{align}
    \widetilde{\mathbb{M}} &= \int \vec{\boldsymbol{\sigma}} \otimes \vec{\boldsymbol{\sigma}}~ \eta\left(\vec{\boldsymbol{\sigma}},t\right)~d\Omega = \int \vec{\boldsymbol{\sigma}} \otimes \vec{\boldsymbol{\sigma}}~ \Big(\vec{\boldsymbol{\sigma}}\cdot\mathbf{B}\vec{\boldsymbol{\sigma}}\Big)~d\Omega\nonumber\\
    &= \frac{2\Omega_D}{D(D+2)}\mathbf{B}.
\end{align}
Using this in Eq.~\eqref{eq: DM DT almost}, we obtain the evolution equation of the $\widetilde{\mathbb{M}}$, which reads as
\begin{align}
    \frac{d\widetilde{\mathbb{M}}}{dt} &= \lambda_2\widetilde{\mathbb{M}}+\bar{\beta}\Bigg[\widetilde{\mathbb{M}}^2-\frac{\mathrm{Tr}\big[\widetilde{\mathbb{M}}^2\big]}{D}\mathbf{I}_D\Bigg] \nonumber\\
    &+\bar{\alpha}b_{40}\Bigg[2\mathrm{Tr}\big[\widetilde{\mathbb{M}}^3\big]\mathbf{I}_D-\alpha_1 \mathrm{Tr}\big[\widetilde{\mathbb{M}}^2\big]\widetilde{\mathbb{M}}-2D\widetilde{\mathbb{M}}^3\Bigg],\label{eq: DM DT}
\end{align}
where $\bar{\beta} = \beta D(D+2)/(2\Omega_D)$ and $\bar{\alpha} = \alpha [D(D+2)/(2\Omega_D)]^2$. When $\lambda_2<0$, the order parameter $\widetilde{\mathbb{M}}$ becomes $\mathbb{O}_D$, a matrix with all elements equal to zero at long times. On the other hand, when $\lambda_2>0$, the order parameter $\widetilde{\mathbb{M}}$ starts having non-zero matrix elements at long times. Since $\widetilde{\mathbb{M}}$ is always a traceless matrix, $\mathrm{Tr}\big[\widetilde{\mathbb{M}}\big]$ cannot capture this effect. The simplest scalar that captures this transition is
\begin{equation}
    \mathbb{B} \equiv \mathrm{Tr}\Big[\widetilde{\mathbb{M}}^2\Big].
\end{equation}
Multiplying both sides of Eq.~\eqref{eq: DM DT almost} with $2\widetilde{\mathbb{M}}$ and taking trace, we obtain the evolution equation of $\mathbb{B}$, which reads
\begin{equation}
    \frac{d\mathbb{B}}{dt} = 2\lambda_2\mathbb{B}+2\bar{\beta}\mathrm{Tr}\big[\widetilde{\mathbb{M}}^3\big]-2\bar{\alpha}b_{40}\Big[\alpha_1\mathbb{B}^2+2D\mathrm{Tr}\big[\widetilde{\mathbb{M}}^4\big]\Big], \label{eq: D BB DT}
\end{equation}
where we have used that $\widetilde{\mathbb{M}}$ is traceless.

Below, we consider successively the cases $D=2$ and $D>2$.

\subsubsection{\texorpdfstring{ $D=2$ Case}{D=2 Case}}

For $D=2$, the most general traceless symmetric matrix has the form
\begin{equation}
    \widetilde{\mathbb{M}}|_{D=2} = 
    \begin{bmatrix}
        a & b\\
        b & -a
    \end{bmatrix}.
\end{equation}
This immediately gives the eigenvalues of the matrix $\widetilde{\mathbb{M}}$ as $\pm \sqrt{a^2+b^2}$. Hence, for $D=2$, we have $\mathrm{Tr}\Big[\widetilde{\mathbb{M}}^{2m+1}\Big] = 0$ and $\mathrm{Tr}\Big[\widetilde{\mathbb{M}}^{2m}\Big] = \big(\mathrm{Tr}\big[\widetilde{\mathbb{M}}^{2}\big]\big)^m/2^{m-1}$. Thus, for $D=2$, we have from their definitions, $\Omega_D = 2\pi,~\bar{\alpha} = K_2/(3\pi),~\alpha_1=7,~b_{40} = 3\pi K_2/[3(K_2+4T)]$, which reduces Eq.~\eqref{eq: D BB DT} to
\begin{equation}
     \frac{d\mathbb{B}}{dt} = 2\lambda_2\mathbb{B}-\frac{4K_2^2}{(K_2+4T)}\mathbb{B}^2. \label{eq: D BB DT D=2}
\end{equation}
Since the coefficient of $\mathbb{B}^2$ is positive, the order parameter $\mathbb{B}$ shows a continuous transition. Note that, unlike the transition in $\vec{\mathbb{P}}$, where the nature of the transition depends on the interaction strength $K_2$, here the nature of the transition is independent of the interaction strength $K_1$. The leading non-linear exponent in the reduced equation~\eqref{eq: D BB DT D=2} is $2$, which turns out to be a special property of two dimensions. Indeed, we will show in the next section that for any other dimension $D>2$, this exponent is $3/2$ (see Eq.~\eqref{eq: D BB DT D}).

Simulation results for the stationary-state value of the order parameters $\mathbb{P}_\mathrm{st}$ and $\mathbb{B}_\mathrm{st}$ as a function of the the second-harmonic coupling $K_2$ is shown in Fig.~\ref{fig: 411}(a) for $D=2$ with fixed values $K_{1}=1.0=T$ and system size $N=10^{4}$. Clearly, $\mathbb{P}_\mathrm{st}$ does not show a transition, whereas $\mathbb{B}_\mathrm{st}$ shows a continuous transition as predicted by our theory. A representative configuration 
in the symmetry-broken phase at $K_2=6.0$ is shown in Fig.~\ref{fig: 411}(d), 
where the oscillators spontaneously organize into two macroscopic clusters.

\subsubsection{\texorpdfstring{ $D>2$ Case}{D=2 Case}}

For general $D$, the condition $\mathrm{Tr}\big[\widetilde{\mathbb{M}}\big] = 0$ does not automatically guarantee $\mathrm{Tr}\Big[\widetilde{\mathbb{M}}^{2m+1}\Big] = 0$. Hence, the second leading order term in Eq.~\eqref{eq: D BB DT} is $\mathrm{Tr}\big[\widetilde{\mathbb{M}}^3\big]$. Note that Eq.~\eqref{eq: D BB DT} is still not closed in $\mathbb{B}$ since it contains terms $\mathrm{Tr}\big[\widetilde{\mathbb{M}}^3\big]$ and $\mathrm{Tr}\big[\widetilde{\mathbb{M}}^4\big]$. To obtain a closed evolution equation of $\mathbb{B}$ from Eq.~\eqref{eq: D BB DT}, we need to use the symmetry of the problem. Since our system as given in Eq.~\eqref{eq: 2D vectorisation model1 2 final} does not have any preferential direction, the correlation between different axes should be identical for all possible pairs of axes. Hence, in the symmetry-adapted basis, the matrix $\widetilde{\mathbb{M}}$ should have the following form
\begin{eqnarray}
    \widetilde{\mathbb{M}} = 
    \begin{bmatrix}
        0 & b & b & \cdots & b\\
        b & 0 & b & \cdots & b\\
         b & b & 0 & \cdots & b\\
        \vdots& \vdots&&\ddots&  \vdots\\
        b & b &b & \cdots& 0
    \end{bmatrix}.
\end{eqnarray}
This matrix has one non-degenerate eigenvalue $(D-1)b$ and $(D-1)$-fold degenerate eigenvalue $-b$. Hence, we can immediately write $\mathrm{Tr}\big[\widetilde{\mathbb{M}}^2\big] = D(D-1)b^2$. Similarly, we can write for any $m$ that $\mathrm{Tr}\Big[\widetilde{\mathbb{M}}^{2m}\Big] = (D-1)\Big[(D-1)^{2m-1}+1\Big]b^{2m}$ and $\mathrm{Tr}\Big[\widetilde{\mathbb{M}}^{2m+1}\Big] = (D-1)\Big[(D-1)^{2m}-1\Big]b^{2m+1}$. Using these, we may write
\begin{align}
    \mathrm{Tr}\Big[\widetilde{\mathbb{M}}^3\Big] = \frac{(D-2)}{\sqrt{D(D-1)}} \mathbb{B}^{3/2}
\end{align}
and
\begin{align}
    \mathrm{Tr}\Big[\widetilde{\mathbb{M}}^4\Big] = \frac{1}{D}\left(\frac{D^2-3D+3}{D-1}\right) \mathbb{B}^{2}.
\end{align}
Up to second leading order, Eq.~\eqref{eq: D BB DT} reads as
\begin{equation}
    \frac{d\mathbb{B}}{dt} = 2\lambda_2\mathbb{B}+ \frac{2(D-2)\bar{\beta}}{\sqrt{D(D-1)}} \mathbb{B}^{3/2}. \label{eq: D BB DT D}
\end{equation}
Since $\bar{\beta}>0$, the order parameter $\mathbb{B}$ shows a discontinuous transition for any $D\geq 3$. In contrast to the case $D=2$, we see that the leading non-linear exponent in the reduced equation~\eqref{eq: D BB DT D} is $3/2$.

Simulation results for the stationary-state order parameters 
$\mathbb{P}_{\mathrm{st}}$ and $\mathbb{B}_{\mathrm{st}}$ as functions of the 
second-harmonic coupling $K_2$ are shown in Fig.~\ref{fig: 411}(b) for $D=3$ 
and Fig.~\ref{fig: 411}(c) for $D=4$, for fixed $K_1=1.0=T$ and system size 
$N=10^4$. While $\mathbb{P}_{\mathrm{st}}$ does not show a transition as predicted, 
$\mathbb{B}_{\mathrm{st}}$ exhibits a sharp discontinuous jump, in agreement 
with our theoretical prediction. A representative configuration in the 
symmetry-broken phase for $D=3$ at $K_2=11.0$ is shown in 
Fig.~\ref{fig: 411}(e), where the system clearly splits into two macroscopic 
clusters, resulting in nonzero correlation based order parameter $\widetilde{\mathbb{M}}$ and correspondingly $\mathbb{B}$.

\begin{figure*}
\includegraphics[width=1.0\linewidth]{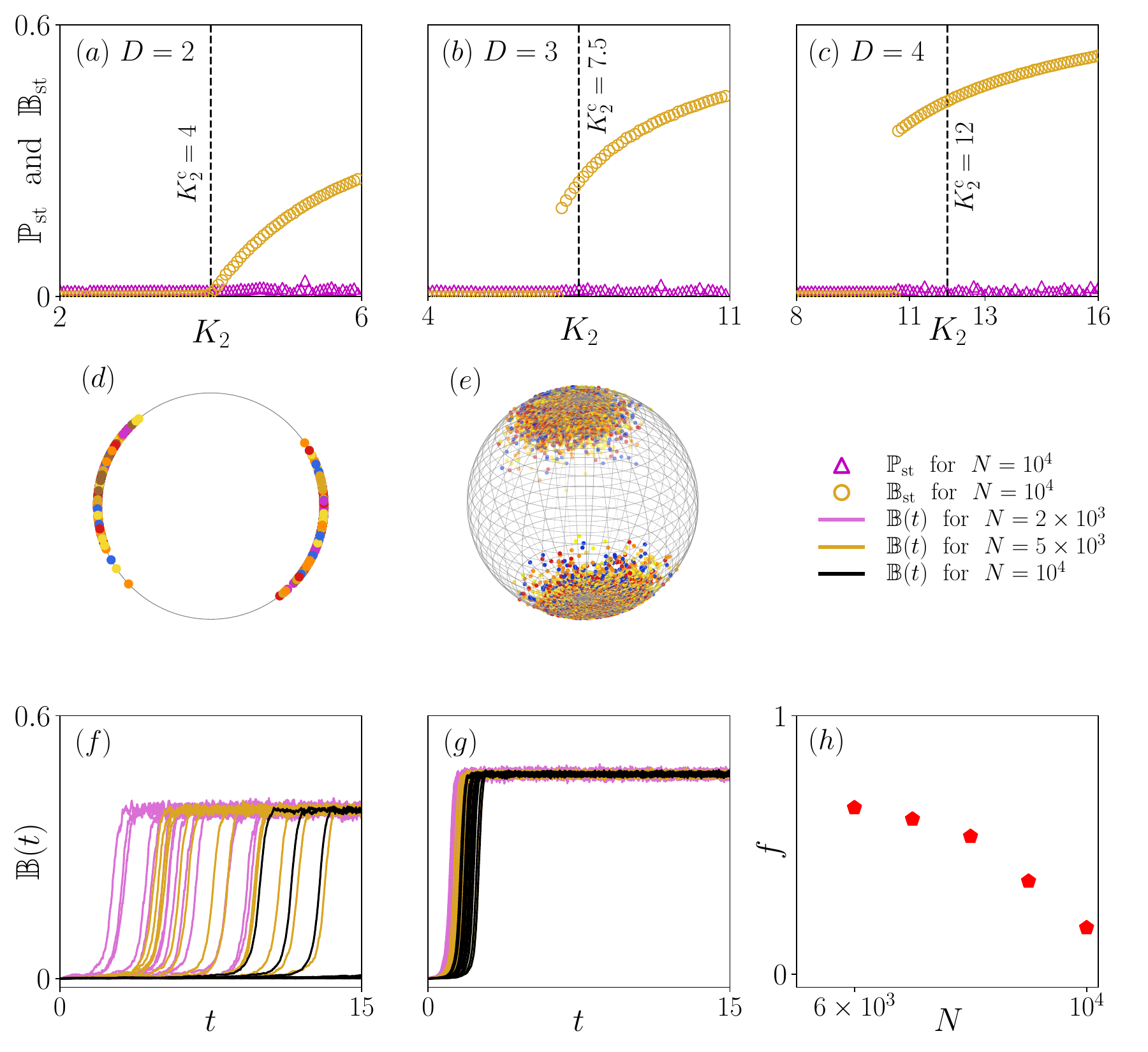}
   \caption{Stationary value of the order parameters $\mathbb{P}_{\mathrm{st}}$ (triangular symbols) and $\mathbb{B}_{\mathrm{st}}$ (circular symbols) are shown as functions of the second–harmonic coupling $K_{2}$ for fixed $K_{1}=1.0=T$ and system size $N=10^{4}$. Results are displayed for (a) $D=2$, (b) $D=3$, and (c) $D=4$.
Panels (d) and (e) show typical late–time configurations of the oscillators for $D=2$ at $K_{2}=6.0$ and for $D=3$ at $K_{2}=11.0$, respectively.
For $D=4$, panels (f) and (g) display the time evolution of $\mathbb{B}(t)$ for fifteen independent noise realizations at $K_{2}=11.0~(<12.0=K_2^\mathrm{c})$ and $K_{2}=13.0~(>12.0=K_2^\mathrm{c})$, respectively. Panel (h) shows as a function of the system size $N$ at fixed $K_{2}=11.0$ the fraction of realizations (out of a total of forty–five noise realizations) in which the order parameter $\mathbb{B}$ jumps from zero to a nonzero value within an observation time $t=15$.}

    \label{fig: 411}
\end{figure*}

An interesting feature of Figs.~\ref{fig: 411}(b) and (c) is that the jump in 
$\mathbb{B}_{\mathrm{st}}$ does not occur precisely at the point where the 
uniformly incoherent state loses linear stability. To elucidate this 
behavior, we examine the time evolution of $\mathbb{B}(t)$ near the critical 
coupling $K_2=K_2^{\mathrm c}$. Each of the figures~\ref{fig: 411}(f) and (g) shows 
$\mathbb{B}(t)$ for fifteen independent noise realizations, for three system sizes $N=2\times 10^3,5\times 10^3,10^4$, at two 
representative values: $K_2=11.0<K_2^{\mathrm c}$ (panel (f)) and 
$K_2=13.0>K_2^{\mathrm c}$ (panel (g)).

For $K_2>K_2^{\mathrm c}$, the uniformly incoherent state is linearly 
unstable, and in all realizations the system rapidly evolves toward a 
symmetry-broken double clustered state, as seen in Fig.~\ref{fig: 411}(g). In 
contrast, for $K_2<K_2^{\mathrm c}$, the uniformly incoherent state is 
linearly stable, so trajectories initiated close to it should remain 
trapped there. However, Fig.~\ref{fig: 411}(f) shows at long times a relaxation to symmetry-broken clustered state. This is due to the fact that the simulations are done with finite number of oscillators, and the associated finite-size fluctuations induce transitions out of the metastable incoherent state with small but finite probability~\cite{rupakfinite2025}. Consistently, increasing $N$ suppresses these finite-size fluctuations, leading to 
longer residence times near the uniformly-incoherent state. We found that the
fraction $f$ of realizations relaxing to the clustered state within a fixed time decreases with $N$ (see Fig.~\ref{fig: 411}(h)). In the thermodynamic limit $N\to\infty$, these 
fluctuation-induced escapes are suppressed, and the uniformly incoherent 
state remains stable for all $K_2<K_2^{\mathrm c}$, fully consistent with our 
linear-stability analysis.

The main summary of the current section is that in contrast to the synchronization transition, the correlation-driven transition exhibits a dimension-dependent nature, being continuous in two dimensions and discontinuous otherwise.

\section{Conclusion\label{sec: conclusion}}

We have studied the effect of higher-harmonic interactions and annealed disorder in a generalized $D$-dimensional Kuramoto model. The collective dynamics of the system is characterized by two distinct macroscopic descriptors: (1) The conventional synchronization order parameter characterizing how the position vectors of the different oscillators are spread out in space, and is thus a measure of global order in the system. (2) A correlation matrix capturing how different components of an oscillator’s position vector correlate when averaged over the ensemble. Our central finding is that these two measures of collective order need not behave identically and, in fact, display qualitatively different critical behaviors depending on the dimensionality of the system.

To build intuition about the results for the general $D$-dimensional formulation, it is useful to briefly consider the simplest cases. In $D = 1$, the unit vector $\vec{\sigma}_j$ reduces to a scalar taking values $\pm 1$, so the dynamics becomes effectively that of globally-coupled binary variables. In this limit, the order parameter $\vec{P}$ reduces to the average ``magnetization,'' while the matrix $M$ becomes trivial due to the absence of transverse components, and no nontrivial rotational structure exists. As a result, the system does not exhibit a genuine synchronization transition; rather, the only possible ordering corresponds to trivial alignment, and no distinct correlation-driven phase transition arises. By contrast, the case $D = 2$ corresponds to the conventional Kuramoto model, where each oscillator is described by a phase $\theta_j$ on the unit circle, with $\vec{\sigma}_j = (\cos\theta_j, \sin\theta_j)$. In this case, the order parameter $P = |\vec{P}|$ coincides with the standard Kuramoto order parameter measuring global phase coherence, while the matrix $M$ encodes second-harmonic correlations. In the presence of annealed disorder and fundamental coupling alone, one recovers a continuous synchronization transition with universal mean-field scaling; Upon introducing higher-harmonic interactions, the transition can become discontinuous depending on the interaction strength, see Eq.~\eqref{eq: dp/dt final} and the discussion following the equation. At the same time, the correlation-driven transition associated with $M$ remains continuous in two dimensions, see Eq.~\eqref{eq: D BB DT D=2}, highlighting the special role of $D = 2$ as the boundary between qualitatively different nonlinear behaviors observed in higher dimensions, Eq.~\eqref{eq: D BB DT D}. These low-dimensional limits illustrate how the general framework interpolates between a minimal scalar system and the familiar phase-oscillator picture, thereby clarifying the physical meaning of the vector and tensor order parameters introduced for arbitrary $D$.

Remarkably, the macroscopic state of the system cannot be fully characterized by the magnitude of the synchronization order parameter alone. The uniformly incoherent state can lose stability in two qualitatively distinct ways: either through the emergence of global synchronization, signaled by both order parameters becoming nonzero, or through the formation of multiple clusters that break rotational symmetry without generating any global order. The latter transition is captured by the correlation matrix becoming nonzero, while the global synchronization order parameter remains strictly zero. While the nature of the first kind of transition is determined by the strength of the higher harmonic interaction, the nature of the second kind of transition is determined entirely by the dimensionality of the system. In particular, the transition of the correlation matrix is continuous for $D=2$ and becomes discontinuous for any $D>2$ with distinct dynamical exponents.

On the analytical side, we put forward a non-trivial generalization of the standard centre-manifold reduction used for the two-dimensional Kuramoto model to arbitrary dimensions. This extension allows us to go beyond linear stability and derive closed evolution equations for the macroscopic order parameters, providing direct access to the nonlinear dynamics near the transition. Importantly, the method is not restricted to the specific model studied here, and its principles can be applied to a broad class of mean-field higher-dimensional models, opening the door to systematic analytical studies of nonlinear synchronization phenomena in high dimensions.

\begin{figure*}
\includegraphics[width=1.0\linewidth]{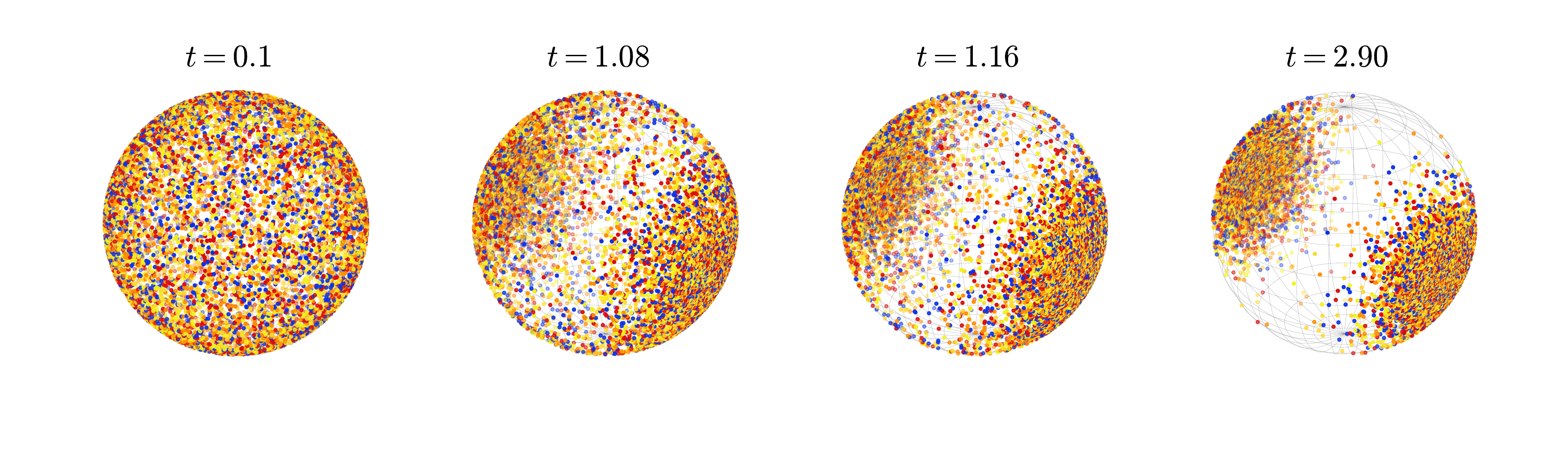}
   \caption{The figure shows for $D=3$ the dynamics of the phases of $10^4$ oscillators for parameter values $K_1=1.0,~K_2=11.0,~T=1.0$ and integration timestep $0.01$. One may observe how starting with a state with $\mathbb{P}=\mathbb{B}=0$ the system evolves to a state with $\mathbb{B}\ne 0$, yet $\mathbb{P}=0$.}
    \label{fig: 411}
\end{figure*}

Our results show that higher-harmonic interactions fundamentally enrich the phase structure of generalized Kuramoto models by allowing for transitions that are invisible to standard order parameters but are captured by correlation-based measures. Several natural extensions of this work remain open. An immediate direction is to study the robustness of these phenomena in the presence of quenched disorder, where fluctuations are expected to play a more prominent role. It would also be of interest to investigate network-based coupling, where heterogeneity in interaction topology may further differentiate the roles of synchronization and inter-axis correlations. Another interesting direction would be to understand how the nature and location of the transition can be manipulated through dynamical interventions applied either to the entire system~\cite{sarkar2022synchronization, 3lvs-3xsy} or to a subset of its degrees of freedom~\cite{PhysRevE.109.064137, np7q-hxld}, without tuning the interaction strengths or introducing additional couplings, and how the effectiveness of such interventions depends on the system dimension. Finally, understanding finite-size effects near the observed transitions will be essential for connecting our results to realistic systems and for developing a comprehensive theory of collective behavior in high-dimensional oscillator models.

Coming to real-world applications, two issues of paramount importance that we have not considered are time delays in the interaction (we have considered the oscillator interactions to be instantaneous in time) and network topologies (we have considered the case of fully-connected networks). We briefly comment on how our formalism may be applied to tackle such issues. Time-delayed interactions in the standard Kuramoto model ($D=2$) have already been analyzed in Ref.~\cite{MetivierGupta2019} using a center-manifold expansion, by invoking the theory of delay differential equations. Although this formalism can, in principle, be generalized to higher dimensions ($D>2$), deriving concrete results requires an explicit and nontrivial implementation of the analysis, as key features, such as phase boundaries and the nature of the resulting phase transitions, can only be determined through a detailed computation. Such a computation is deferred to a future publication. To move beyond fully connected networks, a natural first step is to consider a one-dimensional lattice with Kuramoto oscillators residing on its sites. For the case $D=2$, Ref.~\cite{sarkar2020} demonstrates that an approximate time-averaged theory, originally introduced in Ref.~\cite{rodrigues2016kuramoto}, provides an accurate description of the order parameter, particularly in regimes exhibiting continuous transitions. Building on this framework, one can systematically investigate how the phenomena discussed in this work, which are expected to hold in the dense-network limit, are modified as the connectivity is progressively reduced. Such an approach offers a controlled way to assess the robustness of our results against increasing network dilution.

\section{Acknowledgements} 
RM acknowledges useful discussions with Souparna Nath and Akashdeep Roy. We acknowledge generous allocation of computational resources of the Department of Theoretical Physics, TIFR, assistance of Kapil Ghadiali and Ajay
Salve, and the financial support of the Department of Atomic Energy,
Government of India under Project Identification No. RTI 4002.

\appendix

\section{Derivation of \texorpdfstring{Eq.~\eqref{eq: noise relation rot}}{Eq.~(2)}}
\label{app: -1}

The $lm$-th element $\big[\widetilde{\mathbf{N}}_j(t)\big]_{lm}$ of the rotated noise matrix $\widetilde{\mathbf{N}}_j(t)$ is related to the elements of the original noise matrix $\mathbf{N}_j(t)$ as
\begin{equation}
    \big[\widetilde{\mathbf{N}}_j(t)\big]_{lm} = \sum_{n}\sum_{p} R_{ln}\big[\mathbf{N}_j(t)\big]_{np}R^\top_{pm},
\end{equation}
where $R_{ln}$ is the $ln$-th element of $\mathbf{R}$, $\big[\mathbf{N}_j(t)\big]_{np} = \mathrm{sgn}(n-p)~\eta^{(j)}_{np}(t)$ is the $np$-th element of $\mathbf{N}_j(t)$ and $R^\top_{pm}$ is the $pm$-th element of $\mathbf{R}^\top$. Here, $\mathrm{sgn}(x)$ is the standard sign function, or, the signum function. Since $\mathbf{R}$ is a constant matrix, we obtain
\begin{equation}
    \big\langle\big[\widetilde{\mathbf{N}}_j(t)\big]_{lm} \big\rangle = \sum_{n}\sum_{p} R_{ln}\big\langle\big[\mathbf{N}_j(t)\big]_{np}\big \rangle R^\top_{pm} = 0, \label{eq: rotated noise correlation 3}
\end{equation}
since $\big\langle\big[\mathbf{N}_j(t)\big]_{np} \big \rangle= \mathrm{sgn}(n-p)\big\langle\eta^{(j)}_{np}(t)\big \rangle = 0$. Hence we have
\begin{equation}
    \big\langle\widetilde{\mathbf{N}}_j(t)  \big\rangle = \mathbb{O}_D.
\end{equation}

Similarly, the $lm$-th element of $\widetilde{\mathbf{N}}_i(t)\widetilde{\mathbf{N}}_j(t') = \mathbf{R} \mathbf{N}_i(t)\mathbf{R}^\top\mathbf{R}\mathbf{N}_j(t')\mathbf{R}^\top= \mathbf{R} \mathbf{N}_i(t)\mathbf{N}_j(t')\mathbf{R}^\top$ is
\begin{equation}
    \Big[\widetilde{\mathbf{N}}_i(t)\widetilde{\mathbf{N}}_j(t')\Big]_{lm} = \sum_{n}\sum_{p}R_{ln}\big[\mathbf{N}_i(t)\mathbf{N}_j(t')\big]_{np}R^\top_{pm},
\end{equation}
which, upon averaging over the noise, gives
\begin{equation}
    \Big\langle \Big[\widetilde{\mathbf{N}}_i(t)\widetilde{\mathbf{N}}_j(t')\Big]_{lm}\Big\rangle = \sum_{n}\sum_{p}R_{ln}\Big\langle\big[\mathbf{N}_i(t)\mathbf{N}_j(t')\big]_{np}\Big\rangle R^\top_{pm}. \label{eq: rotated noise correlation 1}
\end{equation}
Since for any matrix $\mathbf{B}$, we may write $\big\langle \big[\mathbf{B}\big]_{ab}\big\rangle = \big[\big\langle\mathbf{B}\big \rangle\big]_{ab}$, i.e.,  the average of the $ab$-th element  is equal to the $ab$-th element of the average matrix, we may rewrite the noise terms in Eq.~\eqref{eq: rotated noise correlation 1} as 
\begin{eqnarray}
    \Big\langle\big[\mathbf{N}_i(t)\mathbf{N}_j(t')\big]_{np}\Big\rangle  &=& \Big[\Big\langle\mathbf{N}_i(t)\mathbf{N}_j(t')\Big\rangle \Big]_{np}\nonumber\\
    &=&-(D-1)\delta_{ij}\delta(t-t')\big[\mathbf{I}_D \big]_{np} \nonumber\\
    &=& -(D-1)\delta_{ij}\delta(t-t')\delta_{np}, \label{eq: rotated noise correlation 2}
\end{eqnarray}
where in obtaining the last equality, we have used the properties of the noise matrix from Eq.~\eqref{eq: noise relation}. Putting Eq.~\eqref{eq: rotated noise correlation 2} back into Eq.~\eqref{eq: rotated noise correlation 1}, we obtain
\begin{eqnarray}
    \Big\langle \Big[\widetilde{\mathbf{N}}_i(t)\widetilde{\mathbf{N}}_j(t')\Big]_{lm}\Big\rangle &=& -(D-1)\delta_{ij}\delta(t-t') \sum_{n}R_{ln} R^\top_{nm}\nonumber\\
    &=& -(D-1)\delta_{ij}\delta(t-t')\delta_{lm},
\end{eqnarray}
where, in obtaining the last equality, we have used $ \sum_{n}R_{ln} R^\top_{nm} =  \big[\mathbf{R}\mathbf{R}^\top\big]_{lm} = \big[\mathbf{I}_D\big]_{lm} = \delta_{lm}$. Hence we have
\begin{equation}
    \big\langle \widetilde{\mathbf{N}}_i(t)\widetilde{\mathbf{N}}_j(t')  \big\rangle = -(D-1)\delta_{ij}\delta(t-t')\mathbf{I}_D. \label{eq: rotated noise correlation 21}
\end{equation}

Equations~\eqref {eq: rotated noise correlation 3}~and~\eqref{eq: rotated noise correlation 21} derived here are the Eq.~\eqref{eq: noise relation rot} of the main text. We conclude that the properties of the rotated noise matrix $\widetilde{\mathbf{N}}_j(t)$ are identical to those of  $\mathbf{N}_j(t)$, which are given in Eq.~\eqref{eq: noise relation}.

\section{D-dimensional spherical polar coordinate system \label{app: 0}}

In $D$ dimensions, let a point be represented in the Cartesian coordinate system as $(x_1,x_2,\ldots,x_D)$. In spherical coordinates, the same point may be expressed as $(r, \theta_1, \theta_2, \ldots, \theta_{D-2},\phi)$, where we have $\theta_i \in [0,\pi]$ and $\phi \in [0,2\pi)$. The relation between these coordinates is given by
\begin{eqnarray}
    x_1 &=& r \cos{\theta_1},\\
    x_2 &=& r \sin{\theta_1} \cos{\theta_2},\\
    x_3 &=& r \sin{\theta_1} \sin{\theta_2} \cos{\theta_3},\\
    &\vdots&\\
    x_{D-2} &=& r \sin{\theta_1} \sin{\theta_2} \cdots \sin{\theta_{D-3}} \cos{\theta_{D-2}},\\
    x_{D-1} &=& r \sin{\theta_1} \sin{\theta_2} \cdots \sin{\theta_{D-3}} \sin{\theta_{D-2}} \cos{\phi},\\
    x_D &=& r \sin{\theta_1} \sin{\theta_2} \cdots \sin{\theta_{D-3}} \sin{\theta_{D-2}} \sin{\phi}.
\end{eqnarray}

\section{Derivation of Fokker-Planck Equation\label{app: FP model 1}}

We start by considering a continuous and differentiable test function $h_1(\vec{\boldsymbol{\sigma}}(t))$ defined on the surface $\mathcal{S}$. Similar to how $F\left(\vec{\mathbf{r}},t\right)$ was extended to the entire $\mathbb{R}^D$ space in Eq.~\eqref{eq: F and curly F main text}, we may now extend this function to the entire $\mathbb{R}^D$ space by defining
\begin{eqnarray}
    h(\vec{\boldsymbol{r}}(t)) &=&h_1(\vec{\boldsymbol{\sigma}}(t))\delta(r-1)\nonumber \\&=& h_1\left(\frac{\vec{\boldsymbol{r}}(t)}{r}\right)\delta(r-1). \label{eq: def h}
\end{eqnarray}
As argued in the main text, the dynamics~\eqref{eq: dr/dt model1 main text} will keep the value of any initial $r$ constant throughout the evolution. Note that the function $h$ has no explicit dependence either on time $t$ or on the underlying stochastic evolution. However, since the evolution of $\vec{\mathbf{r}}(t)$ is inherently stochastic (see Eq.~\eqref{eq: dr/dt model1 main text}), the value of the function $h$ at time $t$ will also be random, even when the dynamics of $\vec{\mathbf{r}}$ is initiated from a given initial condition. We will be interested in the average of $h(\vec{\mathbf{r}}(t))$ over all possible stochastic trajectories of $\vec{\mathbf{r}}(t)$ generated by the evolution~\eqref{eq: dr/dt model1 main text}; we denote this average by $\langle h(\vec{\mathbf{r}}(t))\rangle$. This quantity can also be represented as an average with respect to the distribution of $\vec{\mathbf{r}}$, which reads as
\begin{equation}
    \langle h(\vec{\mathbf{r}}(t))\rangle = \int  h(\vec{\mathbf{r}}) \mathcal{F}\left(\vec{\mathbf{r}},t\right)~dV_D. \label{eq: Test function avg}
\end{equation}
Taking the time derivative on both sides of Eq.~\eqref{eq: Test function avg}, we obtain
\begin{equation}
   \left \langle \frac{dh(\vec{\mathbf{r}}(t))}{dt}\right\rangle = \int  h(\vec{\mathbf{r}}) \frac{\partial\mathcal{F}\left(\vec{\mathbf{r}},t\right)}{\partial t}~dV_D. \label{eq: Test function avg d/dt}
\end{equation}

Now, using the standard multidimensional Taylor expansion, we obtain
\begin{eqnarray}
    dh(\vec{\mathbf{r}}(t)) &=& h(\vec{\mathbf{r}}(t)+d\vec{\mathbf{r}})-h(\vec{\mathbf{r}}(t)) \nonumber\\
    &=& \left(d\vec{\mathbf{r}}\cdot\vec{\nabla} \right)h(\vec{\mathbf{r}}(t)) + \frac{1}{2!}\left(d\vec{\mathbf{r}}\cdot\vec{\nabla} \right)^2h(\vec{\mathbf{r}}(t)) + \cdots.\nonumber \\ \label{eq: test function taylor expansion}
\end{eqnarray}
We are interested in terms up to order $dt$ from the right hand side of  Eq.~\eqref{eq: test function taylor expansion}. From Eq.~\eqref{eq: dr/dt model1 main text}, the quantity $d\vec{\mathbf{r}}$ may be expressed as
\begin{equation}
    d\vec{\mathbf{r}} = \vec{\mathbf{v}}dt + \frac{\sqrt{2T}}{r} ~d\mathbf{N}(t)\vec{\mathbf{r}}, \label{eq: dr}
\end{equation}
where we have $\vec{\mathbf{v}} = K_1  \left[\vec{\mathbb{P}}-\left(\vec{\mathbf{r}}\cdot \vec{\mathbb{P}}\right)\vec{\mathbf{r}}/r^2\right]+2 K_2\Big[\mathbb{M}/r-\left(\vec{\mathbf{r}} \cdot \mathbb{M}\vec{\mathbf{r}}\right)\mathbf{I}_D/r^3\Big]\vec{\mathbf{r}}$ and $d\mathbf{N}(t) = \mathbf{N}(t)dt$. Clearly, in the Cartesian coordinates as in Eq.~\eqref{eq: noise matrix}, the independent elements of the matrix $d\mathbf{N}(t)$ are given by independent Wiener processes, as $\big[d\mathbf{N}(t)\big]_{ij} = dW_{ij} = \eta_{ij}(t)dt~\forall~i<j$, where the quantities $W_{ij}$ are Wiener processes. Well-known properties of the Wiener processes give $\langle dW_{ij}dW_{kl}\rangle = \delta_{ik}\delta_{jl}dt$, which further implies that $\langle d\mathbf{N}(t)d\mathbf{N}(t)\rangle = -dt(D-1)\mathbf{I}_D$.

Let us now focus on the first term on the right hand side of the second equality in Eq.~\eqref{eq: test function taylor expansion}. Its average with respect to the stochastic noise gives
\begin{eqnarray}
   && \left\langle \left(d\vec{\mathbf{r}}\cdot\vec{\nabla} \right)h(\vec{\mathbf{r}}(t)) \right \rangle \nonumber \\
    &&= \left\langle \left(\vec{\mathbf{v}}\cdot\vec{\nabla} \right)h(\vec{\mathbf{r}}(t)) \right \rangle dt +  \sqrt{2T}\left\langle \frac{1}{r}\left(d\mathbf{N}(t)\vec{\mathbf{r}}\cdot\vec{\nabla} \right)h(\vec{\mathbf{r}}(t)) \right \rangle, \nonumber \\ \label{eq: first term}
\end{eqnarray}
where we have used Eq.~\eqref{eq: dr}. In the It\^{o} interpretation of stochastic differential equations, the Wiener increment $d\mathbf{N}(t)$ at time $t$ is statistically independent of the state $\vec{\mathbf{r}}(t)$. Hence, we may write
\begin{eqnarray}
     \left\langle\frac{1}{r} \left(d\mathbf{N}(t)\vec{\mathbf{r}}\cdot\vec{\nabla} \right)h(\vec{\mathbf{r}}(t)) \right \rangle &=&  \left\langle \frac{1}{r}\left(\left\langle d\mathbf{N}(t)\right \rangle\vec{\mathbf{r}}\cdot\vec{\nabla} \right)h(\vec{\mathbf{r}}(t)) \right \rangle \nonumber \\
     &=& 0,
\end{eqnarray}
where in obtaining the first equality, we perform the averaging in two steps, by first fixing the value of $\vec{\mathbf{r}}(t)$ and averaging over $d\mathbf{N}(t)$, and finally averaging over $\vec{\mathbf{r}}(t)$. Equation~\eqref{eq: first term} then simplifies to
\begin{equation}
    \left\langle \left(d\vec{\mathbf{r}}\cdot\vec{\nabla} \right)h(\vec{\mathbf{r}}(t)) \right \rangle =\left\langle \left(\vec{\mathbf{v}}\cdot\vec{\nabla} \right)h(\vec{\mathbf{r}}(t)) \right \rangle dt. \label{eq: first term final}
\end{equation}

Let us now turn to the second term in the second equality on the right hand side of Eq.~\eqref{eq: test function taylor expansion}. Using Eq.~\eqref{eq: dr}, the only $\mathcal{O}(dt)$ contribution from that term is then obtained as
\begin{equation}
    \frac{1}{2!}\left(d\vec{\mathbf{r}}\cdot\vec{\nabla} \right)^2h(\vec{\mathbf{r}}(t)) \approx  T\left(\frac{1}{r}~d\mathbf{N}(t)\vec{\mathbf{r}}\cdot\vec{\nabla} \right)^2h(\vec{\mathbf{r}}(t)). \label{eq: second term}
\end{equation}
Using vector calculus, this expression can be further simplified (see Appendix~\ref{app: 1}); we get
\begin{eqnarray}
    &&\left(\frac{1}{r}d\mathbf{N}(t)\vec{\mathbf{r}}\cdot\vec{\nabla} \right)^2h(\vec{\mathbf{r}}(t))\nonumber \\
    &&= \frac{1}{r^2}d\mathbf{N}(t) d\mathbf{N}(t)\vec{\mathbf{r}}\cdot\vec{\nabla}h(\vec{\mathbf{r}}(t)) \nonumber \\
    &&-\frac{1}{r^2}\mathrm{Tr}\left[d\mathbf{N}(t)\left\{\vec{\nabla}\otimes\vec{\nabla}h(\vec{\mathbf{r}}(t))\right\}d\mathbf{N}(t) \left\{ \vec{\mathbf{r}}\otimes \vec{\mathbf{r}}\right\} \right].\label{eq: second term 2}
\end{eqnarray}
Performing the average of Eq.~\eqref{eq: second term 2} with respect to the stochastic noise gives
\begin{eqnarray}
     &&\left\langle \left(\frac{1}{r}d\mathbf{N}(t)\vec{\mathbf{r}}\cdot\vec{\nabla} \right)^2h(\vec{\mathbf{r}}(t)) \right \rangle \nonumber \\
     &&= \left \langle\frac{1}{r^2}\left \langle d\mathbf{N}(t) d\mathbf{N}(t)\right\rangle \vec{\mathbf{r}}\cdot\vec{\nabla}h(\vec{\mathbf{r}}(t)) \right\rangle\nonumber \\
    &&-\left\langle\frac{1}{r^2}\mathrm{Tr}\left[ d\mathbf{N}(t)\left\{\vec{\nabla}\otimes\vec{\nabla}h(\vec{\mathbf{r}}(t)) \right\} d\mathbf{N}(t) \left\{ \vec{\mathbf{r}}\otimes \vec{\mathbf{r}} \right\} \right]\right\rangle.\nonumber \\
    \label{eq: second term 3}
\end{eqnarray}
Using the properties of $d\mathbf{N}(t)$, we may simplify Eq.~\eqref{eq: second term 3} into (see Appendix~\ref{app: 2})
\begin{eqnarray}
     &&\left\langle\left(\frac{1}{r}d\mathbf{N}(t)\vec{\mathbf{r}}\cdot\vec{\nabla} \right)^2h(\vec{\mathbf{r}}(t)) \right \rangle  \nonumber \\
     &&= dt\left \langle \nabla^2 h\right \rangle-dt(D-2)\left \langle\frac{1}{r^2}\vec{\mathbf{r}}\cdot\vec{\nabla}h \right\rangle - dt~\left \langle \left(\hat{\mathbf{r}}\cdot\vec{\nabla}\right)^2h\right \rangle. \nonumber \\ \label{eq: second term expansion}
\end{eqnarray}

We now express the gradient operator $\vec{\nabla}$ in terms of its radial and angular components on the sphere $\mathcal{S}$ as
\begin{eqnarray}
    \vec{\nabla} = \hat{\mathbf{r}}\frac{\partial}{\partial r} + \frac{1}{r}\vec{\nabla}_\mathcal{S}, \label{eq: nabla}
\end{eqnarray}
where $\vec{\nabla}_\mathcal{S}$ is the surface divergence on the surface of the unit sphere $\mathcal{S}$ in $D$ dimensions. Clearly, $\vec{\mathbf{r}}\cdot\vec{\nabla}_\mathcal{S} = 0.$ Hence, we have
\begin{align}
\left\langle\frac{1}{r^2}\vec{\mathbf{r}}\cdot\vec{\nabla}h \right\rangle &= \left\langle \frac{1}{r}\frac{\partial h}{\partial r} \right\rangle,\\
\left \langle \left(\hat{\mathbf{r}}\cdot\vec{\nabla}\right)^2h\right \rangle&= \left \langle   \frac{\partial^2h}{\partial r^2}\right \rangle,\\
\left \langle \nabla^2 h\right \rangle&=  \left \langle\frac{\partial^2h}{\partial r^2}+\frac{D-1}{r}\frac{\partial h}{\partial r} +\frac{1}{r^2}\nabla^2_\mathcal{S} h\right \rangle, \label{eq: laplacian to laplace beltrami}
\end{align}
 where $\nabla^2_\mathcal{S}$ is the Laplace-Beltrami operator on $\mathcal{S}$. For a proof of Eq.~\eqref{eq: laplacian to laplace beltrami}, see Appendix~\ref{app: lap}. Hence, Eq.~\eqref{eq: second term expansion} simplifies to
\begin{equation}
    \left\langle\left(\frac{1}{r}d\mathbf{N}(t)\vec{\mathbf{r}}\cdot\vec{\nabla} \right)^2h(\vec{\mathbf{r}}(t)) \right \rangle  = dt \left \langle \frac{1}{r^2}\nabla^2_\mathcal{S} h\right \rangle+dt \left\langle \frac{1}{r}\frac{\partial h}{\partial r} \right\rangle. 
\end{equation}
Since $h$ is a function defined on the unit hypersphere, and the dynamics of the particles always confines the particles on the hypersphere with same radius, the second term must be zero, giving
\begin{equation}
    \left\langle\left(\frac{1}{r}d\mathbf{N}(t)\vec{\mathbf{r}}\cdot\vec{\nabla} \right)^2h(\vec{\mathbf{r}}(t)) \right \rangle  = dt \left \langle \frac{1}{r^2}\nabla^2_\mathcal{S} h\right \rangle. \label{eq: second term last}
\end{equation}

Putting Eqs.~\eqref{eq: first term final},~\eqref{eq: second term}~and~\eqref{eq: second term last} back into Eq.~\eqref{eq: test function taylor expansion}, we obtain
\begin{eqnarray}
    \left \langle \frac{dh(\vec{\mathbf{r}}(t))}{dt}\right\rangle = \left\langle \left(\vec{\mathbf{v}}\cdot\vec{\nabla} \right)h(\vec{\mathbf{r}}(t)) \right \rangle+\left \langle \frac{1}{r^2}\nabla^2_\mathcal{S} h(\vec{\mathbf{r}}(t))\right \rangle. \nonumber \\\label{eq: dh/dt right hand side}
\end{eqnarray}
Using the same argument as the one used to obtain Eq.~\eqref{eq: Test function avg}, we may also write the first term on the right hand side of Eq.~\eqref{eq: dh/dt right hand side} as
\begin{eqnarray}
    \left\langle \left(\vec{\mathbf{v}}\cdot\vec{\nabla} \right)h(\vec{\mathbf{r}}(t)) \right \rangle  &=& \int  \mathcal{F}\left(\vec{\mathbf{r}},t\right)\vec{\mathbf{v}}\cdot\vec{\nabla} h(\vec{\mathbf{r}}) ~dV_D \nonumber \\
    &=& -\int  h(\vec{\mathbf{r}})\vec{\nabla}  \cdot \left[\mathcal{F}\left(\vec{\mathbf{r}},t\right)\vec{\mathbf{v}}\right]~dV_D, \nonumber \\ \label{eq: dh/dt rhs first term}
\end{eqnarray}
where in the last step, we have performed integration by parts and have taken the boundary contributions to vanish. Similarly, the second term on the right hand side of Eq.~\eqref{eq: dh/dt right hand side} may also be simplified as
\begin{equation}
    \left \langle \frac{1}{r^2}\nabla^2_\mathcal{S} h(\vec{\mathbf{r}}(t))\right \rangle=\int  \mathcal{F}\left(\vec{\mathbf{r}},t\right) \frac{1}{r^2}\nabla^2_\mathcal{S}h (\vec{\mathbf{r}})~dV_D.\label{eq: dh/dt rhs second term}
\end{equation}
Using the definition of $h$ from Eq.~\eqref{eq: def h} in Eq.~\eqref{eq: dh/dt rhs second term}, we obtain
\begin{eqnarray}
   && \left \langle \frac{1}{r^2}\nabla^2_\mathcal{S} h(\vec{\mathbf{r}}(t))\right \rangle \nonumber \\
   &&= \int_0^\infty \frac{1}{r^2}\delta(r-1)\left[\int\mathcal{F}\left(\vec{\mathbf{r}},t\right) \nabla^2_\mathcal{S}h _1(\vec{\boldsymbol{\sigma}})~d\Omega\right]r^{D-1} dr\nonumber \\
   &&= \int_0^\infty \frac{1}{r^2}\delta(r-1)\left[\int h_1(\vec{\boldsymbol{\sigma}}) \nabla^2_\mathcal{S}\mathcal{F}\left(\vec{\mathbf{r}},t\right)~d\Omega\right]r^{D-1}dr \nonumber\\
   &&= \int h(\vec{\mathbf{r}}) \left[\frac{1}{r^2}\nabla^2_\mathcal{S}\mathcal{F}\left(\vec{\mathbf{r}},t\right)\right]dV_D, \label{eq: dh/dt rhs second term final}
\end{eqnarray}
where in obtaining the second equality, we have performed integration by parts for the term within the square brackets.

Using Eqs.~\eqref{eq: Test function avg d/dt}~and~\eqref{eq: dh/dt right hand side}~along with Eqs.~\eqref{eq: dh/dt rhs first term}~and~\eqref{eq: dh/dt rhs second term final} and noting that $h$ is an arbitrary function on $\mathcal{S}$, we arrive at the desired Fokker-Planck equation, which reads as
\begin{eqnarray}
    \frac{\partial\mathcal{F}}{\partial t} &=&-\vec{\nabla}  \cdot \left[\mathcal{F}\vec{\mathbf{v}}\right]+\frac{T}{r^2}\nabla^2_\mathcal{S}\mathcal{F}.\label{eq: FP first form}
\end{eqnarray}

Equation~\eqref{eq: FP first form} is the desired Eq.~\eqref{eq: FP first form model 2} of the main text.

\section{Derivation of \texorpdfstring{Eq.~\eqref{eq: second term 2}}{Eq.~(2)}}
\label{app: 1}

Here, we prove the expression given in Eq.~\eqref{eq: second term 2}. The left hand side is of the form $\left(r^{-1}\mathbf{M}\vec{\mathbf{r}}\cdot \vec{\nabla}\right)^2h(\vec{\mathbf{r}})$, where $\mathbf{M}$ is a constant ($\vec{\mathbf{r}}$-independent) antisymmetric matrix. The given quantity rewrites as
\begin{eqnarray}
    \left(\frac{1}{r}\mathbf{M}\vec{\mathbf{r}}\cdot \vec{\nabla}\right)^2h(\vec{\mathbf{r}}) = \frac{1}{r}\mathbf{M}\vec{\mathbf{r}}\cdot \vec{\nabla}\left[\frac{1}{r}\mathbf{M}\vec{\mathbf{r}}\cdot \vec{\nabla}h\right].\label{eq: app1 eq 1}
\end{eqnarray}
Using the vector calculus identity $\vec{\nabla}(fg) = f\vec{\nabla}g+g\vec{\nabla}f$ for any two scalar functions $f$ and $g$, we may simplify Eq.~\eqref{eq: app1 eq 1} into
\begin{eqnarray}
    \vec{\nabla}\left[\frac{1}{r}\mathbf{M}\vec{\mathbf{r}}\cdot \vec{\nabla}h\right] &=& \left(\mathbf{M}\vec{\mathbf{r}}\cdot \vec{\nabla}h\right) \vec{\nabla}\left(\frac{1}{r}\right)+\frac{1}{r}\vec{\nabla}\left[\mathbf{M}\vec{\mathbf{r}}\cdot \vec{\nabla}h \right]\nonumber \\
    &=& -\frac{1}{r^3} \vec{\mathbf{r}}\left(\mathbf{M}\vec{\mathbf{r}}\cdot \vec{\nabla}h\right) +\frac{1}{r}\vec{\nabla}\left[\mathbf{M}\vec{\mathbf{r}}\cdot \vec{\nabla}h\right]. \nonumber \\
\end{eqnarray}
Using this in Eq.~\eqref{eq: app1 eq 1}, we obtain
\begin{eqnarray}
    \left(\frac{1}{r}\mathbf{M}\vec{\mathbf{r}}\cdot \vec{\nabla}\right)^2h(\vec{\mathbf{r}}) &=& -\frac{1}{r^4} \vec{\mathbf{r}}\left(\mathbf{M}\vec{\mathbf{r}}\cdot \vec{\nabla}h\right)\left(\mathbf{M}\vec{\mathbf{r}}\cdot \vec{\mathbf{r}}\right)\nonumber \\
    &&+\frac{1}{r^2}\left(\mathbf{M}\vec{\mathbf{r}}\cdot \vec{\nabla}\right)^2h(\vec{\mathbf{r}}). \label{eq: app1 eq 2}
\end{eqnarray}
Since $\mathbf{M}$ is an antisymmetric matrix, we have $\mathbf{M}\vec{\mathbf{r}}\cdot \vec{\mathbf{r}} = 0$. Hence, Eq.~\eqref{eq: app1 eq 2} reduces to
\begin{eqnarray}
    \left(\frac{1}{r}\mathbf{M}\vec{\mathbf{r}}\cdot \vec{\nabla}\right)^2h(\vec{\mathbf{r}}) = \frac{1}{r^2}\left(\mathbf{M}\vec{\mathbf{r}}\cdot \vec{\nabla}\right)^2h(\vec{\mathbf{r}}). \label{eq: app1 eq 3}
\end{eqnarray}

Let us now focus on computing $\left(\mathbf{M}\vec{\mathbf{r}}\cdot \vec{\nabla}\right)^2h(\vec{\mathbf{r}})$. We will do further computations in the Cartesian coordinate system. Let $M_{ij}$ denote the $ij$-th element of the matrix $\mathbf{M}$. Similarly, let the $i$-th component of the vector $\vec{\mathbf{r}}$ be $r_i$. Hence, we may write $\vec{\mathbf{r}} = \sum_{i=1}^D x_i \hat{\mathbf{r}}_i$, where $\hat{\mathbf{r}}_i$ is the unit vector along the $i$-th Cartesian coordinate axis. Using this, we may write
\begin{eqnarray}
   \mathbf{M}\vec{\mathbf{r}}\cdot \vec{\nabla}h(\vec{\mathbf{r}}) = \sum_{k = 1}^D\sum_{j=1}^D M_{kj}x_j \frac{\partial h}{\partial x_k}. \label{eq: app1 eq 4}
\end{eqnarray}
Proceeding in a similar way, we obtain
\begin{eqnarray}
     &&\mathbf{M}\vec{\mathbf{r}}\cdot \vec{\nabla}\left[\mathbf{M}\vec{\mathbf{r}}\cdot \vec{\nabla}h(\vec{\mathbf{r}})\right]\nonumber \\
     &&= \sum_{l = 1}^D\sum_{m=1}^D M_{lm}x_m \frac{\partial }{\partial x_l}\left[\sum_{k = 1}^D\sum_{j=1}^D M_{kj}x_j \frac{\partial h}{\partial x_k}\right]\nonumber \\
     &&=\sum_{l = 1}^D\sum_{m=1}^D\sum_{k = 1}^D\sum_{j=1}^D \delta_{lj} M_{lm}M_{kj} x_m \frac{\partial h}{\partial x_k} \nonumber \\
     &&+ \sum_{l = 1}^D\sum_{m=1}^D\sum_{k = 1}^D\sum_{j=1}^D  M_{lm}M_{kj} x_m x_j \frac{\partial^2 h}{\partial x_l\partial x_k}. \label{eq: app1 eq 5}
\end{eqnarray}
Now, the first term may be simplified as follows:
\begin{eqnarray}
&& \sum_{l = 1}^D\sum_{m=1}^D\sum_{k = 1}^D\sum_{j=1}^D \delta_{lj} M_{lm}M_{kj} x_m \frac{\partial h}{\partial x_k} \nonumber \\
     &&=\sum_{m=1}^D\sum_{k = 1}^D \left(\sum_{j=1}^D M_{kj}M_{jm}\right) x_m \frac{\partial h}{\partial x_k} \nonumber \\
      &&=\sum_{k = 1}^D\sum_{m=1}^D \left(\mathbf{M}^2\right)_{km} x_m \frac{\partial h}{\partial x_k} \nonumber \\
      &&= \mathbf{M}^2\vec{\mathbf{r}}\cdot \vec{\nabla}h(\vec{\mathbf{r}}).\label{eq: app1 eq 6}
\end{eqnarray}
Using the anti-symmetry property of the $\mathbf{M}$ matrix, the second term simplifies to
\begin{eqnarray}
    && \sum_{l = 1}^D\sum_{m=1}^D\sum_{k = 1}^D\sum_{j=1}^D  M_{lm}M_{kj} x_m x_j \frac{\partial^2 h}{\partial x_l\partial x_k} \nonumber \\
    &&= - \sum_{m=1}^D\sum_{l = 1}^D\sum_{k = 1}^D\sum_{j=1}^D  M_{ml} \left[\frac{\partial^2 h}{\partial x_l\partial x_k}\right]M_{kj} \left[x_j x_m \right] \nonumber \\
    &&= - \sum_{m=1}^D\sum_{l = 1}^D\sum_{k = 1}^D\sum_{j=1}^D  \mathbf{M}_{ml} \left[\vec{\nabla}\otimes \vec{\nabla}h\right]_{lk}\mathbf{M}_{kj} \left[\vec{\mathbf{r}}\otimes \vec{\mathbf{r}} \right]_{jm}\nonumber\\
    &&= -\mathrm{Tr}\left[\mathbf{M}\left(\vec{\nabla}\otimes \vec{\nabla}h\right)\mathbf{M}\left(\vec{\mathbf{r}}\otimes \vec{\mathbf{r}} \right)\right].\label{eq: app1 eq 7}
\end{eqnarray}
Hence, combining Eqs.~\eqref{eq: app1 eq 5},~\eqref{eq: app1 eq 6}~and~\eqref{eq: app1 eq 7}, we finally obtain
\begin{eqnarray}
&&\left(\mathbf{M}\vec{\mathbf{r}}\cdot \vec{\nabla}\right)^2h(\vec{\mathbf{r}}) \nonumber \\ &&= \mathbf{M}^2\vec{\mathbf{r}}\cdot \vec{\nabla}h(\vec{\mathbf{r}}) -\mathrm{Tr}\left[\mathbf{M}\left(\vec{\nabla}\otimes \vec{\nabla}h\right)\mathbf{M}\left(\vec{\mathbf{r}}\otimes \vec{\mathbf{r}} \right)\right].
\end{eqnarray}
Using this in Eq.~\eqref{eq: app1 eq 3}, we finally obtain
\begin{eqnarray}
    &&\left(\frac{1}{r}\mathbf{M}\vec{\mathbf{r}}\cdot \vec{\nabla}\right)^2h(\vec{\mathbf{r}}) \nonumber \\ &&= \frac{1}{r^2}\mathbf{M}^2\vec{\mathbf{r}}\cdot \vec{\nabla}h(\vec{\mathbf{r}}) -\frac{1}{r^2}\mathrm{Tr}\left[\mathbf{M}\left(\vec{\nabla}\otimes \vec{\nabla}h\right)\mathbf{M}\left(\vec{\mathbf{r}}\otimes \vec{\mathbf{r}} \right)\right], \nonumber \\
\end{eqnarray}
which is used to obtain Eq.~\eqref{eq: second term 2} of Appendix~\ref{app: FP model 1}.

\section{Derivation of \texorpdfstring{Eq.~\eqref{eq: second term expansion}}{Eq.~(41)}}
\label{app: 2}

The independent matrix elements of $d\mathbf{N}(t)$, represented by $\left[d\mathbf{N}(t)\right]_{ij}$, are independent differential Wiener processes. We represent them by $\left[d\mathbf{N}(t)\right]_{ij} = -\left[d\mathbf{N}(t)\right]_{ji} = dW_{ij}~\forall~i>j$. This further gives us
\begin{eqnarray}
    \left\langle \left[d\mathbf{N}(t)\right]_{ij}\left[d\mathbf{N}(t)\right]_{kl} \right \rangle =  dt \left(\delta_{ik}\delta_{jl}-\delta_{il}\delta_{jk}\right). \label{eq: app2 eq 1}
\end{eqnarray}
Using $\langle d\mathbf{N}(t)d\mathbf{N}(t)\rangle = -dt(D-1)\mathbf{I} $ in the first term on the right hand side of Eq.~\eqref{eq: second term 3}, we obtain
\begin{eqnarray}
    \left \langle\frac{1}{r^2}\left \langle d\mathbf{N}(t) d\mathbf{N}(t)\right\rangle \vec{\mathbf{r}}\cdot\vec{\nabla}h \right\rangle = -dt (D-1)\left \langle\frac{1}{r^2}\vec{\mathbf{r}}\cdot\vec{\nabla}h \right\rangle. \nonumber \\ \label{eq: app2 eq 2}
\end{eqnarray}
We now focus on the second term on the right hand side of Eq.~\eqref{eq: second term 3}. We will do all further computations in the Cartesian coordinate system. Using Eq.~\eqref{eq: app2 eq 1} in the second term, we obtain
\begin{eqnarray}
&& \left\langle\frac{1}{r^2}\mathrm{Tr}\left[d\mathbf{N}(t)\left(\vec{\nabla}\otimes \vec{\nabla}h\right)d\mathbf{N}(t)\left(\vec{\mathrm{r}}\otimes \vec{\mathrm{r}} \right)\right] \right\rangle\nonumber \\
&&= \sum_{m=1}^D\sum_{l = 1}^D\sum_{k = 1}^D\sum_{j=1}^D  \left \langle\frac{1}{r^2} \left[d\mathbf{N}(t)\right]_{ml}\frac{\partial^2 h}{\partial x_l\partial x_k}\left[d\mathbf{N}(t)\right]_{kj} x_jx_m\right\rangle  \nonumber \\
&&=- \sum_{l = 1}^D\sum_{m=1}^D\sum_{k = 1}^D\sum_{j=1}^D  \left \langle \left[d\mathbf{N}(t)\right]_{lm}\left[d\mathbf{N}(t)\right]_{kj}\right\rangle \left \langle \frac{x_m x_j}{r^2} \frac{\partial^2 h}{\partial x_l\partial x_k}\right\rangle \nonumber \\
&&= -dt \sum_{l = 1}^D\sum_{m=1}^D\sum_{k = 1}^D\sum_{j=1}^D  \left(\delta_{lk}\delta_{mj}-\delta_{lj}\delta_{mk}\right) \left \langle \frac{x_m x_j}{r^2} \frac{\partial^2 h}{\partial x_l\partial x_k}\right\rangle \nonumber\\
&&= -dt \sum_{l = 1}^D\sum_{m=1}^D  \left[\left \langle \frac{x_m^2}{r^2} \frac{\partial^2 h}{\partial x_l^2}\right\rangle  -\left \langle \frac{x_m x_l}{r^2} \frac{\partial^2 h}{\partial x_l\partial x_m}\right\rangle \right]\nonumber \\
&&= -dt~ \left \langle \frac{1}{r^2} r^2 \nabla^2 h\right \rangle+dt \left \langle\sum_{l = 1}^D\sum_{m=1}^D \frac{x_m x_l}{r^2} \frac{\partial^2 h}{\partial x_l\partial x_m}\right\rangle \nonumber \\
&&= -dt~ \left \langle  \nabla^2 h\right \rangle+dt \left \langle\sum_{l = 1}^D\sum_{m=1}^D \frac{x_m x_l}{r^2} \frac{\partial^2 h}{\partial x_l\partial x_m}\right\rangle . \label{eq: app2 eq 3}
\end{eqnarray}
In the semifinal step, we have used $r^2 = \sum_{m=1}^Dx_m^2$.  Let us now focus on the second term. From definition
\begin{align}
    \left(\vec{\mathbf{r}}\cdot\vec{\nabla}\right)^2 &=\sum_{l=1}^D\sum_{m=1}^D \left[x_l\frac{\partial}{\partial x_l}\right]\left[x_m\frac{\partial}{\partial x_m}\right]\nonumber\\
    &=\sum_{l=1}^D\sum_{m=1}^D \left[\delta_{lm}x_l\frac{\partial}{\partial x_m}+x_lx_m\frac{\partial^2}{\partial x_l\partial x_m}\right]\nonumber\\
    &=\sum_{l=1}^Dx_l\frac{\partial}{\partial x_l}+\sum_{l=1}^D\sum_{m=1}^Dx_lx_m\frac{\partial^2}{\partial x_l\partial x_m}
\end{align}

This finally gives
\begin{equation}
    \sum_{l=1}^D\sum_{m=1}^Dx_lx_m\frac{\partial^2}{\partial x_l\partial x_m}=\left(\vec{\mathbf{r}}\cdot\vec{\nabla}\right)^2 -\vec{\mathbf{r}}\cdot\vec{\nabla}
\end{equation}

Using Eqs.~\eqref{eq: app2 eq 2}~and~\eqref{eq: app2 eq 3} in Eq.~\eqref{eq: second term 3}, we finally obtain
\begin{eqnarray}
     &&\left\langle\frac{1}{r}\left(d\mathbf{N}(t)\vec{\mathbf{r}}\cdot\vec{\nabla} \right)^2h(\vec{\mathbf{r}}(t)) \right \rangle  \nonumber \\
     &&= -dt~(D-2)\left \langle\frac{1}{r^2}\vec{\mathbf{r}}\cdot\vec{\nabla}h \right\rangle - dt\left \langle \left(\hat{\mathbf{r}}\cdot\vec{\nabla}\right)^2h\right \rangle+dt\left \langle \nabla^2 h\right \rangle, \nonumber \\
\end{eqnarray}
which is Eq.~\eqref{eq: second term expansion} of the main text.

\section{Derivation of \texorpdfstring{Eq.~\eqref{eq: laplacian to laplace beltrami}}{Eq.~(45)}}
\label{app: lap}

We start with the expression of $\vec{\nabla}$ in $D$-dimensional spherical polar coordinates, which reads as
\begin{equation}
    \vec{\nabla} = \hat{\mathbf{r}}\frac{\partial}{\partial r} + \frac{1}{r}\vec{\nabla}_\mathcal{S}. \label{eq: nabla app}
\end{equation}
Now, our quantity of interest may be evaluated as follows:
\begin{align}
   &\nabla^2h \nonumber \\
    \nonumber &= \vec{\nabla} \cdot\vec{\nabla}h =  \vec{\nabla} \cdot \bigg[\hat{\mathbf{r}} \frac{\partial h}{\partial r}\bigg]+\vec{\nabla} \cdot \bigg[ \frac{1}{r}\vec{\nabla}_\mathcal{S}h\bigg]\\
    \nonumber &= \big(\vec{\nabla} \cdot \hat{\mathbf{r}} \big)\frac{\partial h}{\partial r} + \hat{\mathbf{r}} \cdot \vec{\nabla}\bigg[ \frac{\partial h}{\partial r}\bigg] + \vec{\nabla}\bigg[\frac{1}{r}\bigg] \cdot \vec{\nabla}_\mathcal{S}h + \frac{1}{r} \vec{\nabla}\cdot  \vec{\nabla}_\mathcal{S}h\\
    &= \frac{D-1}{r} \frac{\partial h}{\partial r} + \frac{\partial^2h}{\partial r^2} + \frac{1}{r^2} \nabla^2_\mathcal{S}h. \label{eq: D3}
\end{align}
In the last step, we have used the fact that $\hat{\mathbf{r}} \cdot \vec{\nabla}_\mathcal{S} = 0$. Also, we have used $\vec{\nabla} \cdot \hat{\mathbf{r}} = (D-1)/r$. Equation~\eqref{eq: D3} is Eq.~\eqref{eq: laplacian to laplace beltrami} of Appendix~\ref{app: FP model 1}.

\section{Derivation of \texorpdfstring{Eq.~\eqref{eq: continuity final model 3}}{Eq.~(52)}}
\label{app: 11}

We start with the first term on the right hand side of Eq.~\eqref{eq: FP first form model 2}, which upon using standard vector identity gives
\begin{equation}
    \vec{\nabla}  \cdot \left[\mathcal{F}\vec{\mathbf{v}}\right] = \vec{\mathbf{v}}\cdot \vec{\nabla}\mathcal{F}+\mathcal{F}\vec{\nabla}\cdot\vec{\mathbf{v}}, \label{eq: app e 1 model 2}
\end{equation}
where we have $\vec{\mathbf{v}} = K_1  \left[\vec{\mathbb{P}}-\left(\vec{\mathbf{r}}\cdot \vec{\mathbb{P}}\right)\vec{\mathbf{r}}/r^2\right]+2 K_2\Bigg[\mathbb{M}/r-\left(\vec{\mathbf{r}} \cdot \mathbb{M}\vec{\mathbf{r}}\right)\mathbf{I}_D/r^3\Bigg]\vec{\mathbf{r}}$. Now, using Eq.~\eqref{eq: nabla}, we have
\begin{equation}
    \vec{\nabla}\mathcal{F} = \hat{\mathbf{r}}\frac{\partial \mathcal{F}}{\partial r} + \frac{1}{r}\vec{\nabla}_\mathcal{S}\mathcal{F}.
\end{equation}
Since $\vec{\mathbf{v}}$ is along the tangent to the sphere centered at the origin, we have $\vec{\mathbf{r}}\cdot \vec{\mathbf{v}}=0$. This is also evident from the expression of $\vec{\mathbf{v}}$ on using the result $\vec{\mathbf{r}}\cdot \vec{\mathbf{r}} = r^2$. Hence, we have
\begin{eqnarray}
      \vec{\mathbf{v}}\cdot \vec{\nabla}\mathcal{F} &=& \frac{1}{r} \vec{\mathbf{v}} \cdot \vec{\nabla}_\mathcal{S}\mathcal{F}\nonumber\\
      &=& \frac{1}{r}\Big(K_1\vec{\mathbb{P}}+2K_2\mathbb{M}\vec{\boldsymbol{\sigma}}\Big)\cdot \vec{\nabla}_\mathcal{S}\mathcal{F}\nonumber\\
      &=&  \frac{\delta(r-1)}{r}\bigg[\Big(K_1\vec{\mathbb{P}}+2K_2\mathbb{M}\vec{\boldsymbol{\sigma}}\Big)\cdot \vec{\nabla}_\mathcal{S}F\bigg],\label{eq: app e 7 model 2}
\end{eqnarray}
where in deriving the second equality, we have used $\vec{\mathbf{r}} \cdot \vec{\nabla}_\mathcal{S}\mathcal{F}=0$ by the definition of $\vec{\nabla}_\mathcal{S}\mathcal{F}$.

We now focus on the second term on the right hand side of Eq.~\eqref{eq: app e 1 model 2}. Using the expression of $\vec{\mathbf{v}}$, we obtain
\begin{align}
    \vec{\nabla}\cdot\vec{\mathbf{v}} &= K_1  \vec{\nabla}\cdot \Bigg[\vec{\mathbb{P}}-\frac{1}{r^2}\left(\vec{\mathbf{r}}\cdot \vec{\mathbb{P}}\right)\vec{\mathbf{r}}\Bigg]\nonumber\\
    &+2 K_2\vec{\nabla}\cdot \Bigg[\frac{1}{r}\mathbb{M\vec{\mathbf{r}}}-\frac{1}{r^3}\left(\vec{\mathbf{r}} \cdot \mathbb{M}\vec{\mathbf{r}}\right)\vec{\mathbf{r}}\Bigg]. \label{eq: grad vv model 2}
\end{align}
Focusing on the first term on the right hand side of Eq.~\eqref{eq: grad vv model 2},
in terms of Cartesian components, we may write
\begin{eqnarray}
    &&  \vec{\nabla}\cdot \Bigg[\vec{\mathbb{P}}-\frac{1}{r^2}\left(\vec{\mathbf{r}}\cdot \vec{\mathbb{P}}\right)\vec{\mathbf{r}}\Bigg] \nonumber\\
    &=& -\vec{\nabla}\cdot\Big[\left(\vec{\mathbb{P}} \cdot \vec{\boldsymbol{\sigma}}\right)\vec{\boldsymbol{\sigma}}\Big] \nonumber \\
    &=&  -\bigg[\left(\vec{\mathbb{P}} \cdot \vec{\boldsymbol{\sigma}}\right)\vec{\nabla}\cdot\vec{\boldsymbol{\sigma}}+\vec{\nabla} \left(\vec{\mathbb{P}} \cdot \vec{\boldsymbol{\sigma}}\right)\cdot\vec{\boldsymbol{\sigma}}\bigg]\nonumber\\
&=& -\bigg[\left(\vec{\mathbb{P}} \cdot \vec{\boldsymbol{\sigma}}\right) \bigg[\frac{D-1}{r}\bigg]+ \frac{1}{r}\Big[\vec{\mathbb{P}}-\left(\vec{\mathbb{P}} \cdot \vec{\boldsymbol{\sigma}}\right)\vec{\boldsymbol{\sigma}}\Big]\cdot \vec{\boldsymbol{\sigma}} \bigg]\nonumber \\
&=&-\frac{1}{r}(D-1)\vec{\mathbb{P}} \cdot \vec{\boldsymbol{\sigma}}. \label{eq: we need later}
\end{eqnarray}
Here, in obtaining the fourth equality, we have used the identity $\vec{\nabla} \left(\vec{\mathbf{A}} \cdot \vec{\boldsymbol{\sigma}}\right)  = \frac{1}{r}\Big[\vec{\mathbf{A}}-\left(\vec{\mathbf{A}} \cdot \vec{\boldsymbol{\sigma}}\right)\vec{\boldsymbol{\sigma}}\Big]$ for any constant vector $\vec{\mathbf{A}}$, which we derive below (see Eq.~\eqref{eq: G1}).

We now focus on simplifying the second bracketed term in Eq.~\eqref{eq: grad vv model 2}. From Eq.~\eqref{eq: nabla b sigma}, we obtain
\begin{equation}
    \vec{\nabla}\cdot \Bigg[\frac{1}{r}\mathbb{M}\vec{\mathbf{r}}\Bigg]= \frac{1}{r}\bigg[\mathrm{Tr}\big[\mathbb{M}\big]-\vec{\boldsymbol{\sigma}}\cdot\mathbb{M}\vec{\boldsymbol{\sigma}}\bigg],
\end{equation}
and from Eq.~\eqref{eq: we will need later 2}, we obtain
\begin{equation}
    \vec{\nabla}\cdot \Bigg[\frac{1}{r^3}\left(\vec{\mathbf{r}} \cdot \mathbb{M}\vec{\mathbf{r}}\right)\vec{\mathbf{r}}\Bigg] =  \frac{(D-1)}{r}\left(\vec{\boldsymbol{\sigma}} \cdot \mathbb{M}\vec{\boldsymbol{\sigma}}\right).
\end{equation}
Combining these results, we obtain
\begin{equation}
    \vec{\nabla}\cdot \Bigg[\frac{1}{r}\mathbb{M\vec{\mathbf{r}}}-\frac{1}{r^3}\left(\vec{\mathbf{r}} \cdot \mathbb{M}\vec{\mathbf{r}}\right)\vec{\mathbf{r}}\Bigg] = \frac{1}{r}\bigg[1-D\vec{\boldsymbol{\sigma}}\cdot\mathbb{M}\vec{\boldsymbol{\sigma}}\bigg],
\end{equation}
where we have used the fact that $\mathrm{Tr}\big[\mathbb{M}\big]=1$. Hence, from Eq.~\eqref{eq: grad vv model 2}, we get
\begin{align}
    \mathcal{F}\vec{\nabla}\cdot\vec{\mathbf{v}} &= -\frac{\delta(r-1)}{r}K_1(D-1)F~\vec{\mathbb{P}} \cdot \vec{\boldsymbol{\sigma}}\nonumber\\
    &+\frac{\delta(r-1)}{r}2K_2F-\frac{\delta(r-1)}{r}2K_2DF~\vec{\boldsymbol{\sigma}}\cdot\mathbb{M}\vec{\boldsymbol{\sigma}}.\label{eq: app e 5 model 2}
\end{align}

Using Eqs.~\eqref{eq: F and curly F main text},~\eqref{eq: app e 1 model 2},~\eqref{eq: app e 7 model 2},~and~\eqref{eq: app e 5 model 2} in Eq.~\eqref{eq: FP first form model 2}, we obtain
\begin{align}
    \delta(r-1)\frac{\partial F}{\partial t} &= -\frac{\delta(r-1)}{r}\bigg[\Big(K_1\vec{\mathbb{P}}+2K_2\mathbb{M}\vec{\boldsymbol{\sigma}}\Big)\cdot \vec{\nabla}_\mathcal{S}F\bigg]\nonumber\\
    &+\frac{\delta(r-1)}{r}K_1(D-1)F~\vec{\mathbb{P}} \cdot \vec{\boldsymbol{\sigma}}\nonumber\\
    &+\frac{\delta(r-1)}{r}2K_2DF~\vec{\boldsymbol{\sigma}}\cdot\mathbb{M}\vec{\boldsymbol{\sigma}}\nonumber\\
    & -\frac{\delta(r-1)}{r}2K_2F+\frac{\delta(r-1)}{r^2}T\nabla^2_\mathcal{S}F.
\end{align}
Integrating both sides over $r$ from $1-\epsilon$ to $1+\epsilon$ for small $\epsilon$, we finally obtain
\begin{align}
    \frac{\partial F}{\partial t}=&-K_1 \left[ \vec{\nabla}_\mathcal{S}F-(D-1)F \vec{\boldsymbol{\sigma}}\right]\cdot \vec{\mathbb{P}}-2K_2 F \nonumber \\
    & -2K_2 \left[\vec{\nabla}_\mathcal{S}F -D F \vec{\boldsymbol{\sigma}}\right]\cdot \mathbb{M} \vec{\boldsymbol{\sigma}} +T~\nabla^2_\mathcal{S}F. \label{eq: FP final app e model 2}
\end{align}
Equation~\eqref{eq: FP final app e model 2} is the desired Eq.~\eqref{eq: continuity final model 3} of the main text.

\section{Derivation of some useful identities}
\label{app: 6}

We start from the following expression:
\begin{equation}
    \vec{\nabla} \left(\vec{\mathbf{A}} \cdot \vec{\boldsymbol{\sigma}}\right) =  \vec{\nabla} \left(\vec{\mathbf{A}} \cdot \frac{\vec{\mathbf{r}}}{r}\right).
\end{equation}
Now, in the Cartesian coordinates, if we denote the $j$-th component of $\vec{\mathbf{A}}$ by $A_j$ and that of $\vec{\mathbf{r}}$ by $x_j$, then we may write
\begin{equation}
    \vec{\nabla} \left(\vec{\mathbf{A}} \cdot \vec{\boldsymbol{\sigma}}\right) = \vec{\nabla} \left(\sum_{j=1}^D \frac{A_jx_j}{r}
    \right) .
\end{equation}
Hence, the $k$-th component of $\vec{\nabla} \left(\vec{\mathbf{A}} \cdot \vec{\boldsymbol{\sigma}}\right) $ may be written as
\begin{eqnarray}
    \Big[\vec{\nabla} \left(\vec{\mathbf{A}} \cdot \vec{\boldsymbol{\sigma}}\right) \Big]_k &=& \frac{\partial}{\partial x_k}\left(\sum_{j=1}^D \frac{A_jx_j}{r}
    \right) \nonumber \\
    &=& \sum_{j=1}^D \frac{A_j\delta_{jk}}{r}-\sum_{j=1}^D \frac{A_jx_jx_k}{r^3}\nonumber\\
    &=& \frac{A_k}{r} - \frac{x_k}{r^2}\left(\vec{\mathbf{A}} \cdot \vec{\boldsymbol{\sigma}}\right).
\end{eqnarray}
Hence, we may write
\begin{equation}
    \vec{\nabla} \left(\vec{\mathbf{A}} \cdot \vec{\boldsymbol{\sigma}}\right)  = \frac{1}{r}\Big[\vec{\mathbf{A}}-\left(\vec{\mathbf{A}} \cdot \vec{\boldsymbol{\sigma}}\right)\vec{\boldsymbol{\sigma}}\Big].\label{eq: G1}
\end{equation}
Now, from Eq.~\eqref{eq: nabla}, we may rewrite the left hand side of the equation as
\begin{eqnarray}
    \vec{\nabla} \left(\vec{\mathbf{A}} \cdot \vec{\boldsymbol{\sigma}}\right) = \Big[\hat{\mathbf{r}}\frac{\partial}{\partial r} + \frac{1}{r}\vec{\nabla}_\mathcal{S}\Big]\left(\vec{\mathbf{A}} \cdot \vec{\boldsymbol{\sigma}}\right).
\end{eqnarray}
Since  $\vec{\mathbf{A}} \cdot \vec{\boldsymbol{\sigma}}$ is independent of $r$, we may immediately write
\begin{equation}
     \vec{\nabla} \left(\vec{\mathbf{A}} \cdot \vec{\boldsymbol{\sigma}}\right) = \frac{1}{r} \vec{\nabla}_\mathcal{S}\left(\vec{\mathbf{A}} \cdot \vec{\boldsymbol{\sigma}}\right). \label{eq: G2}
\end{equation}
Comparing Eqs.~\eqref{eq: G1} and~\eqref{eq: G2}, we finally obtain
\begin{equation}
    \vec{\nabla}_\mathcal{S}\left(\vec{\mathbf{A}} \cdot \vec{\boldsymbol{\sigma}}\right) = \vec{\mathbf{A}}-\left(\vec{\mathbf{A}} \cdot \vec{\boldsymbol{\sigma}}\right)\vec{\boldsymbol{\sigma}}.\label{eq: ans rel 2 app}
\end{equation}

Taking divergence on both sides of Eq.~\eqref{eq: ans rel 2 app}, we obtain
\begin{equation}
    \vec{\nabla}\cdot  \vec{\nabla}_\mathcal{S}\left(\vec{\mathbf{A}} \cdot \vec{\boldsymbol{\sigma}}\right) = -\vec{\nabla}\cdot\Big[\left(\vec{\mathbf{A}} \cdot \vec{\boldsymbol{\sigma}}\right)\vec{\boldsymbol{\sigma}}\Big], \label{eq: ans rel 3 app 1}
\end{equation}
where we have used $\vec{\nabla}\cdot\vec{\mathbf{A}}=0$, since $\vec{\mathbf{A}}$ is a constant vector. The left hand side of Eq.~\eqref{eq: ans rel 3 app 1} gives
\begin{eqnarray}
    \vec{\nabla}\cdot  \vec{\nabla}_\mathcal{S}\left(\vec{\mathbf{A}} \cdot \vec{\boldsymbol{\sigma}}\right) &=& \Big[\hat{\mathbf{r}}\frac{\partial}{\partial r} + \frac{1}{r}\vec{\nabla}_\mathcal{S}\Big]\cdot \vec{\nabla}_\mathcal{S}\left(\vec{\mathbf{A}} \cdot \vec{\boldsymbol{\sigma}}\right) \nonumber \\
    &=& \frac{1}{r}\nabla^2_\mathcal{S}\left(\vec{\mathbf{A}} \cdot \vec{\boldsymbol{\sigma}}\right),\label{eq: ans rel 3 app 2}
\end{eqnarray}
where we have used $\hat{\mathbf{r}} \cdot\vec{\nabla}_\mathcal{S}=0 $. Let us now simplify the right hand side of Eq.~\eqref{eq: ans rel 3 app 1}. Using standard vector identity, we obtain
\begin{eqnarray}
&&\vec{\nabla}\cdot\Big[\left(\vec{\mathbf{A}} \cdot \vec{\boldsymbol{\sigma}}\right)\vec{\boldsymbol{\sigma}}\Big] \nonumber\\ 
&&=  \left(\vec{\mathbf{A}} \cdot \vec{\boldsymbol{\sigma}}\right)\vec{\nabla}\cdot\vec{\boldsymbol{\sigma}}+\vec{\nabla} \left(\vec{\mathbf{A}} \cdot \vec{\boldsymbol{\sigma}}\right)\cdot\vec{\boldsymbol{\sigma}}\nonumber\\
&&= \left(\vec{\mathbf{A}} \cdot \vec{\boldsymbol{\sigma}}\right) \bigg[\frac{D-1}{r}\bigg]+ \frac{1}{r}\Big[\vec{\mathbf{A}}-\left(\vec{\mathbf{A}} \cdot \vec{\boldsymbol{\sigma}}\right)\vec{\boldsymbol{\sigma}}\Big]\cdot \vec{\boldsymbol{\sigma}} \nonumber \\
&&=\frac{1}{r}(D-1)\vec{\mathbf{A}} \cdot \vec{\boldsymbol{\sigma}}, \label{eq: ans rel 3 app 3}
\end{eqnarray}
where in obtaining the second equality, we have used Eq.~\eqref{eq: G1}, while in arriving at the last step of Eq.~\eqref{eq: ans rel 3 app 3}, we have used $\vec{\boldsymbol{\sigma}} \cdot \vec{\boldsymbol{\sigma}} =1$ which makes the second term on the right hand side of the second equality equal to zero. Putting Eqs.~\eqref{eq: ans rel 3 app 2} and~\eqref{eq: ans rel 3 app 3} back into Eq.~\eqref{eq: ans rel 3 app 1}, we finally obtain
\begin{equation}
    \nabla^2_\mathcal{S}\left(\vec{\mathbf{A}} \cdot \vec{\boldsymbol{\sigma}}\right) = -(D-1)\vec{\mathbf{A}} \cdot \vec{\boldsymbol{\sigma}}.\label{eq: ans rel 3 app 4}
\end{equation}

Another expression we want to compute is
\begin{eqnarray}
    \vec{\nabla}\cdot\mathbf{B}\vec{\boldsymbol{\sigma}} &=& \sum_{j=1}^D\frac{\partial}{\partial x_j} \bigg[\sum_{m=1}^DB_{jm}\bigg(\frac{x_m}{r}\bigg)\bigg]\nonumber \\
    &=& \sum_{j=1}^D\sum_{m=1}^DB_{jm} \bigg[\frac{\delta_{jm}}{r}-\frac{x_mx_j}{r^3}\bigg]\nonumber\\
    &=& \frac{1}{r}\bigg[\sum_{m=1}^DB_{mm}-\frac{1}{r^2}\sum_{m=1}^Dx_j\sum_{m=1}^DB_{jm}x_m\bigg]\nonumber\\
    &=& \frac{1}{r}\bigg[\mathrm{Tr}\big[\mathbf{B}\big]-\frac{1}{r^2}\vec{\mathbf{r}}\cdot\mathbf{B}\vec{\mathbf{r}}\bigg]\nonumber\\
    &=&  \frac{1}{r}\bigg[\mathrm{Tr}\big[\mathbf{B}\big]-\vec{\boldsymbol{\sigma}}\cdot\mathbf{B}\vec{\boldsymbol{\sigma}}\bigg].\label{eq: nabla b sigma}
\end{eqnarray}

We now present another computation of interest:
\begin{equation}
    \vec{\nabla}\big(\vec{\boldsymbol{\sigma}} \cdot \mathbf{B}\vec{\boldsymbol{\sigma}}\big) = \vec{\nabla}\bigg(\mathbf{B}\frac{\vec{\mathbf{r}}}{r}\cdot \frac{\vec{\mathbf{r}}}{r}\bigg).
\end{equation}
Now, in the Cartesian coordinates, if we denote the $j$-th component of $\vec{\mathbf{r}}$ by $x_j$ and $ij$-th element of $\mathbf{B}$ by $B_{ij}$, then we may write
\begin{equation}
    \vec{\nabla}\big(\vec{\boldsymbol{\sigma}} \cdot \mathbf{B}\vec{\boldsymbol{\sigma}}\big) = \vec{\nabla}\bigg(\sum_{i=1}^D\sum_{j=1}^DB_{ij}\frac{x_ix_j}{r^2}\bigg).
\end{equation}
If we focus on the $k$-th component of $\vec{\nabla}\big(\vec{\boldsymbol{\sigma}} \cdot \mathbf{B}\vec{\boldsymbol{\sigma}}\big) $, we obtain
\begin{eqnarray}
    &&\Big[\vec{\nabla}\big(\vec{\boldsymbol{\sigma}} \cdot \mathbf{B}\vec{\boldsymbol{\sigma}}\big)\Big]_k \nonumber \\
    &&= \frac{\partial}{\partial x_k}\bigg(\sum_{i=1}^D\sum_{j=1}^DB_{ij}\frac{x_ix_j}{r^2}\bigg)\nonumber\\
    &&= \sum_{i=1}^D\sum_{j=1}^DB_{ij}\delta_{ik}\frac{x_j}{r^2}+\sum_{i=1}^D\sum_{j=1}^DB_{ij}\delta_{jk}\frac{x_i}{r^2}\nonumber\\
    && -\sum_{i=1}^D\sum_{j=1}^DB_{ij}x_ix_j\frac{2x_k}{r^4}\nonumber \\
    &&=\frac{1}{r} \sum_{i=1}^D\sum_{j=1}^DB_{kj}\frac{x_j}{r}+\frac{1}{r}\sum_{i=1}^D\sum_{j=1}^DB_{ik}\frac{x_i}{r} \nonumber \\
    &&-\frac{2}{r}\bigg[\sum_{i=1}^D \frac{x_i}{r}\sum_{j=1}^DB_{ij}\frac{x_j}{r}\bigg]\frac{x_k}{r}.
\end{eqnarray}
Since $\mathbf{B}$ is a symmetric matrix, we may simplify the previous expression as
\begin{eqnarray}
&&\Big[\vec{\nabla}\big(\vec{\boldsymbol{\sigma}} \cdot \mathbf{B}\vec{\boldsymbol{\sigma}}\big)\Big]_k \nonumber \\
       &&=\frac{1}{r} \sum_{i=1}^D\sum_{j=1}^DB_{jk}\frac{x_j}{r}+\frac{1}{r}\sum_{i=1}^D\sum_{j=1}^DB_{ki}\frac{x_i}{r} \nonumber \\
    &&-\frac{2}{r}\bigg[\sum_{i=1}^D \frac{x_i}{r}\sum_{j=1}^DB_{ij}\frac{x_j}{r}\bigg]\frac{x_k}{r} \nonumber\\
    &&= \frac{2}{r}\bigg[\mathbf{B}\frac{\vec{\mathbf{r}}}{r}\bigg]_k - \frac{2}{r}\bigg(\frac{\vec{\mathbf{r}}}{r}\cdot \mathbf{B}\frac{\vec{\mathbf{r}}}{r}\bigg)\bigg[\frac{\vec{\mathbf{r}}}{r}\bigg]_k \nonumber\\
    &&= \frac{2}{r}\bigg[\mathbf{B}\vec{\boldsymbol{\sigma}}-\big(\mathbf{B}\vec{\boldsymbol{\sigma}} \cdot \vec{\boldsymbol{\sigma}}\big)\vec{\boldsymbol{\sigma}}\bigg]_k.
\end{eqnarray}
Hence, we may write
\begin{equation}
    \vec{\nabla}\big(\vec{\boldsymbol{\sigma}} \cdot \mathbf{B}\vec{\boldsymbol{\sigma}}\big) = \frac{2}{r}\bigg[\mathbf{B}\vec{\boldsymbol{\sigma}}-\big(\mathbf{B}\vec{\boldsymbol{\sigma}} \cdot \vec{\boldsymbol{\sigma}}\big)\vec{\boldsymbol{\sigma}}\bigg].\label{eq: H1}
\end{equation}
Now, from Eq.~\eqref{eq: nabla}, we may rewrite the left hand side of the equation as
\begin{eqnarray}
    \vec{\nabla}\big(\vec{\boldsymbol{\sigma}} \cdot \mathbf{B}\vec{\boldsymbol{\sigma}}\big)= \Big[\hat{\mathbf{r}}\frac{\partial}{\partial r} + \frac{1}{r}\vec{\nabla}_\mathcal{S}\Big]\big(\vec{\boldsymbol{\sigma}} \cdot \mathbf{B}\vec{\boldsymbol{\sigma}}\big).
\end{eqnarray}
Since  $\big(\vec{\boldsymbol{\sigma}} \cdot \mathbf{B}\vec{\boldsymbol{\sigma}}\big)$ is independent of $r$, we may immediately write
\begin{equation}
     \vec{\nabla} \big(\vec{\boldsymbol{\sigma}} \cdot \mathbf{B}\vec{\boldsymbol{\sigma}}\big)= \frac{1}{r} \vec{\nabla}_\mathcal{S}\big(\vec{\boldsymbol{\sigma}} \cdot \mathbf{B}\vec{\boldsymbol{\sigma}}\big). \label{eq: H2}
\end{equation}
Comparing Eqs.~\eqref{eq: H1} and~\eqref{eq: H2}, we finally obtain
\begin{equation}
    \vec{\nabla}_\mathcal{S}\big(\vec{\boldsymbol{\sigma}} \cdot \mathbf{B}\vec{\boldsymbol{\sigma}}\big)= 2\bigg[\mathbf{B}\vec{\boldsymbol{\sigma}}-\big(\mathbf{B}\vec{\boldsymbol{\sigma}} \cdot \vec{\boldsymbol{\sigma}}\big)\vec{\boldsymbol{\sigma}}\bigg].\label{eq: ans rel 2 app H}
\end{equation}

Another quantity we want to compute is
\begin{align}
   & \vec{\nabla}\cdot \Bigg[\left(\vec{\boldsymbol{\sigma}} \cdot \mathbf{B}\vec{\boldsymbol{\sigma}}\right)\vec{\boldsymbol{\sigma}}\Bigg] = \vec{\boldsymbol{\sigma}}\cdot \vec{\nabla}\Big[\vec{\boldsymbol{\sigma}} \cdot \mathbf{B}\vec{\boldsymbol{\sigma}}\Big] +\left(\vec{\boldsymbol{\sigma}} \cdot \mathbf{B}\vec{\boldsymbol{\sigma}}\right)   \vec{\nabla}\cdot\vec{\boldsymbol{\sigma}}\nonumber \\
    &= \vec{\boldsymbol{\sigma}}\cdot \bigg[\frac{2}{r}\mathbf{B}\vec{\boldsymbol{\sigma}}-\frac{2}{r}\big(\mathbf{B}\vec{\boldsymbol{\sigma}} \cdot \vec{\boldsymbol{\sigma}}\big)\vec{\boldsymbol{\sigma}}\bigg]+\left(\vec{\boldsymbol{\sigma}} \cdot \mathbf{B}\vec{\boldsymbol{\sigma}}\right)   \frac{(D-1)}{r}\nonumber\\
    &= \frac{(D-1)}{r}\left(\vec{\boldsymbol{\sigma}} \cdot \mathbf{B}\vec{\boldsymbol{\sigma}}\right) . \label{eq: we will need later 2}
\end{align}

\section{Derivation of \texorpdfstring{Eq.~\eqref{eq: eig eq tensor n}}{Eq.~(52)}}
\label{app: 7}

In the Cartesian coordinates in $D$ dimensions, we may write a general position vector $\vec{\mathbf{r}} = r\vec{\boldsymbol{\sigma}} = (x_1,x_2,\ldots,x_D)$ and the corresponding unit vector $\vec{\boldsymbol{\sigma}} = (\sigma_1,\sigma_2,\ldots,\sigma_D)$ as defined in Sec.~\ref{sec: Gen D}. Using this, we may define a new function
\begin{equation}
    \Big( \mathbb{T}^{(n)},\vec{\mathbf{r}}^{\otimes n}\Big) \equiv \sum_{i_1=1}^D\sum_{i_2=1}^D \cdots \sum_{i_n=1}^DT^{(n)}_{i_1,i_2,\ldots,i_n}  x_{i_1} x_{i_2} \ldots x_{i_n},
\end{equation}
where $\mathbb{T}^{(n)}$ is a symmetric traceless tensor of order $n$. Since $x_j = r\sigma_j$ for any $j = 1,2,\dots,D$, we may write
\begin{equation}
    \Big( \mathbb{T}^{(n)},\vec{\mathbf{r}}^{\otimes n}\Big) = r^n \Big( \mathbb{T}^{(n)},\vec{\boldsymbol{\sigma}}^{\otimes n}\Big), \label{eq: F2}
\end{equation}
where $\Big( \mathbb{T}^{(n)},\vec{\boldsymbol{\sigma}}^{\otimes n}\Big)$ is defined in Eq.~\eqref{eq: T sigma expan D DIM}.

We now study the effect of the full Laplacian $\nabla^2$ on $\Big( \mathbb{T}^{(n)},\vec{\mathbf{r}}^{\otimes n}\Big)$. We first compute the $k$-th component of $\vec{\nabla}\Big( \mathbb{T}^{(n)},\vec{\mathbf{r}}^{\otimes n}\Big)$ in the Cartesian coordinates, which reads as
\begin{align}
    &\frac{\partial}{\partial x_k}\Big( \mathbb{T}^{(n)},\vec{\mathbf{r}}^{\otimes n}\Big) = \sum_{i_1=1}^D \cdots \sum_{i_n=1}^DT^{(n)}_{i_1,i_2,\ldots,i_n}\Big[ \delta_{i_1,k} x_{i_2} \ldots x_{i_n}\nonumber\\
    &+x_{i_1}\delta_{i_2,k}x_{i_3}\ldots x_{i_n} + \cdots+x_{i_1}\ldots x_{i_{n-1}}\delta_{i_n,k}\Big]. \label{eq: F3}
\end{align}
Since $\mathbb{T}^{(n)}$ is a symmetric tensor, we may simplify Eq.~\eqref{eq: F3} into
\begin{equation}
    \frac{\partial}{\partial x_k}\Big(\mathbb{T}^{(n)},\vec{\mathbf{r}}^{\otimes n}\Big) = n \sum_{i_2=1}^D \cdots \sum_{i_n=1}^DT^{(n)}_{k,i_2,i_3,\ldots,i_n}x_{i_2}x_{i_3} \ldots x_{i_n}. \label{eq: G4}
\end{equation}
Repeating the same argument, we obtain
\begin{equation}
    \frac{\partial^2}{\partial x^2_k}\Big( \mathbb{T}^{(n)},\vec{\mathbf{r}}^{\otimes n}\Big) = n(n-1) \sum_{i_2=1}^D \cdots \sum_{i_n=1}^DT^{(n)}_{k,k,i_3,\ldots,i_n}x_{i_3} \ldots x_{i_n}.
\end{equation}
Now computing $\nabla^2\Big\langle \mathbb{T}^{(n)},\vec{\mathbf{r}}^{\otimes n}\Big\rangle$ in the Cartesian components, we obtain
\begin{align}
    &\nabla^2\Big(\mathbb{T}^{(n)},\vec{\mathbf{r}}^{\otimes n}\Big)  = \sum_{k=1}^D\frac{\partial^2}{\partial x^2_k}\Big( \mathbb{T}^{(n)},\vec{\mathbf{r}}^{\otimes n}\Big), \nonumber\\
    &= n(n-1) \sum_{i_2=1}^D \cdots \sum_{i_n=1}^D\bigg[\sum_{k=1}^DT^{(n)}_{k,k,i_3,\ldots,i_n}\bigg]x_{i_3} \ldots x_{i_n} = 0, \label{eq: F6}
\end{align}
where in the last step, we have used the traceless property of $\mathbb{T}^{(n)}$ as given in Eq.~\eqref{eq: traceless D}.

We now compute $\nabla^2\Big(\mathbb{T}^{(n)},\vec{\mathbf{r}}^{\otimes n}\Big) $ in $D$-dimensional spherical-polar coordinates as defined in Appendix~\ref{app: 0}. Using Eq.~\eqref{eq: D3} along with Eq.~\eqref{eq: F2} and noting the fact that $\Big\langle \mathbb{T}^{(n)},\vec{\boldsymbol{\sigma}}^{\otimes n}\Big\rangle$ is independent of $r$, we obtain
\begin{align}
    &\nabla^2\Big( \mathbb{T}^{(n)},\vec{\mathbf{r}}^{\otimes n}\Big), \nonumber\\
    &= \bigg[\frac{D-1}{r} \frac{\partial }{\partial r} + \frac{\partial^2}{\partial r^2} + \frac{1}{r^2} \nabla^2_\mathcal{S}\bigg] \bigg[r^n \Big( \mathbb{T}^{(n)},\vec{\boldsymbol{\sigma}}^{\otimes n}\Big)\bigg],\nonumber\\
    &= r^{n-2}\bigg[n(n+D-2)\Big(\mathbb{T}^{(n)},\vec{\boldsymbol{\sigma}}^{\otimes n}\Big)+\nabla^2_\mathcal{S}\Big( \mathbb{T}^{(n)},\vec{\boldsymbol{\sigma}}^{\otimes n}\Big)\bigg].\label{eq: F7}
\end{align}

Comparing Eqs.~\eqref{eq: F6}~and~\eqref{eq: F7}, we obtain
\begin{equation}
    \nabla^2_\mathcal{S}\Big( \mathbb{T}^{(n)},\vec{\boldsymbol{\sigma}}^{\otimes n}\Big) = -n(n+D-2)\Big( \mathbb{T}^{(n)},\vec{\boldsymbol{\sigma}}^{\otimes n}\Big).\label{eq: F8}
\end{equation}
Equation~\eqref{eq: F8} is the desired Eq.~\eqref{eq: eig eq tensor n} of the main text.

\section{Derivation of \texorpdfstring{Eq.~\eqref{eq: eig eq ten K1 INT}}{Eq.~(58)}}
\label{app: 8}

Using the definition of $\big( \mathbb{T}^{(n)},\vec{\boldsymbol{\sigma}}^{\otimes n}\big)$ from Eq.~\eqref{eq: T sigma expan D DIM}, we may write the $k$-th component of the integral $\int \vec{\boldsymbol{\sigma}}\big(\mathbb{T}^{(n)},\vec{\boldsymbol{\sigma}}^{\otimes n}\big) d\Omega$ as
\begin{align}
    &\bigg[\int \vec{\boldsymbol{\sigma}}\Big( \mathbb{T}^{(n)},\vec{\boldsymbol{\sigma}}^{\otimes n}\Big)d\Omega\bigg]_k \nonumber\\
    &=\sum_{i_1=1}^D\sum_{i_2=1}^D \cdots \sum_{i_n=1}^DT^{(n)}_{i_1,i_2,\ldots,i_n}\int\sigma_k \sigma_{i_1}\sigma_{i_2} \ldots\sigma_{i_n}d\Omega.
\end{align}
If $n$ is even, then the integral contains the product of odd number of $\sigma$'s. Hence, using Eq.~\eqref{eq: TEN INT}, we immediately get $\int \vec{\boldsymbol{\sigma}}\Big( \mathbb{T}^{(n)},\vec{\boldsymbol{\sigma}}^{\otimes n}\Big)d\Omega=0$ when $n$ is even.

When $n$ is odd and $n \geq 3$, we compute integrals of product of even number of $\sigma_i$'s. Using Eq.~\eqref{eq: TEN INT}, we obtain
\begin{align}
    &\bigg[\int \vec{\boldsymbol{\sigma}}\Big( \mathbb{T}^{(n)},\vec{\boldsymbol{\sigma}}^{\otimes n}\Big)d\Omega\bigg]_k =\sum_{i_1=1}^D\sum_{i_2=1}^D \cdots \sum_{i_n=1}^DT^{(n)}_{i_1,i_2,\ldots,i_n} \nonumber\\
    &\times\bigg[\frac{\Omega_D}{\prod_{k=0}^{m-1}(D+2k)}\sum_{\{\mathrm{pairings}~P\}}\prod_{(a,b)\in P}\delta_{i_ai_b}\bigg]\nonumber\\
    &=\frac{\Omega_D}{\prod_{k=0}^{m-1}(D+2k)}\sum_{\mathrm{pairings}~P}\sum_{i_1=1}^D\cdots \sum_{i_{\alpha-1}=1}^D \sum_{i_{\alpha+1}=1}^D\cdots\sum_{i_{\beta-1}=1}^D  \nonumber\\
    &\sum_{i_{\beta+1}=1}^D\cdots \sum_{i_n=1}^D \prod_{\substack{(a,b)\in P \\(a,b) \neq (\alpha,\beta)}}\delta_{i_ai_b}\bigg[ \sum_{i_\alpha =1}^D\sum_{i_\beta =1}^D\delta_{i_\alpha i_\beta} \nonumber\\
    &\times T^{(n)}_{i_1,\ldots,i_\alpha,\ldots, i_\beta,\ldots,i_n}\bigg].
\end{align}
Now the term inside the bracket in the last step is a trace of $\mathbb{T}^{(n)}$, and from Eq.~\eqref{eq: traceless D}, we know it is zero since $\mathbb{T}^{(n)}$ is a traceless tensor. Hence, $\int \vec{\boldsymbol{\sigma}}\Big( \mathbb{T}^{(n)},\vec{\boldsymbol{\sigma}}^{\otimes n}\Big)d\Omega = 0$ when $n$ is odd and $n \geq 3$.

The only case left to compute is $n=1$. Focusing on the $k$-th component and using Eq.~\eqref{eq: TEN INT}, we obtain
\begin{align}
    &\bigg[\int \vec{\boldsymbol{\sigma}}\Big( \mathbb{T}^{(1)},\vec{\boldsymbol{\sigma}}\Big)d\Omega\bigg]_k =\sum_{i_1=1}^DT^{(1)}_{i_1}  \int \sigma_k \sigma_{i_1}d\Omega \nonumber\\
    &= \sum_{i_1=1}^DT^{(1)}_{i_1} \bigg[\frac{\Omega_D}{D}\delta_{i_{1}k}\bigg] =\frac{\Omega_D}{D} T^{(1)}_{k},
\end{align}
which immediately gives
\begin{equation}
    \vec{\boldsymbol{\sigma}}\cdot \int \vec{\boldsymbol{\sigma}}'\Big( \mathbb{T}^{(1)},\vec{\boldsymbol{\sigma}}'\Big) d\Omega'= \frac{\Omega_D}{D}\sum_{k=1}^DT^{(1)}_k\sigma_k= \frac{\Omega_D}{D}\Big( \mathbb{T}^{(1)},\vec{\boldsymbol{\sigma}}\Big).
\end{equation}

Combining everything, we thus obtain
\begin{eqnarray}
    \vec{\boldsymbol{\sigma}}\cdot \int \vec{\boldsymbol{\sigma}}'\Big( \mathbb{T}^{(n)},\vec{\boldsymbol{\sigma}}'\Big) d\Omega' = \delta_{n,1} \frac{\Omega_D}{D}\Big( \mathbb{T}^{(n)},\vec{\boldsymbol{\sigma}}'\Big).\label{eq: eig eq ten K1 INT app}
\end{eqnarray}
Equation~\eqref{eq: eig eq ten K1 INT app} is the desired Eq.~\eqref{eq: eig eq ten K1 INT} of the main text.

\section{Tensor Integral Formula}
\label{app: 9}

Let $\vec{\boldsymbol{\sigma}}$ be a $D$-dimensional unit vector having the form $\vec{\boldsymbol{\sigma}} = (\sigma_1,\sigma_2,\ldots, \sigma_D)$ in Cartesian coordinates, then it is known that~\cite{Weyl1939, Calderon2025,enwiki:1323001941}
\begin{align}
    &\int \sigma_{i_1}\sigma_{i_2} \ldots \sigma_{i_n}~d\Omega \nonumber\\
    &= 0,~~~\mathrm{if}~n=2m+1,\nonumber\\
    &= \frac{\Omega_D}{\prod_{k=0}^{m-1}(D+2k)}\sum_{\{\mathrm{pairings}\,P\}}\prod_{(a,b)\in P}\delta_{i_ai_b},~~~\mathrm{if}~n = 2m. \label{eq: TEN INT}
\end{align}

\section{Derivation of some useful tensor integrals}
\label{app: 10}

The first integral we compute is $\int \vec{\boldsymbol{\sigma}}\Big(\vec{\mathbf{A}} \cdot \vec{\boldsymbol{\sigma}} \Big)d\Omega$. Note that the result after performing the integration is a vector, whose $k$-th element may be written as
\begin{equation}
    \left[\int \vec{\boldsymbol{\sigma}}\Big(\vec{\mathbf{A}} \cdot \vec{\boldsymbol{\sigma}} \Big)d\Omega\right]_k=\sum_{j=1}^DA_j\int \sigma_k \sigma_j~d\Omega = \sum_{j=1}^DA_j\bigg[\frac{\Omega_D}{D}\delta_{kj}\bigg],
\end{equation}
where in the last step, we have used Eq.~\eqref{eq: TEN INT}. Performing the summation over $j$, we straightforwardly obtain $\left[\int \vec{\boldsymbol{\sigma}}\big(\vec{\mathbf{A}} \cdot \vec{\boldsymbol{\sigma}} \big)d\Omega\right]_k = (\Omega_D/D)A_k$. Hence, we have 
\begin{equation}
    \int \vec{\boldsymbol{\sigma}}\Big(\vec{\mathbf{A}} \cdot \vec{\boldsymbol{\sigma}} \Big)d\Omega = \frac{\Omega_D}{D}\vec{\mathbf{A}}.
\end{equation}

The second integral we compute is $\int \vec{\boldsymbol{\sigma}}\Big(\vec{\boldsymbol{\sigma}} \cdot \mathbf{B}\vec{\boldsymbol{\sigma}} \Big)d\Omega$. Focusing on the $k$-th component, we obtain
\begin{equation}
    \bigg[\int \vec{\boldsymbol{\sigma}}\Big(\vec{\boldsymbol{\sigma}} \cdot \mathbf{B}\vec{\boldsymbol{\sigma}} \Big)d\Omega\bigg]_k = \sum_{i=1}^D\sum_{j=1}^DB_{ij}\int \sigma_i\sigma_j\sigma_k~d\Omega = 0,
\end{equation}
where in the last step, we have used Eq.~\eqref{eq: TEN INT}. Hence, we obtain
\begin{equation}
    \int \vec{\boldsymbol{\sigma}}\Big(\vec{\boldsymbol{\sigma}} \cdot \mathbf{B}\vec{\boldsymbol{\sigma}} \Big)d\Omega = 0.
\end{equation}

The third integral we compute is $\int \vec{\boldsymbol{\sigma}}\Big(\vec{\mathbf{A}}\cdot  \vec{\boldsymbol{\sigma}}\Big)^2~d\Omega $. Focusing on the $k$-th component, we obtain
\begin{equation}
    \bigg[\int \vec{\boldsymbol{\sigma}}\Big(\vec{\mathbf{A}}\cdot  \vec{\boldsymbol{\sigma}}\Big)^2d\Omega \bigg]_k = \sum_{i=1}^D\sum_{j=1}^DA_iA_j\int \sigma_i\sigma_j\sigma_k~d\Omega = 0,
\end{equation}
where in the last step, we have used Eq.~\eqref{eq: TEN INT}. Hence, we obtain
\begin{equation}
    \int \vec{\boldsymbol{\sigma}}\Big(\vec{\mathbf{A}}\cdot  \vec{\boldsymbol{\sigma}}\Big)^2d\Omega = 0.
\end{equation}

The fourth integral we compute is $\int \vec{\boldsymbol{\sigma}}\Big( \vec{\boldsymbol{\sigma}}\cdot \mathbf{B}\vec{\boldsymbol{\sigma}} \Big)^2~d\Omega $. Focusing on the $k$-th component, we obtain
\begin{align}
    &\bigg[\int \vec{\boldsymbol{\sigma}}\Big( \vec{\boldsymbol{\sigma}}\cdot \mathbf{B}\vec{\boldsymbol{\sigma}} \Big)^2d\Omega \bigg]_k \nonumber\\
    &= \sum_{i=1}^D\sum_{j=1}^D\sum_{l=1}^D\sum_{m=1}^DB_{ij}B_{lm}\int \sigma_i\sigma_j\sigma_l\sigma_m\sigma_k~d\Omega = 0,
\end{align}
where in the last step, we have used Eq.~\eqref{eq: TEN INT}. Hence, we obtain
\begin{equation}
    \int \vec{\boldsymbol{\sigma}}\Big( \vec{\boldsymbol{\sigma}}\cdot \mathbf{B}\vec{\boldsymbol{\sigma}} \Big)^2d\Omega = 0.
\end{equation}

The fifth integral we compute is $\int \vec{\boldsymbol{\sigma}}\Big(\vec{\mathbf{A}}\cdot  \vec{\boldsymbol{\sigma}}\Big)\Big(\vec{\boldsymbol{\sigma}} \cdot \mathbf{B} \vec{\boldsymbol{\sigma}}\Big)~d\Omega $. Focusing on the $k$-th component, we obtain
\begin{align}
    &\bigg[\int \vec{\boldsymbol{\sigma}}\Big(\vec{\mathbf{A}}\cdot  \vec{\boldsymbol{\sigma}}\Big)\Big(\vec{\boldsymbol{\sigma}} \cdot \mathbf{B} \vec{\boldsymbol{\sigma}}\Big)d\Omega\bigg]_k \nonumber\\
    &= \sum_{i=1}^D\sum_{j=1}^D\sum_{m=1}^DA_{m}
    B_{ij}\int \sigma_i\sigma_j\sigma_m\sigma_k~d\Omega =\sum_{i=1}^D\sum_{j=1}^D\sum_{m=1}^DA_{m}
    B_{ij}\nonumber\\
    &\times \bigg[\frac{\Omega_D}{D(D+2)}\bigg(\delta_{ij}\delta_{mk}+\delta_{im}\delta_{jk}+\delta_{ik}\delta_{jm}\bigg)\bigg],
\end{align}
where in the last step we have used Eq.~\eqref{eq: TEN INT}. Summing over the Kronecker deltas and using the fact that $\mathrm{Tr}\big[\mathbf{B}\big]=0$, we obtain
\begin{equation}
    \int \vec{\boldsymbol{\sigma}}\Big(\vec{\mathbf{A}}\cdot  \vec{\boldsymbol{\sigma}}\Big)\Big(\vec{\boldsymbol{\sigma}} \cdot \mathbf{B} \vec{\boldsymbol{\sigma}}\Big)d\Omega = \frac{2\Omega_D}{D(D+2)}\mathbf{B}\vec{\mathbf{A}}.
\end{equation}

The sixth integral we compute is $\int \vec{\boldsymbol{\sigma}}\Big(\vec{\mathbf{A}} \cdot \mathbf{B}\vec{\boldsymbol{\sigma}}\Big)~d\Omega $. Focusing on the $k$-th component, we obtain
\begin{align}
    &\bigg[\int \vec{\boldsymbol{\sigma}}\Big(\vec{\mathbf{A}} \cdot \mathbf{B}\vec{\boldsymbol{\sigma}}\Big)~d\Omega\bigg]_k = \sum_{i=1}^D\sum_{j=1}^DA_{i}
    B_{ij}\int \sigma_j\sigma_k~d\Omega  \nonumber\\
    &=\sum_{i=1}^D\sum_{j=1}^DA_{i}
    B_{ij}\bigg[\frac{\Omega_D}{D}\delta_{jk}\bigg] =\frac{\Omega_D}{D} \sum_{i=1}^DA_{i}B_{ik}.
\end{align}
Since, $\mathbf{B}$ is a symmetric matrix, we may write $\big[\int \vec{\boldsymbol{\sigma}}\Big(\vec{\mathbf{A}} \cdot \mathbf{B}\vec{\boldsymbol{\sigma}}\Big)~d\Omega\big]_k  = (\Omega_D/D) \sum_{i=1}^DB_{ki}A_{i} = (\Omega_D/D) \big[\mathbf{B}\vec{\mathbf{A}}\big]_k$. Hence, we obtain
\begin{equation}
    \int \vec{\boldsymbol{\sigma}}\Big(\vec{\mathbf{A}} \cdot \mathbf{B}\vec{\boldsymbol{\sigma}}\Big)~d\Omega = \frac{\Omega_D}{D}\mathbf{B}\vec{\mathbf{A}}.
\end{equation}

The seventh integral we compute is $\int \vec{\boldsymbol{\sigma}}\otimes\vec{\boldsymbol{\sigma}}\big(\vec{\boldsymbol{\sigma}}\cdot \mathbf{B}\vec{\boldsymbol{\sigma}}\big)d\Omega $. Focusing on the $mn$-th component, we obtain
\begin{align}
   &\bigg[\int \vec{\boldsymbol{\sigma}}\otimes\vec{\boldsymbol{\sigma}}\Big(\vec{\boldsymbol{\sigma}}\cdot \mathbf{B}\vec{\boldsymbol{\sigma}}\Big)d\Omega \bigg]_{mn} \nonumber\\
   &= \sum_{i=1}^D\sum_{j=1}^D B_{ij}\int \sigma_m\sigma_n\sigma_i\sigma_j~d\Omega. =\sum_{i=1}^D\sum_{j=1}^D B_{ij}\nonumber\\
   &\times \bigg[\frac{\Omega_D}{D(D+2)}\bigg(\delta_{mn}\delta_{ij}+\delta_{mi}\delta_{nj}+\delta_{mj}\delta_{ni}\bigg)\bigg],\label{eq: s t s sB^2s}
\end{align}
where in the last step, we have used Eq.~\eqref{eq: TEN INT}. Summing over the Kronecker deltas, we finally obtain
\begin{equation}
    \int \vec{\boldsymbol{\sigma}}\otimes\vec{\boldsymbol{\sigma}}\Big(\vec{\boldsymbol{\sigma}}\cdot \mathbf{B}\vec{\boldsymbol{\sigma}}\Big)d\Omega=\frac{\Omega_D}{D(D+2)}\bigg[\mathrm{Tr}\big[\mathbf{B}\big]\mathbf{I}_D+2\mathbf{B}\bigg] \label{eq: s t s s B s}.
\end{equation}
Using the traceless property of $\mathbf{B}$, we may simplify Eq.~\eqref{eq: s t s s B s} into
\begin{equation}
    \int \vec{\boldsymbol{\sigma}}\otimes\vec{\boldsymbol{\sigma}}\Big(\vec{\boldsymbol{\sigma}}\cdot \mathbf{B}\vec{\boldsymbol{\sigma}}\Big)d\Omega=\frac{2\Omega_D}{D(D+2)}\mathbf{B}.
\end{equation}

Replacing $\mathbf{B}$ by $\mathbf{B}^2$ and $\mathbf{B}^3$ in Eq.~\eqref{eq: s t s s B s}, we get
\begin{align}
    \int \vec{\boldsymbol{\sigma}}\otimes\vec{\boldsymbol{\sigma}}\Big(\vec{\boldsymbol{\sigma}}\cdot \mathbf{B}^2\vec{\boldsymbol{\sigma}}\Big)d\Omega&=\frac{\Omega_D}{D(D+2)}\bigg[\mathrm{Tr}\big[\mathbf{B}^2\big]\mathbf{I}_D+2\mathbf{B}^2\bigg],\\
    \int \vec{\boldsymbol{\sigma}}\otimes\vec{\boldsymbol{\sigma}}\Big(\vec{\boldsymbol{\sigma}}\cdot \mathbf{B}^3\vec{\boldsymbol{\sigma}}\Big)d\Omega&=\frac{\Omega_D}{D(D+2)}\bigg[\mathrm{Tr}\big[\mathbf{B}^3\big]\mathbf{I}_D+2\mathbf{B}^3\bigg].
\end{align}

The eighth integral we compute is $\int \vec{\boldsymbol{\sigma}}\otimes\vec{\boldsymbol{\sigma}}\big(\vec{\boldsymbol{\sigma}}\cdot \mathbf{B}\vec{\boldsymbol{\sigma}}\big)^2d\Omega $. Focusing on the $mn$-th component, we obtain
\begin{align}
   &\bigg[\int \vec{\boldsymbol{\sigma}}\otimes\vec{\boldsymbol{\sigma}}\Big(\vec{\boldsymbol{\sigma}}\cdot \mathbf{B}\vec{\boldsymbol{\sigma}}\Big)^2d\Omega \bigg]_{mn} \nonumber\\
   &= \sum_{i=1}^D\sum_{j=1}^D\sum_{k=1}^D\sum_{l=1}^D B_{ij}B_{kl}\int \sigma_m\sigma_n\sigma_i\sigma_j\sigma_k\sigma_l~d\Omega. \label{eq: s t s sBs square}
\end{align}
We now use Eq.~\eqref{eq: TEN INT} to compute the integral. Due to the traceless property of $\mathbf{B}$, any term containing $\delta_{ij}$ or $\delta_{kl}$ will not contribute after the summation. Considering only those terms which will contribute, we may write Eq.~\eqref{eq: s t s sBs square} as
\begin{align}
   &\bigg[\int \vec{\boldsymbol{\sigma}}\otimes\vec{\boldsymbol{\sigma}}\Big(\vec{\boldsymbol{\sigma}}\cdot \mathbf{B}\vec{\boldsymbol{\sigma}}\Big)^2d\Omega \bigg]_{mn} =\frac{\Omega_D}{D(D+2)(D+4)}\sum_{i=1}^D \nonumber\\
   &\sum_{j=1}^D\sum_{k=1}^D\sum_{l=1}^D B_{ij}B_{kl}\bigg[\delta_{mn}\delta_{ik}\delta_{jl}+\delta_{mn}\delta_{il}\delta_{jk}+\delta_{mi}\delta_{nk}\delta_{jl}\nonumber\\
   &+\delta_{mi}\delta_{nl}\delta_{jk}+\delta_{mj}\delta_{nk}\delta_{il}+ \delta_{mj}\delta_{nl}\delta_{ik}+\delta_{mk}\delta_{ni}\delta_{jl} \nonumber\\
   &+\delta_{mk}\delta_{nj}\delta_{il}+\delta_{ml}\delta_{ni}\delta_{jk}+\delta_{ml}\delta_{nj}\delta_{ik}\bigg] \nonumber\\
   &=\frac{\Omega_D}{D(D+2)(D+4)}\bigg[2\delta_{mn}\sum_{i=1}^D\sum_{j=1}^DB^2_{ij}+8\big[\mathbf{B}^2\big]_{mn}\bigg]\nonumber\\
   &=\frac{\Omega_D}{D(D+2)(D+4)}\bigg[2\delta_{mn}\mathrm{Tr}\big[\mathbf{B}^2\big]+8\big[\mathbf{B}^2\big]_{mn}\bigg].
\end{align}
Hence, we obtain
\begin{align}
    &\int \vec{\boldsymbol{\sigma}}\otimes\vec{\boldsymbol{\sigma}}\Big(\vec{\boldsymbol{\sigma}}\cdot \mathbf{B}\vec{\boldsymbol{\sigma}}\Big)^2d\Omega  \nonumber\\
    &= \frac{2\Omega_D}{D(D+2)(D+4)}\bigg[\mathrm{Tr}\big[\mathbf{B}^2\big]\mathbf{I}_D+4\mathbf{B}^2\bigg].
\end{align}

The ninth integral we compute is $\int \vec{\boldsymbol{\sigma}}\otimes\vec{\boldsymbol{\sigma}}\big(\vec{\boldsymbol{\sigma}}\cdot \mathbf{B}\vec{\boldsymbol{\sigma}}\big)\big(\vec{\boldsymbol{\sigma}}\cdot \mathbf{B}^2\vec{\boldsymbol{\sigma}}\big)d\Omega$. Focusing on the $mn$-th component, we obtain
\begin{align}
    &\bigg[\int\vec{\boldsymbol{\sigma}}\otimes\vec{\boldsymbol{\sigma}}\Big(\vec{\boldsymbol{\sigma}}\cdot \mathbf{B}\vec{\boldsymbol{\sigma}}\Big)\Big(\vec{\boldsymbol{\sigma}}\cdot \mathbf{B}^2\vec{\boldsymbol{\sigma}}\Big)d\Omega\bigg]_{mn}\nonumber\\
    &= \sum_{i=1}^D\sum_{j=1}^D\sum_{k=1}^D\sum_{l=1}^D B_{ij}\big[\mathbf{B}\big]_{kl}\int \sigma_m\sigma_n\sigma_i\sigma_j\sigma_k\sigma_l~d\Omega. \label{eq: s t s sBs sB^2s}
\end{align}
We now use Eq.~\eqref{eq: TEN INT} to compute the integral. Due to the traceless property of $\mathbf{B}$, any term containing $\delta_{ij}$ will not contribute after the summation. Considering only those terms which will contribute, we may write Eq.~\eqref{eq: s t s sBs sB^2s} as
\begin{align}
   &\bigg[\int\vec{\boldsymbol{\sigma}}\otimes\vec{\boldsymbol{\sigma}}\Big(\vec{\boldsymbol{\sigma}}\cdot \mathbf{B}\vec{\boldsymbol{\sigma}}\Big)\Big(\vec{\boldsymbol{\sigma}}\cdot \mathbf{B}^2\vec{\boldsymbol{\sigma}}\Big)d\Omega\bigg]_{mn} \nonumber\\
   &=\frac{\Omega_D}{D(D+2)(D+4)}\sum_{i=1}^D\sum_{j=1}^D\sum_{k=1}^D\sum_{l=1}^D B_{ij}\big[\mathbf{B}^2\big]_{kl}\nonumber\\
   &\times\bigg[\delta_{mn}\delta_{ik}\delta_{jl}+\delta_{mn}\delta_{il}\delta_{jk}+\delta_{mi}\delta_{nj}\delta_{kl}+\delta_{mi}\delta_{nk}\delta_{jl}\nonumber\\
   &+\delta_{mi}\delta_{nl}\delta_{jk}+\delta_{mj}\delta_{ni}\delta_{kl}+\delta_{mj}\delta_{nk}\delta_{il}+ \delta_{mj}\delta_{nl}\delta_{ik}\nonumber\\
   &+\delta_{mk}\delta_{ni}\delta_{jl} +\delta_{mk}\delta_{nj}\delta_{il}+\delta_{ml}\delta_{ni}\delta_{jk}+\delta_{ml}\delta_{nj}\delta_{ik}\bigg].
\end{align}
Summing over the Kronecker deltas, we obtain
\begin{align}
&\bigg[\int\vec{\boldsymbol{\sigma}}\otimes\vec{\boldsymbol{\sigma}}\Big(\vec{\boldsymbol{\sigma}}\cdot \mathbf{B}\vec{\boldsymbol{\sigma}}\Big)\Big(\vec{\boldsymbol{\sigma}}\cdot \mathbf{B}^2\vec{\boldsymbol{\sigma}}\Big)d\Omega\bigg]_{mn}\nonumber\\
   &=\frac{\Omega_D}{D(D+2)(D+4)}\bigg[2\delta_{mn}\sum_{i=1}^D\sum_{j=1}^DB_{ij}\big[\mathbf{B}^2\big]_{ji}\nonumber\\
   &+2B_{mn}\mathrm{Tr}\big[\mathbf{B}^2\big]+8\big[\mathbf{B}^3\big]_{mn}\bigg]\nonumber\\
   &=\frac{2\Omega_D}{D(D+2)(D+4)}\bigg[\delta_{mn}\mathrm{Tr}\big[\mathbf{B}^3\big]+B_{mn}\mathrm{Tr}\big[\mathbf{B}^2\big] \nonumber\\
   &+4\big[\mathbf{B}^3\big]_{mn}\bigg].
\end{align}
Hence, we obtain
\begin{align}
    &\int\vec{\boldsymbol{\sigma}}\otimes\vec{\boldsymbol{\sigma}}\Big(\vec{\boldsymbol{\sigma}}\cdot \mathbf{B}\vec{\boldsymbol{\sigma}}\Big)\Big(\vec{\boldsymbol{\sigma}}\cdot \mathbf{B}^2\vec{\boldsymbol{\sigma}}\Big)d\Omega =\frac{2\Omega_D}{D(D+2)(D+4)}
    \nonumber\\
    &\times\bigg[\mathrm{Tr}\big[\mathbf{B}^3\big]\mathbf{I}_D+\mathrm{Tr}\big[\mathbf{B}^2\big] \mathbf{B}+4\mathbf{B}^3\bigg].
\end{align}

The tenth integral we compute is $\int \vec{\boldsymbol{\sigma}}\otimes\vec{\boldsymbol{\sigma}}\big(\vec{\boldsymbol{\sigma}}\cdot \mathbf{B}\vec{\boldsymbol{\sigma}}\big)^3d\Omega$. Focusing on the $mn$-th component, we obtain
\begin{align}
    &\bigg[\int\vec{\boldsymbol{\sigma}}\otimes\vec{\boldsymbol{\sigma}}\Big(\vec{\boldsymbol{\sigma}}\cdot \mathbf{B}\vec{\boldsymbol{\sigma}}\Big)^3d\Omega\bigg]_{mn}=\sum_{g=1}^D\sum_{h=1}^D\sum_{i=1}^D\sum_{j=1}^D\sum_{k=1}^D\sum_{l=1}^D\nonumber\\
    &  B_{gh}B_{ij}B_{kl}\int \sigma_m\sigma_n\sigma_g\sigma_h\sigma_i\sigma_j\sigma_k\sigma_l~d\Omega. \label{eq: s t s (sBs)^3}
\end{align}
From~\eqref{eq: TEN INT} we have
\begin{align}
    \int \sigma_m\sigma_n\sigma_g\sigma_h\sigma_i\sigma_j\sigma_k\sigma_l~d\Omega &= \frac{\Omega_D}{D(D+2)(D+4)(D+6)}\nonumber\\
    &\times\sum_{\{\mathrm{pairings}\,P\}}\delta_{i_ai_b}\delta_{i_ci_d}\delta_{i_ei_f}\delta_{i_oi_p}.
\end{align}
Total number of such pairs is $8!/(2^44!) = 105$. Since it is tedious to write down all the combinations of the Kronecker deltas, we will resort to a diagrammatic method to compute the integral.
\begin{equation}
 B_{ab} =~
\mathrel{
\tikz[baseline={(B.base)}]{
  \node[
    draw,
    thick,
    rectangle,
    rounded corners=1pt,
    minimum width=0.5cm,
    minimum height=0.5cm
  ] (B) {$\mathbf{B}$};

  % upper leg: index i
  \draw[thick, shorten <=-0.6pt]
    (B.north east) -- ++(0.3,0.3)
    node[circle, fill=black, inner sep=1.2pt] {}
    node[above right=-2pt] {$a$};

  % lower leg: index j
  \draw[thick, shorten <=-0.6pt]
    (B.south east) -- ++(0.3,-0.3)
    node[circle, fill=black, inner sep=1.2pt] {}
    node[below right=-2pt] {$b$};
}},
~~
\delta_{ab} = ~~
\mathrel{
\tikz[baseline={(c.base)}]{
  \node[circle, fill=black, inner sep=1.2pt] (a) at (0,0.5) {};\node[right=2pt] at (a) {$a$};
  \node[circle, fill=black, inner sep=1.2pt] (b) at (0,-0.5) {};\node[right=2pt] at (b) {$b$};
  \draw[blue, dashed, thick] (a) -- (b);
  \node[inner sep=2pt] (c) at (0,0) {};
}},
~~
\sum_{a=1}^D\sum_{b=1}^D\delta_{ab}B_{ab} = 
~\mathrel{
\tikz[baseline={(B.base)}]{
  \node[
    draw,
    thick,
    rectangle,
    rounded corners=1pt,
    minimum width=0.5cm,
    minimum height=0.5cm
  ] (B) {$\mathbf{B}$};

  % upper leg: index i
  \draw[thick, shorten <=-0.6pt]
    (B.north east) -- ++(0.3,0.3)
    node[circle, fill=black, inner sep=1.2pt](u){};

  % lower leg: index j
  \draw[thick, shorten <=-0.6pt]
    (B.south east) -- ++(0.3,-0.3)
    node[circle, fill=black, inner sep=1.2pt](d){};
    \draw[blue, dashed, thick] (u) -- (d);
}}.
\label{eq: fig diagram intro1}
\end{equation}
In Eq.~\eqref{eq: fig diagram intro1}, we have denoted symbolically the basic rules of the diagrams. If we have a term $B_{ab}$, we will add $\left(\mathrel{
\tikz[baseline={(B.base)}]{
  \node[
    draw,
    thick,
    rectangle,
    rounded corners=1pt,
    minimum width=0.5cm,
    minimum height=0.5cm
  ] (B) {$\mathbf{B}$};

  % upper leg: index i
  \draw[thick, shorten <=-0.6pt]
    (B.north east) -- ++(0.3,0.3)
   {};

  % lower leg: index j
  \draw[thick, shorten <=-0.6pt]
    (B.south east) -- ++(0.3,-0.3) {};
}}\right)$
between the indices $a$ and $b$. Similarly, if we have $\delta_{ab}$, we will add $\left(\mathrel{
\tikz[baseline={(c.base)}]{
  \node[circle, inner sep=1.2pt] (a) at (0,0.4) {};
  \node[circle, inner sep=1.2pt] (b) at (0,-0.3) {};
  \draw[blue, dashed, thick] (a) -- (b);
}}\right)$ between the indices $a$ and $b$. If there are two lines (one solid and one dashed) connected to the same index, then we will sum over that index. 

To evaluate the integral in Eq.~\eqref{eq: s t s (sBs)^3}, there are seven distinct diagrams possible, as shown below.
\begin{equation}
   \mathbf{I} =~~  \mathrel{
\tikz[baseline={(B1.base)}]{
  \node[
    draw,
    thick,
    rectangle,
    rounded corners=1pt,
    minimum width=0.5cm,
    minimum height=0.5cm
  ] (B1) {$\mathbf{B}$};
  \draw[thick, shorten <=-0.6pt]
    (B1.north east) -- ++(0.2,0.2)
    node[circle, fill=black, inner sep=1.2pt](B11) {};
  \draw[thick, shorten <=-0.6pt]
    (B1.south east) -- ++(0.2,-0.2)
    node[circle, fill=black, inner sep=1.2pt](B12) {};
%=====================================================
    \node[
    draw,
    thick,
    rectangle,
    rounded corners=1pt,
    minimum width=0.5cm,
    minimum height=0.5cm
  ] (B2) at ($(B1)+(1.2cm,1.2cm)$) {$\mathbf{B}$};
  \draw[thick, shorten <=-0.6pt]
    (B2.south west) -- ++(-0.2,-0.2)
    node[circle, fill=black, inner sep=1.2pt] (B21){};
  \draw[thick, shorten <=-0.6pt]
    (B2.south east) -- ++(0.2,-0.2)
    node[circle, fill=black, inner sep=1.2pt](B22) {};
%=====================================================
    \node[
    draw,
    thick,
    rectangle,
    rounded corners=1pt,
    minimum width=0.5cm,
    minimum height=0.5cm
  ] (B3) at ($(B1)+(1.2cm,-1.2cm)$) {$\mathbf{B}$};
  \draw[thick, shorten <=-0.6pt]
    (B3.north west) -- ++(-0.2,0.2)
    node[circle, fill=black, inner sep=1.2pt](B31) {};
  \draw[thick, shorten <=-0.6pt]
    (B3.north east) -- ++(0.2,0.2)
    node[circle, fill=black, inner sep=1.2pt] (B32) {};
%=====================================================
    \node[] (B4) at ($(B1)+(2.4cm,0.0cm)$) {};
    \draw[](B4)++(-0.5,0.4)
    node[circle, fill=black, inner sep=1.2pt] (m) {}
    node[above right=0pt] {$m$};
  \draw[](B4) ++(-0.5,-0.4)
    node[circle, fill=black, inner sep=1.2pt] (n) {}
    node[below right=0pt] {$n$};
%=====================================================
    \draw[blue, dashed, thick] (B11) -- (B12);
    \draw[blue, dashed, thick] (B21) -- (B22);
    \draw[blue, dashed, thick] (B31) -- (B32);
    \draw[blue, dashed, thick] (m) -- (n);
}}
~~= \left(\mathrm{Tr}\big[\mathbf{B}\big]\right)^3\delta_{mn} = 0,
\end{equation}

\begin{equation}
   \mathbf{II} =~~  \mathrel{
\tikz[baseline={(B1.base)}]{
  \node[
    draw,
    thick,
    rectangle,
    rounded corners=1pt,
    minimum width=0.5cm,
    minimum height=0.5cm
  ] (B1) {$\mathbf{B}$};
  \draw[thick, shorten <=-0.6pt]
    (B1.north east) -- ++(0.2,0.2)
    node[circle, fill=black, inner sep=1.2pt](B11) {};
  \draw[thick, shorten <=-0.6pt]
    (B1.south east) -- ++(0.2,-0.2)
    node[circle, fill=black, inner sep=1.2pt](B12) {};
%=====================================================
    \node[
    draw,
    thick,
    rectangle,
    rounded corners=1pt,
    minimum width=0.5cm,
    minimum height=0.5cm
  ] (B2) at ($(B1)+(1.2cm,1.2cm)$) {$\mathbf{B}$};
  \draw[thick, shorten <=-0.6pt]
    (B2.south west) -- ++(-0.2,-0.2)
    node[circle, fill=black, inner sep=1.2pt] (B21){};
  \draw[thick, shorten <=-0.6pt]
    (B2.south east) -- ++(0.2,-0.2)
    node[circle, fill=black, inner sep=1.2pt](B22) {};
%=====================================================
    \node[
    draw,
    thick,
    rectangle,
    rounded corners=1pt,
    minimum width=0.5cm,
    minimum height=0.5cm
  ] (B3) at ($(B1)+(1.2cm,-1.2cm)$) {$\mathbf{B}$};
  \draw[thick, shorten <=-0.6pt]
    (B3.north west) -- ++(-0.2,0.2)
    node[circle, fill=black, inner sep=1.2pt](B31) {};
  \draw[thick, shorten <=-0.6pt]
    (B3.north east) -- ++(0.2,0.2)
    node[circle, fill=black, inner sep=1.2pt] (B32) {};
%=====================================================
    \node[] (B4) at ($(B1)+(2.4cm,0.0cm)$) {};
    \draw[](B4)++(-0.5,0.4)
    node[circle, fill=black, inner sep=1.2pt] (m) {}
    node[above right=0pt] {$m$};
  \draw[](B4) ++(-0.5,-0.4)
    node[circle, fill=black, inner sep=1.2pt] (n) {}
    node[below right=0pt] {$n$};
%=====================================================
    \draw[blue, dashed, thick] (B11) -- (B12);
    \draw[blue, dashed, thick] (B21) -- (B31);
    \draw[blue, dashed, thick] (B22) -- (B32);
    \draw[blue, dashed, thick] (m) -- (n);
}}
~~= \mathrm{Tr}\big[\mathbf{B}\big] \mathrm{Tr}\big[\mathbf{B}^2\big]\delta_{mn} = 0,
\end{equation}

\begin{equation}
   \mathbf{III} =~~  \mathrel{
\tikz[baseline={(B1.base)}]{
  \node[
    draw,
    thick,
    rectangle,
    rounded corners=1pt,
    minimum width=0.5cm,
    minimum height=0.5cm
  ] (B1) {$\mathbf{B}$};
  \draw[thick, shorten <=-0.6pt]
    (B1.north east) -- ++(0.2,0.2)
    node[circle, fill=black, inner sep=1.2pt](B11) {};
  \draw[thick, shorten <=-0.6pt]
    (B1.south east) -- ++(0.2,-0.2)
    node[circle, fill=black, inner sep=1.2pt](B12) {};
%=====================================================
    \node[
    draw,
    thick,
    rectangle,
    rounded corners=1pt,
    minimum width=0.5cm,
    minimum height=0.5cm
  ] (B2) at ($(B1)+(1.2cm,1.2cm)$) {$\mathbf{B}$};
  \draw[thick, shorten <=-0.6pt]
    (B2.south west) -- ++(-0.2,-0.2)
    node[circle, fill=black, inner sep=1.2pt] (B21){};
  \draw[thick, shorten <=-0.6pt]
    (B2.south east) -- ++(0.2,-0.2)
    node[circle, fill=black, inner sep=1.2pt](B22) {};
%=====================================================
    \node[
    draw,
    thick,
    rectangle,
    rounded corners=1pt,
    minimum width=0.5cm,
    minimum height=0.5cm
  ] (B3) at ($(B1)+(1.2cm,-1.2cm)$) {$\mathbf{B}$};
  \draw[thick, shorten <=-0.6pt]
    (B3.north west) -- ++(-0.2,0.2)
    node[circle, fill=black, inner sep=1.2pt](B31) {};
  \draw[thick, shorten <=-0.6pt]
    (B3.north east) -- ++(0.2,0.2)
    node[circle, fill=black, inner sep=1.2pt] (B32) {};
%=====================================================
    \node[] (B4) at ($(B1)+(2.4cm,0.0cm)$) {};
    \draw[](B4)++(-0.4,0.4)
    node[circle, fill=black, inner sep=1.2pt] (m) {}
    node[above right=0pt] {$m$};
  \draw[](B4) ++(-0.4,-0.4)
    node[circle, fill=black, inner sep=1.2pt] (n) {}
    node[below right=0pt] {$n$};
%=====================================================
    \draw[blue, dashed, thick] (B11) -- (B12);
    \draw[blue, dashed, thick] (B21) -- (n);
    \draw[blue, dashed, thick] (B31) -- (B32);
    \draw[blue, dashed, thick] (B22) -- (m);
}}
~~= \left(\mathrm{Tr}\big[\mathbf{B}\big]\right)^2B_{mn} = 0,
\end{equation}

\begin{equation}
   \mathbf{IV} =~~  \mathrel{
\tikz[baseline={(B1.base)}]{
  \node[
    draw,
    thick,
    rectangle,
    rounded corners=1pt,
    minimum width=0.5cm,
    minimum height=0.5cm
  ] (B1) {$\mathbf{B}$};
  \draw[thick, shorten <=-0.6pt]
    (B1.north east) -- ++(0.2,0.2)
    node[circle, fill=black, inner sep=1.2pt](B11) {};
  \draw[thick, shorten <=-0.6pt]
    (B1.south east) -- ++(0.2,-0.2)
    node[circle, fill=black, inner sep=1.2pt](B12) {};
%=====================================================
    \node[
    draw,
    thick,
    rectangle,
    rounded corners=1pt,
    minimum width=0.5cm,
    minimum height=0.5cm
  ] (B2) at ($(B1)+(1.2cm,1.2cm)$) {$\mathbf{B}$};
  \draw[thick, shorten <=-0.6pt]
    (B2.south west) -- ++(-0.2,-0.2)
    node[circle, fill=black, inner sep=1.2pt] (B21){};
  \draw[thick, shorten <=-0.6pt]
    (B2.south east) -- ++(0.2,-0.2)
    node[circle, fill=black, inner sep=1.2pt](B22) {};
%=====================================================
    \node[
    draw,
    thick,
    rectangle,
    rounded corners=1pt,
    minimum width=0.5cm,
    minimum height=0.5cm
  ] (B3) at ($(B1)+(1.2cm,-1.2cm)$) {$\mathbf{B}$};
  \draw[thick, shorten <=-0.6pt]
    (B3.north west) -- ++(-0.2,0.2)
    node[circle, fill=black, inner sep=1.2pt](B31) {};
  \draw[thick, shorten <=-0.6pt]
    (B3.north east) -- ++(0.2,0.2)
    node[circle, fill=black, inner sep=1.2pt] (B32) {};
%=====================================================
    \node[] (B4) at ($(B1)+(2.4cm,0.0cm)$) {};
    \draw[](B4)++(-0.4,0.4)
    node[circle, fill=black, inner sep=1.2pt] (m) {}
    node[above right=0pt] {$m$};
  \draw[](B4) ++(-0.4,-0.4)
    node[circle, fill=black, inner sep=1.2pt] (n) {}
    node[below right=0pt] {$n$};
%=====================================================
    \draw[blue, dashed, thick] (B11) -- (B12);
    \draw[blue, dashed, thick] (B21) -- (B31);
    \draw[blue, dashed, thick] (B32) -- (n);
    \draw[blue, dashed, thick] (B22) -- (m);
}}
~~= \mathrm{Tr}\big[\mathbf{B}\big]\big[\mathbf{B}^2\big]_{mn} = 0,
\end{equation}

\begin{equation}
   \mathbf{V} =~~  \mathrel{
\tikz[baseline={(B1.base)}]{
  \node[
    draw,
    thick,
    rectangle,
    rounded corners=1pt,
    minimum width=0.5cm,
    minimum height=0.5cm
  ] (B1) {$\mathbf{B}$};
  \draw[thick, shorten <=-0.6pt]
    (B1.north east) -- ++(0.2,0.2)
    node[circle, fill=black, inner sep=1.2pt](B11) {};
  \draw[thick, shorten <=-0.6pt]
    (B1.south east) -- ++(0.2,-0.2)
    node[circle, fill=black, inner sep=1.2pt](B12) {};
%=====================================================
    \node[
    draw,
    thick,
    rectangle,
    rounded corners=1pt,
    minimum width=0.5cm,
    minimum height=0.5cm
  ] (B2) at ($(B1)+(1.5cm,1.5cm)$) {$\mathbf{B}$};
  \draw[thick, shorten <=-0.6pt]
    (B2.south west) -- ++(-0.2,-0.2)
    node[circle, fill=black, inner sep=1.2pt] (B21){};
  \draw[thick, shorten <=-0.6pt]
    (B2.south east) -- ++(0.2,-0.2)
    node[circle, fill=black, inner sep=1.2pt](B22) {};
%=====================================================
    \node[
    draw,
    thick,
    rectangle,
    rounded corners=1pt,
    minimum width=0.5cm,
    minimum height=0.5cm
  ] (B3) at ($(B1)+(1.5cm,-1.5cm)$) {$\mathbf{B}$};
  \draw[thick, shorten <=-0.6pt]
    (B3.north west) -- ++(-0.2,0.2)
    node[circle, fill=black, inner sep=1.2pt](B31) {};
  \draw[thick, shorten <=-0.6pt]
    (B3.north east) -- ++(0.2,0.2)
    node[circle, fill=black, inner sep=1.2pt] (B32) {};
%=====================================================
    \node[] (B4) at ($(B1)+(3.0cm,0.0cm)$) {};
    \draw[](B4)++(-0.5,0.4)
    node[circle, fill=black, inner sep=1.2pt] (m) {}
    node[above right=0pt] {$m$};
  \draw[](B4) ++(-0.5,-0.4)
    node[circle, fill=black, inner sep=1.2pt] (n) {}
    node[below right=0pt] {$n$};
%=====================================================
    \draw[blue, dashed, thick] (B11) -- (B21);
    \draw[blue, dashed, thick] (B12) -- (B31);
    \draw[blue, dashed, thick] (B22) -- (B32);
    \draw[blue, dashed, thick] (m) -- (n);
}}
~~= \mathrm{Tr}\big[\mathbf{B}^3\big]\delta_{mn} ,
\end{equation}

\begin{equation}
   \mathbf{VI} =~~  \mathrel{
\tikz[baseline={(B1.base)}]{
  \node[
    draw,
    thick,
    rectangle,
    rounded corners=1pt,
    minimum width=0.5cm,
    minimum height=0.5cm
  ] (B1) {$\mathbf{B}$};
  \draw[thick, shorten <=-0.6pt]
    (B1.north east) -- ++(0.2,0.2)
    node[circle, fill=black, inner sep=1.2pt](B11) {};
  \draw[thick, shorten <=-0.6pt]
    (B1.south east) -- ++(0.2,-0.2)
    node[circle, fill=black, inner sep=1.2pt](B12) {};
%=====================================================
    \node[
    draw,
    thick,
    rectangle,
    rounded corners=1pt,
    minimum width=0.5cm,
    minimum height=0.5cm
  ] (B2) at ($(B1)+(1.4cm,1.4cm)$) {$\mathbf{B}$};
  \draw[thick, shorten <=-0.6pt]
    (B2.south west) -- ++(-0.2,-0.2)
    node[circle, fill=black, inner sep=1.2pt] (B21){};
  \draw[thick, shorten <=-0.6pt]
    (B2.south east) -- ++(0.2,-0.2)
    node[circle, fill=black, inner sep=1.2pt](B22) {};
%=====================================================
    \node[
    draw,
    thick,
    rectangle,
    rounded corners=1pt,
    minimum width=0.5cm,
    minimum height=0.5cm
  ] (B3) at ($(B1)+(1.4cm,-1.4cm)$) {$\mathbf{B}$};
  \draw[thick, shorten <=-0.6pt]
    (B3.north west) -- ++(-0.2,0.2)
    node[circle, fill=black, inner sep=1.2pt](B31) {};
  \draw[thick, shorten <=-0.6pt]
    (B3.north east) -- ++(0.2,0.2)
    node[circle, fill=black, inner sep=1.2pt] (B32) {};
%=====================================================
    \node[] (B4) at ($(B1)+(2.8cm,0.0cm)$) {};
    \draw[](B4)++(-0.4,0.4)
    node[circle, fill=black, inner sep=1.2pt] (m) {}
    node[above right=0pt] {$m$};
  \draw[](B4) ++(-0.4,-0.6)
    node[circle, fill=black, inner sep=1.2pt] (n) {}
    node[below right=0pt] {$n$};
%=====================================================
    \draw[blue, dashed, thick] (B11) -- (B32);
    \draw[blue, dashed, thick] (B21) -- (n);
    \draw[blue, dashed, thick] (B12) -- (B31);
    \draw[blue, dashed, thick] (B22) -- (m);
}}
~~= \mathrm{Tr}\big[\mathbf{B}^2\big]B_{mn} ,
\end{equation}

\begin{equation}
   \mathbf{VII} =~~  \mathrel{
\tikz[baseline={(B1.base)}]{
  \node[
    draw,
    thick,
    rectangle,
    rounded corners=1pt,
    minimum width=0.5cm,
    minimum height=0.5cm
  ] (B1) {$\mathbf{B}$};
  \draw[thick, shorten <=-0.6pt]
    (B1.north east) -- ++(0.2,0.2)
    node[circle, fill=black, inner sep=1.2pt](B11) {};
  \draw[thick, shorten <=-0.6pt]
    (B1.south east) -- ++(0.2,-0.2)
    node[circle, fill=black, inner sep=1.2pt](B12) {};
%=====================================================
    \node[
    draw,
    thick,
    rectangle,
    rounded corners=1pt,
    minimum width=0.5cm,
    minimum height=0.5cm
  ] (B2) at ($(B1)+(1.4cm,1.4cm)$) {$\mathbf{B}$};
  \draw[thick, shorten <=-0.6pt]
    (B2.south west) -- ++(-0.2,-0.2)
    node[circle, fill=black, inner sep=1.2pt] (B21){};
  \draw[thick, shorten <=-0.6pt]
    (B2.south east) -- ++(0.2,-0.2)
    node[circle, fill=black, inner sep=1.2pt](B22) {};
%=====================================================
    \node[
    draw,
    thick,
    rectangle,
    rounded corners=1pt,
    minimum width=0.5cm,
    minimum height=0.5cm
  ] (B3) at ($(B1)+(1.4cm,-1.4cm)$) {$\mathbf{B}$};
  \draw[thick, shorten <=-0.6pt]
    (B3.north west) -- ++(-0.2,0.2)
    node[circle, fill=black, inner sep=1.2pt](B31) {};
  \draw[thick, shorten <=-0.6pt]
    (B3.north east) -- ++(0.2,0.2)
    node[circle, fill=black, inner sep=1.2pt] (B32) {};
%=====================================================
    \node[] (B4) at ($(B1)+(2.8cm,0.0cm)$) {};
    \draw[](B4)++(-0.4,0.4)
    node[circle, fill=black, inner sep=1.2pt] (m) {}
    node[above right=0pt] {$m$};
  \draw[](B4) ++(-0.4,-0.4)
    node[circle, fill=black, inner sep=1.2pt] (n) {}
    node[below right=0pt] {$n$};
%=====================================================
    \draw[blue, dashed, thick] (B11) -- (B21);
    \draw[blue, dashed, thick] (B12) -- (B31);
    \draw[blue, dashed, thick] (B32) -- (n);
    \draw[blue, dashed, thick] (B22) -- (m);
}}
~~= \big[\mathbf{B}^3\big]_{mn}.
\end{equation}
Hence, only the diagrams $\textbf{V},~\textbf{VI}$ and $\textbf{VII}$ contribute the integral in Eq.~\eqref{eq: s t s (sBs)^3}. 

We now compute the degeneracy of the aforementioned diagrams. For the diagram $\textbf{V}$, one hand of any $\mathbf{B}$ can be contracted with one hand of another $\mathbf{B}$ in $2\times2 = 4$ ways. For each of these cases, the third $\mathbf{B}$ can be contracted with the open hands of the first two $\mathbf{B}$'s in $2$ ways. Hence, the total degeneracy of the diagram is $8$.

For diagram $\mathbf{VI}$, which among the three $\mathbf{B}$'s will be contracted with the free indices $m$ and $n$ can be chosen in $3$ ways. Upon choosing the $\mathbf{B}$ in one of these ways, which of its two hands will be connected with the free indices $m$ and $n$ results in a degeneracy factor of $2$. The rest of the two $\mathbf{B}$'s can be contracted among themselves in $2$ ways. Hence, the total degeneracy of the diagram is $12$.

For diagram $\mathbf{VII}$, which among the three $\mathbf{B}$'s will be contracted with the free indices $m$ can be chosen in $3$ ways. Upon choosing it, which among the rest two $\mathbf{B}$'s will be contracted with the free indices $n$ can be chosen in $2$ ways. Upon choosing the $\mathbf{B}$'s, in each case, we have two choices about which hand to be contracted with the free indices, $m$ or $n$. After doing this, the free hands of these two $\mathbf{B}$ will be contracted with the third $\mathbf{B}$, and it can be done in $2$ ways. Hence, the total degeneracy of the diagram is $3 \times 2 \times 2 \times 2\times 2=48$.

Combining everything, we obtain the value of the integral in Eq.~\eqref{eq: s t s (sBs)^3} as
\begin{align}
    &\int\vec{\boldsymbol{\sigma}}\otimes\vec{\boldsymbol{\sigma}}\Big(\vec{\boldsymbol{\sigma}}\cdot \mathbf{B}\vec{\boldsymbol{\sigma}}\Big)^3d\Omega=\frac{4\Omega_D}{D(D+2)(D+4)(D+6)}
    \nonumber\\
    &\times\bigg[2\mathrm{Tr}\big[\mathbf{B}^3\big]\mathbf{I}_D+3\mathrm{Tr}\big[\mathbf{B}^2\big] \mathbf{B}+12\mathbf{B}^3\bigg]. \label{eq: fig diagram intro}
\end{align}

\section{Derivation of \texorpdfstring{Eq.~\eqref{eq: eig eq ten K2 INT}}{Eq.~(143)}}
\label{app: 12}

Using the definition of $\big( \mathbb{T}^{(n)},\vec{\boldsymbol{\sigma}}^{\otimes n}\big)$ from Eq.~\eqref{eq: T sigma expan D DIM}, we may write the $km$-th component of the integral $\int \vec{\boldsymbol{\sigma}} \otimes \vec{\boldsymbol{\sigma}}\big(\mathbb{T}^{(n)},\vec{\boldsymbol{\sigma}}^{\otimes n}\big) d\Omega$ as
\begin{align}
    &\bigg[\int \vec{\boldsymbol{\sigma}} \otimes \vec{\boldsymbol{\sigma}}\Big( \mathbb{T}^{(n)},\vec{\boldsymbol{\sigma}}^{\otimes n}\Big)d\Omega\bigg]_{km} \nonumber\\
    &=\sum_{i_1=1}^D\sum_{i_2=1}^D \cdots \sum_{i_n=1}^DT^{(n)}_{i_1,i_2,\ldots,i_n}\int\sigma_k \sigma_m \sigma_{i_1}\sigma_{i_2} \ldots\sigma_{i_n}d\Omega.
\end{align}
If $n$ is odd, then the integral contains the product of odd number of $\sigma$'s. Hence, using Eq.~\eqref{eq: TEN INT}, we immediately get $\int \vec{\boldsymbol{\sigma}} \otimes \vec{\boldsymbol{\sigma}}\Big( \mathbb{T}^{(n)},\vec{\boldsymbol{\sigma}}^{\otimes n}\Big)d\Omega=0$ when $n$ is odd. For even $n$ and $n\geq 4$, following a similar argument as given in Appendix~\ref{app: 8}, we may show $\int \vec{\boldsymbol{\sigma}} \otimes \vec{\boldsymbol{\sigma}}\Big( \mathbb{T}^{(n)},\vec{\boldsymbol{\sigma}}^{\otimes n}\Big)d\Omega=0$. The only case left to compute is $n=2$. Focusing on the $km$-th component and using Eq.~\eqref{eq: TEN INT}, we obtain
\begin{align}
    &\bigg[\int \vec{\boldsymbol{\sigma}} \otimes \vec{\boldsymbol{\sigma}}\Big( \mathbb{T}^{(2)},\vec{\boldsymbol{\sigma}}\otimes\vec{\boldsymbol{\sigma}}\Big)d\Omega\bigg]_{km} \nonumber\\
    &=\sum_{i_1=1}^D\sum_{i_2=1}^D T^{(2)}_{i_1,i_2}\int\sigma_k \sigma_m \sigma_{i_1}\sigma_{i_2}d\Omega =\sum_{i_1=1}^D\sum_{i_2=1}^D T^{(2)}_{i_1,i_2}\nonumber\\
    &\times\bigg[\frac{\Omega_D}{D(D+2)}\bigg(\delta_{i_1i_2}\delta_{mk}+\delta_{i_1m}\delta_{i_2k}+\delta_{i_1k}\delta_{i_2m}\bigg)\bigg]. \label{eq: L2}
\end{align}
Using the traceless property of $\mathbb{T}^{(2)}$, we have $\sum_{i_1=1}^D\sum_{i_2=1}^D T^{(2)}_{i_1,i_2}\delta_{i_1i_2}=0$. Using the symmetric property of $\mathbb{T}^{(2)}$, we further obtain $\sum_{i_1=1}^D\sum_{i_2=1}^D T^{(2)}_{i_1,i_2}\delta_{i_1m}\delta_{i_2k}=\sum_{i_1=1}^D\sum_{i_2=1}^D T^{(2)}_{i_1,i_2}\delta_{i_1k}\delta_{i_2m}=T^{(2)}_{km}$. Using these results back into Eq.~\eqref{eq: L2}, we obtain
\begin{equation}
    \bigg[\int \vec{\boldsymbol{\sigma}} \otimes \vec{\boldsymbol{\sigma}}\Big( \mathbb{T}^{(2)},\vec{\boldsymbol{\sigma}}\otimes\vec{\boldsymbol{\sigma}}\Big)d\Omega\bigg]_{km} = \frac{2\Omega_D}{D(D+2)}T^{(2)}_{km},
\end{equation}
which immediately gives
\begin{align}
    &\vec{\boldsymbol{\sigma}} \cdot \bigg[\int \vec{\boldsymbol{\sigma}}'\otimes \vec{\boldsymbol{\sigma}}'\Big( \mathbb{T}^{(2)},\vec{\boldsymbol{\sigma}}'\otimes \vec{\boldsymbol{\sigma}}' \Big) d\Omega'\bigg]\vec{\boldsymbol{\sigma}} \nonumber\\
    &= \frac{2\Omega_D}{D(D+2)} \sum_{k=1}^D\sum_{m=1}^DT^{(2)}_{km}\sigma_k \sigma_m = \frac{2\Omega_D}{D(D+2)} \Big( \mathbb{T}^{(2)},\vec{\boldsymbol{\sigma}}\otimes \vec{\boldsymbol{\sigma}} \Big).
\end{align}

Combining everything, we thus obtain
\begin{equation}
    \vec{\boldsymbol{\sigma}} \cdot \bigg[\int \vec{\boldsymbol{\sigma}}'\otimes \vec{\boldsymbol{\sigma}}'\Big( \mathbb{T}^{(n)},\vec{\boldsymbol{\sigma}}'^{\otimes n}\Big) d\Omega'\bigg]\vec{\boldsymbol{\sigma}}= \frac{\delta_{n,2}2\Omega_D}{D(D+2)}\Big( \mathbb{T}^{(n)},\vec{\boldsymbol{\sigma}}^{\otimes n}\Big). \label{eq: eig eq ten K2 INT app}
\end{equation}
Equation~\eqref{eq: eig eq ten K2 INT app} is the desired Eq.~\eqref{eq: eig eq ten K2 INT} of the main text.

\section{\texorpdfstring{Checking of Eq.~\eqref{eq: TIJKL}}{Checking of Eq. 170}}
\label{app: 13}

Here we show that the expression of the elements of $\mathbb{T}^{(4)}$ as given in Eq.~\eqref{eq: TIJKL} is indeed traceless and symmetric. Since $b_4\Big(\mathrm{Tr}\big[\mathbf{B}^2\big],\mathrm{Tr}\big[\mathbf{B}^3\big],\ldots,\mathrm{Tr}\big[\mathbf{B}^D\big]\Big)$ is simply a scalar, we will represent it as $b_4$ in this section. Hence, Eq.~\eqref{eq: TIJKL} becomes
\begin{align}
    &T^{(4)}_{ijkl} = b_4\Bigg[\Big(B_{ij}B_{kl}+ B_{ik}B_{jl}+B_{il}B_{jk}\Big) \nonumber\\
    &-\frac{2}{(D+2)}\Big[\delta_{ij}\big[\mathbf{B}^2\big]_{kl} +\delta_{ik}\big[\mathbf{B}^2\big]_{jl}+\delta_{il}\big[\mathbf{B}^2\big]_{jk} \nonumber\\
    &+\delta_{jk}\big[\mathbf{B}^2\big]_{il}+\delta_{jl}\big[\mathbf{B}^2\big]_{ik} +\delta_{kl}\big[\mathbf{B}^2\big]_{ij}\Big]\nonumber\\
    &+\frac{2\mathrm{Tr}\big[\mathbf{B}^2\big]}{(D+2)(D+4)} \Big(\delta_{ij}\delta_{kl} + \delta_{ik}\delta_{jl}+\delta_{il}\delta_{jk}\Big)\Bigg]. \label{eq: TIJKL app 1}
\end{align}

Since $\mathbf{B}$ is a symmetric matrix, we have $B_{ab} = B_{ba}$ and $\big[\mathbf{B}^2\big]_{ab}=\big[\mathbf{B}^2\big]_{ba}$. Further, from the definition, we have $\delta_{ab} = \delta_{ba}$. Using these properties, we immediately see that $T^{(4)}_{ijkl}$ remains invariant under interchange of any pair of its indices, i.e., $i \leftrightarrow j$ or $i \leftrightarrow k$ or $i\leftrightarrow
l$ or $j \leftrightarrow
 k$ or $j \leftrightarrow
l$ or $k \leftrightarrow
 l$. Hence, $\mathbb{T}^{(4)}$ is a symmetric tensor. 

 Since $\mathbb{T}^{(4)}$ is symmetric, vanishing of the trace over any one pair of indices implies vanishing of all traces, indicating tracelessness of $\mathbb{T}^{(4)}$. In the following, we evaluate the trace by contracting the $ij$ index pair. By definition, the trace is
 \begin{align}
    &\sum_{i=1}^D\sum_{j=1}^D \delta_{ij}\mathbb{T}^{(4)}_{ijkl} \nonumber\\
    &= b_4\Bigg[\Big(\sum_{i=1}^DB_{ii}\Big)B_{kl}+ \sum_{i=1}^DB_{ik}B_{il}+\sum_{i=1}^DB_{il}B_{ik}  \nonumber\\
    &-\frac{2}{(D+4)}\Big[(D+2)\big[\mathbf{B}^2\big]_{kl}+2\big[\mathbf{B}^2\big]_{lk}+\delta_{kl}\Big(\sum_{i=1}^D\big[\mathbf{B}^2\big]_{ii}\Big)\Big]\nonumber\\
    &+\frac{2\mathrm{Tr}\big[\mathbf{B}^2\big]}{(D+2)(D+4)} \Big(D\delta_{kl} + 2\sum_{i=1}^D\delta_{il}\delta_{ik}\Big)\Bigg]. \label{eq: TIJKL app 2}
\end{align}
Since $\mathbf{B}$ is traceless, we have $\sum_{i=1}^DB_{ii} = 0$. The symmetric property of $\mathbf{B}$ gives $\sum_{i=1}^DB_{ik}B_{il} =\sum_{i=1}^DB_{il}B_{ik}= \big[\mathbf{B}^2\big]_{kl}$. The symmetric property of $\mathbf{B}^2$ gives $\big[\mathbf{B}^2\big]_{kl} =\big[\mathbf{B}^2\big]_{lk}$. Further, from the definition, we have $\sum_{i=1}^D\big[\mathbf{B}^2\big]_{ii} =\mathrm{Tr}\big[\mathbf{B}^2\big] $ and $\sum_{i=1}^D\delta_{il}\delta_{ik} = \delta_{kl}$. Hence, Eq.~\eqref{eq: TIJKL app 2} simplifies to
\begin{align}
    \sum_{i=1}^D\sum_{j=1}^D \delta_{ij}\mathbb{T}^{(4)}_{ijkl} &= b_4\Bigg[2 \big[\mathbf{B}^2\big]_{kl} -\frac{2}{(D+4)}\Big[(D+4)\big[\mathbf{B}^2\big]_{kl}\nonumber\\
    &+\mathrm{Tr}\big[\mathbf{B}^2\big]\delta_{kl}\Big]+\frac{2\mathrm{Tr}\big[\mathbf{B}^2\big]}{(D+4)}\delta_{kl}\Bigg]=0.
\end{align}
We have thus proved that $\mathbb{T}^{(4)}$ is traceless and symmetric.

\section{\texorpdfstring{Derivation of~$\vec{\nabla}_\mathcal{S}\big(\mathbb{T}^{(4)},\vec{\boldsymbol{\sigma}}^{\otimes 4}\big)$}{Derivation of T4 DERIVATIVE}}
\label{app20}

Similar to Appendix~\ref{app: 7}, we start by defining 
\begin{equation}
    \Big( \mathbb{T}^{(4)},\vec{\mathbf{r}}^{\otimes4} \Big) \equiv \sum_{i_1=1}^D\sum_{i_2=1}^D\sum_{i_3=1}^D \sum_{i_4=1}^DT^{(n)}_{i_1,i_2,i_3,i_4}  x_{i_1} x_{i_2} x_{i_3} x_{i_4},
\end{equation}
where $\vec{\mathbf{r}} = r\vec{\boldsymbol{\sigma}} = (x_1,x_2,\ldots,x_D)$, which immediately gives
\begin{equation}
    \Big( \mathbb{T}^{(4)},\vec{\mathbf{r}}^{\otimes4} \Big) = r^4\Big(\mathbb{T}^{(4)},\vec{\boldsymbol{\sigma}}^{\otimes4}\Big). \label{eq: T4 r4 to T4 sigma4}
\end{equation}
Computing the $k$-th component of $\Big( \mathbb{T}^{(4)},\vec{\mathbf{r}}^{\otimes4} \Big) $ in the Cartesian coordinates, we obtain using Eq.~\eqref{eq: G4} that
\begin{equation}
    \left[\vec{\nabla} \Big( \mathbb{T}^{(4)},\vec{\mathbf{r}}^{\otimes4} \Big)\right]_k = 4\sum_{i_2=1}^D\sum_{i_3=1}^D \sum_{i_4=1}^DT^{(4)}_{k,i_2,i_3,i_4} x_{i_2} x_{i_3} x_{i_4}, \label{eq: grad T4 CON k component}
\end{equation}
where $\vec{\nabla}$ is the full gradient operator. The right side of Eq.~\eqref{eq: grad T4 CON k component} can be thought as the $k$-th Cartesian component of the vector obtained when a rank-$4$ tensor $\mathbb{T}^{(4)}$ is contracted with $\vec{\mathbf{r}}\otimes\vec{\mathbf{r}}\otimes\vec{\mathbf{r}}$. In this notation, we may write
\begin{equation}
    \vec{\nabla} \Big( \mathbb{T}^{(4)},\vec{\mathbf{r}}^{\otimes4} \Big)=4 \Big( \mathbb{T}^{(4)},\vec{\mathbf{r}}^{\otimes3} \Big) = 4r^3 \Big( \mathbb{T}^{(4)},\vec{\boldsymbol{\sigma}}^{\otimes3} \Big). \label{eq: grad T4 CON }
\end{equation}

We now compute $\vec{\nabla} \big( \mathbb{T}^{(4)},\vec{\mathbf{r}}^{\otimes4} \big) $ in $D$-dimensional spherical-polar coordinates as defined in Appendix~\ref{app: 0}. Using Eq.~\eqref{eq: nabla app} along with Eq.~\eqref{eq: T4 r4 to T4 sigma4} and noting the fact that $\Big(\mathbb{T}^{(4)},\vec{\boldsymbol{\sigma}}^{\otimes4}\Big)$ is independent of $r$, we obtain
\begin{align}
    &\vec{\nabla} \big( \mathbb{T}^{(4)},\vec{\mathbf{r}}^{\otimes4} \big) = \bigg[\hat{\mathbf{r}}\frac{\partial}{\partial r} + \frac{1}{r}\vec{\nabla}_\mathcal{S}\bigg]\bigg[r^4\Big(\mathbb{T}^{(4)},\vec{\boldsymbol{\sigma}}^{\otimes4}\Big)\bigg]\nonumber\\
    &=4r^3\Big(\mathbb{T}^{(4)},\vec{\boldsymbol{\sigma}}^{\otimes4}\Big)\hat{\mathbf{r}}+r^3\vec{\nabla}_\mathcal{S}\Big(\mathbb{T}^{(4)},\vec{\boldsymbol{\sigma}}^{\otimes4}\Big).
\end{align}

Note that, by definition $\hat{\mathbf{r}} = \vec{\mathbf{r}}/r = \vec{\boldsymbol{\sigma}}$, using which we may write
\begin{equation}
    \vec{\nabla} \big( \mathbb{T}^{(4)},\vec{\mathbf{r}}^{\otimes4} \big) =4r^3\Big(\mathbb{T}^{(4)},\vec{\boldsymbol{\sigma}}^{\otimes4}\Big)\vec{\boldsymbol{\sigma}}+r^3\vec{\nabla}_\mathcal{S}\Big(\mathbb{T}^{(4)},\vec{\boldsymbol{\sigma}}^{\otimes4}\Big). \label{eq: grad T4 CON sph}
\end{equation}
Comparing Eqs.~\eqref{eq: grad T4 CON }~and~\eqref{eq: grad T4 CON sph}, we obtain
\begin{equation}
    \vec{\nabla}_\mathcal{S}\Big(\mathbb{T}^{(4)},\vec{\boldsymbol{\sigma}}^{\otimes4}\Big) = 4\Big[\Big( \mathbb{T}^{(4)},\vec{\boldsymbol{\sigma}}^{\otimes3} \Big)-\Big(\mathbb{T}^{(4)},\vec{\boldsymbol{\sigma}}^{\otimes4}\Big)\vec{\boldsymbol{\sigma}}\Big]. \label{eq: M7}
\end{equation}

We now express the right hand side of Eq.~\eqref{eq: M7} in terms of $\mathbf{B}$ and $\vec{\boldsymbol{\sigma}}$ using Eq.~\eqref{eq: TIJKL}. We will denote $b_4\Big(\mathrm{Tr}\big[\mathbf{B}^2\big],\mathrm{Tr}\big[\mathbf{B}^3\big],\ldots,\mathrm{Tr}\big[\mathbf{B}^D\big]\Big)$ by $b_4$ in the following computation. Focusing on the $ k$-th component of $\Big( \mathbb{T}^{(4)},\vec{\boldsymbol{\sigma}}^{\otimes3} \Big)$ in the Cartesian coordinates, we obtain
\begin{align}
    &\left[\Big( \mathbb{T}^{(4)},\vec{\boldsymbol{\sigma}}^{\otimes3} \Big)\right]_k = \sum_{i_2=1}^D\sum_{i_3=1}^D \sum_{i_4=1}^DT^{(4)}_{k,i_2,i_3,i_4} \sigma_{i_2} \sigma_{i_3} \sigma_{i_4} \nonumber\\
    &=b_4\sum_{i_2=1}^D\sum_{i_3=1}^D \sum_{i_4=1}^D \sigma_{i_2} \sigma_{i_3} \sigma_{i_4}\Bigg[\Big(B_{ki_2}B_{i_3i_4} + B_{ki_3}B_{i_2i_4}\nonumber\\
    &+B_{ki_4}B_{i_2i_3}\Big)-\frac{2}{(D+4)}\Big[\delta_{ki_2}\big[\mathbf{B}^2\big]_{i_3i_4}+\delta_{ki_3}\big[\mathbf{B}^2\big]_{i_2i_4} \nonumber\\
    &+\delta_{ki_4}\big[\mathbf{B}^2\big]_{i_2i_3}+\delta_{i_2i_3}\big[\mathbf{B}^2\big]_{ki_4}+\delta_{i_2i_4}\big[\mathbf{B}^2\big]_{ki_3}+\delta_{i_3i_4}\big[\mathbf{B}^2\big]_{ki_2}\Big]\nonumber\\
    &+\frac{2\mathrm{Tr}\big[\mathbf{B}^2\big]}{(D+2)(D+4)} \Big(\delta_{ki_2}\delta_{i_3i_4} + \delta_{ki_3}\delta_{i_2i_4}+\delta_{ki_4}\delta_{i_2i_3}\Big)\Bigg] \label{eq: M8}
\end{align}
Let us focus on Eq.~\eqref{eq: M8} term by term. Firstly,
\begin{align}
    &\sum_{i_2=1}^D\sum_{i_3=1}^D \sum_{i_4=1}^D \sigma_{i_2} \sigma_{i_3} \sigma_{i_4}B_{ki_2}B_{i_3i_4}= \bigg[\sum_{i_2=1}^DB_{ki_2}\sigma_{i_2} \bigg]\nonumber\\
    &\times\bigg[\sum_{i_3=1}^D \sum_{i_4=1}^D\sigma_{i_3}B_{i_3i_4}\sigma_{i_4}\bigg] = \big[\mathbf{B}\vec{\boldsymbol{\sigma}}\big]_k\big(\vec{\boldsymbol{\sigma}}\cdot\mathbf{B}\vec{\boldsymbol{\sigma}}\big).
\end{align}
Similarly, summing over the remaining two terms of the form $B_{kb}B_{cd}$, we will also get $\big[\mathbf{B}\vec{\boldsymbol{\sigma}}\big]_k\big(\vec{\boldsymbol{\sigma}}\cdot\mathbf{B}\vec{\boldsymbol{\sigma}}\big)$. Focusing on the terms of the form $\delta_{ka}\big[\mathbf{B}^2\big]_{cd}$, we obtain
\begin{align}
    \sum_{i_2=1}^D\sum_{i_3=1}^D \sum_{i_4=1}^D \sigma_{i_2} \sigma_{i_3} \sigma_{i_4}\delta_{ki_2}\big[\mathbf{B}^2\big]_{i_3i_4}= \big[\vec{\boldsymbol{\sigma}}\big]_k\big(\vec{\boldsymbol{\sigma}}\cdot\mathbf{B}^2\vec{\boldsymbol{\sigma}}\big).
\end{align}
Similarly, summing over the remaining two terms of the form $\delta_{ka}\big[\mathbf{B}^2\big]_{cd}$, we will also get $\big[\vec{\boldsymbol{\sigma}}\big]_k\big(\vec{\boldsymbol{\sigma}}\cdot\mathbf{B}^2\vec{\boldsymbol{\sigma}}\big)$. Focusing on the terms of the form $\delta_{ab}\big[\mathbf{B}^2\big]_{kc}$, we obtain
\begin{align}
    \sum_{i_2=1}^D\sum_{i_3=1}^D \sum_{i_4=1}^D \sigma_{i_2} \sigma_{i_3} \sigma_{i_4}\delta_{i_2i_3}\big[\mathbf{B}^2\big]_{ki_4}&= \big[\mathbf{B}^2\vec{\boldsymbol{\sigma}}\big]_k\big(\vec{\boldsymbol{\sigma}}\cdot\vec{\boldsymbol{\sigma}}\big)\nonumber\\
    &= \big[\mathbf{B}^2\vec{\boldsymbol{\sigma}}\big]_k.
\end{align}
Similarly, summing over the remaining two terms of the form $\delta_{ab}\big[\mathbf{B}^2\big]_{kc}$, we will also get $\big[\mathbf{B}^2\vec{\boldsymbol{\sigma}}\big]_k$. The three remaining terms are of the form $\delta_{ka}\delta_{bc}$. Summing over each of them, we obtain $\big[\vec{\boldsymbol{\sigma}}\big]_k$. Combining everything and putting it back into Eq.~\eqref{eq: M8}, we obtain
\begin{align}
    &\Big( \mathbb{T}^{(4)},\vec{\boldsymbol{\sigma}}^{\otimes3} \Big) \nonumber\\
    =&~ 3b_4\bigg[\Big(\vec{\boldsymbol{\sigma}}\cdot\mathbf{B}\vec{\boldsymbol{\sigma}}\Big)\mathbf{B}\vec{\boldsymbol{\sigma}}-\frac{2}{(D+4)}\bigg[\Big(\vec{\boldsymbol{\sigma}}\cdot\mathbf{B}^2\vec{\boldsymbol{\sigma}}\Big)\vec{\boldsymbol{\sigma}}+\mathbf{B}^2\vec{\boldsymbol{\sigma}}\bigg]\nonumber\\
    &+\frac{2\mathrm{Tr}\big[\mathbf{B}^2\big]}{(D+2)(D+4)}\vec{\boldsymbol{\sigma}}\bigg]. \label{eq: M12}
\end{align}

The expansion of $\Big(\mathbb{T}^{(4)},\vec{\boldsymbol{\sigma}}^{\otimes4}\Big)$ in terms of $\mathbf{B}$ and $\vec{\boldsymbol{\sigma}}$ is given in Eq.~\eqref{eq: T4 B4 S4 FULL}, which reads as
\begin{align}
    \Big(\mathbb{T}^{(4)},\vec{\boldsymbol{\sigma}}^{\otimes4}\Big) &= 3b_4\bigg[\Big(\vec{\boldsymbol{\sigma}}\cdot\mathbf{B}\vec{\boldsymbol{\sigma}}\Big)^2-\frac{4}{(D+4)}\Big(\vec{\boldsymbol{\sigma}}\cdot\mathbf{B}^2\vec{\boldsymbol{\sigma}}\Big)\nonumber\\
    &+\frac{2\mathrm{Tr}\big[\mathbf{B}^2\big]}{(D+2)(D+4)}\bigg]. \label{eq: M13}
\end{align}
Substituting Eqs.~\eqref{eq: M12}~and~\eqref{eq: M13} back into Eq.~\eqref{eq: M7}, we finally obtain
\begin{align}
    &\vec{\nabla}_\mathcal{S}\Big(\mathbb{T}^{(4)},\vec{\boldsymbol{\sigma}}^{\otimes4}\Big) \nonumber\\
    &= 12b_4\bigg[\Big(\vec{\boldsymbol{\sigma}}\cdot\mathbf{B}\vec{\boldsymbol{\sigma}}\Big)\mathbf{B}\vec{\boldsymbol{\sigma}}+\frac{2}{(D+4)}\Big(\vec{\boldsymbol{\sigma}}\cdot\mathbf{B}^2\vec{\boldsymbol{\sigma}}\Big)\vec{\boldsymbol{\sigma}}\nonumber\\
    &-\Big(\vec{\boldsymbol{\sigma}}\cdot\mathbf{B}\vec{\boldsymbol{\sigma}}\Big)^2\vec{\boldsymbol{\sigma}}-\frac{2}{(D+4)}\mathbf{B}^2\vec{\boldsymbol{\sigma}}\bigg].
\end{align}

\section{\texorpdfstring{Checking of Eqs.~\eqref{eq: 145}~and~\eqref{eq: 146}}{Checking of Eq. 170}}
\label{app: 21}

Using Eq.~\eqref{eq: F6}, we may write
\begin{equation}
    \nabla^2\Big(\vec{\boldsymbol{\sigma}}\cdot\mathbf{B}^2\vec{\boldsymbol{\sigma}}\Big)  = 2 \sum_{k=1}^D\left[\mathbf{B}^2\right]_{k,k} = 2 \mathrm{Tr}\left[\mathbf{B}^2\right], \label{eq: p1}
\end{equation}
where $\left[\mathbf{B}^2\right]_{k,k}$ is the $k$-th diagonal element of the matrix $\mathbf{B}^2$. Similarly, from Eq.~\eqref{eq: F7}, we obtain
\begin{align}
    \nabla^2\Big(\vec{\boldsymbol{\sigma}}\cdot\mathbf{B}^2\vec{\boldsymbol{\sigma}}\Big) = 2D\Big(\vec{\boldsymbol{\sigma}}\cdot\mathbf{B}^2\vec{\boldsymbol{\sigma}}\Big)+\nabla^2_\mathcal{S}\Big(\vec{\boldsymbol{\sigma}}\cdot\mathbf{B}^2\vec{\boldsymbol{\sigma}}\Big).\label{eq:p2}
\end{align}
Comparing Eqs.~\eqref{eq: p1}~and~\eqref{eq:p2} gives
\begin{equation}
    \nabla^2_\mathcal{S}\Big(\vec{\boldsymbol{\sigma}}\cdot\mathbf{B}^2\vec{\boldsymbol{\sigma}}\Big)=-2D\bigg[\Big(\vec{\boldsymbol{\sigma}}\cdot\mathbf{B}^2\vec{\boldsymbol{\sigma}}\Big)-\frac{1}{D}\mathrm{Tr}\left[\mathbf{B}^2\right]\bigg]. \label{eq: P3}
\end{equation}
Noting that $\mathrm{Tr}\left[\mathbf{B}^2\right]$ is independent of $\vec{\boldsymbol{\sigma}}$ and therefore $\nabla^2_\mathcal{S}\mathrm{Tr}\left[\mathbf{B}^2\right]=0$, we may rewrite Eq.~\eqref{eq: P3} as
\begin{equation}
    \bigg[\nabla^2_\mathcal{S}+2D\bigg]\bigg[\Big(\vec{\boldsymbol{\sigma}}\cdot\mathbf{B}^2\vec{\boldsymbol{\sigma}}\Big)-\frac{1}{D}\mathrm{Tr}\left[\mathbf{B}^2\right]\bigg]=0.
\end{equation}
This immediately gives from the definition~\eqref{eq: proj def} of the projector operator $\mathcal{P}_4$ that
\begin{equation}
    \mathcal{P}_4\bigg[\Big(\vec{\boldsymbol{\sigma}}\cdot\mathbf{B}^2\vec{\boldsymbol{\sigma}}\Big)-\frac{1}{D}\mathrm{Tr}\left[\mathbf{B}^2\right]\bigg]=0. \label{eq: proj 2 spin}
\end{equation}
Equation~\eqref{eq: proj 2 spin} is the desired Eq.~\eqref{eq: 145} of the main text.

From Eq.~\eqref{eq: F8}, we have
\begin{equation}
    \nabla^2_\mathcal{S}\Big(\mathbb{T}^{(4)},\vec{\boldsymbol{\sigma}}^{\otimes 4}\Big) = -4(D+2)\Big( \mathbb{T}^{(n)},\vec{\boldsymbol{\sigma}}^{\otimes n}\Big).
\end{equation}
Hence, we have
\begin{equation}
    \Big[\nabla^2_\mathcal{S}+2D\Big]\Big(\mathbb{T}^{(4)},\vec{\boldsymbol{\sigma}}^{\otimes 4}\Big) = -2(D+4)\Big( \mathbb{T}^{(n)},\vec{\boldsymbol{\sigma}}^{\otimes n}\Big).
\end{equation}
From the definition of the projector operator $\mathcal{P}_4$ from Eq.~\eqref{eq: proj def}, we may write
\begin{align}
    \mathcal{P}_4\Big(\mathbb{T}^{(4)},\vec{\boldsymbol{\sigma}}^{\otimes 4}\Big) &= \frac{\nabla^2_\mathcal{S}\Big[\nabla^2_\mathcal{S}+2D\Big]}{8(D+2)(D+4)}\Big(\mathbb{T}^{(4)},\vec{\boldsymbol{\sigma}}^{\otimes 4}\Big)\nonumber\\
    & =\frac{1}{4(D+2)} \nabla^2_\mathcal{S}\Big(\mathbb{T}^{(4)},\vec{\boldsymbol{\sigma}}^{\otimes 4}\Big) \nonumber\\
    &=\Big(\mathbb{T}^{(4)},\vec{\boldsymbol{\sigma}}^{\otimes 4}\Big). \label{eq: proj 4 spin}
\end{align}
Equation~\eqref{eq: proj 4 spin} is the desired Eq.~\eqref{eq: 146} of the main text.

\bibliography{new}
\bibliographystyle{unsrt}
\end{document}